\PassOptionsToPackage{table,xcdraw}{xcolor}
\documentclass[acmlarge,authorversion]{acmart}

\AtBeginDocument{%
  \providecommand\BibTeX{{%
    \normalfont B\kern-0.5em{\scshape i\kern-0.25em b}\kern-0.8em\TeX}}}
    
\def\BibTeX{{\rm B\kern-.05em{\sc i\kern-.025em b}\kern-.08em
    T\kern-.1667em\lower.7ex\hbox{E}\kern-.125emX}}

\setcopyright{acmcopyright}
\acmJournal{TOSEM}
\acmYear{2021} 
\acmVolume{1} 
\acmNumber{1} 
\acmArticle{1} 
\acmMonth{1} 
\acmPrice{15.00}
\acmDOI{10.1145/3490488}

\usepackage{algorithmic}
\usepackage{graphicx}
\usepackage{textcomp}
\usepackage{xcolor}
\usepackage{color}
\usepackage{multirow}
\usepackage{subfigure}
\usepackage{CJKutf8}
\usepackage{wrapfig}
\usepackage{hyperref}
\usepackage{balance}
\usepackage{mathrsfs}
\usepackage{adjustbox}
\usepackage[framemethod=TikZ]{mdframed}

\definecolor{mygreen}{HTML}{02818a}

\mdfdefinestyle{MyFrame}{%
    linecolor=black,
    outerlinewidth=0.3pt,
    roundcorner=5pt,
    skipabove = 10.5pt,
    innertopmargin=10.5pt, 
    innerbottommargin=10.5pt, 
    innerrightmargin=10.5pt,
    innerleftmargin=10.5pt,
    backgroundcolor=gray!25!white
    }

\begin{document}
\title{SemMT: A Semantic-based Testing Approach for Machine Translation Systems}


\author{Jialun CAO}
\email{jcaoap@cse.ust.hk}
\affiliation{%
  \institution{The Hong Kong University of Science and Technology}
  \city{Hong Kong}
  \country{China}
}

\author{Meiziniu LI}
\email{mlick@cse.ust.hk}
\affiliation{%
  \institution{The Hong Kong University of Science and Technology}
  \city{Hong Kong}
  \country{China}
}

\author{Yeting LI}
\email{liyt@ios.ac.cn}
\affiliation{%
  \institution{State Key Laboratory of Computer Science, Institute of Software,
Chinese Academy of Sciences,
University of Chinese Academy of Sciences}
  \city{Beijing}
  \country{China}
}

\author{Ming Wen*}
\email{mwenaa@hust.edu.cn}
\affiliation{%
  \institution{Huazhong University of Science and Technology}
  \city{Wuhan}
  \country{China}
}

\author{Shing-Chi Cheung*}
\thanks{* Corresponding author.}
\email{scc@cse.ust.hk}
\affiliation{
  \institution{The Hong Kong University of Science and Technology}
  \city{Hong Kong}
  \country{China}
}

\author{Haiming Chen}
\email{chm@ios.ac.cn}
\affiliation{%
  \institution{State Key Laboratory of Computer Science, Institute of Software,
Chinese Academy of Sciences,
University of Chinese Academy of Sciences}
  \city{Beijing}
  \country{China}
}

\renewcommand{\shortauthors}{J. CAO et al.}

\begin{abstract}
Machine translation has wide applications in daily life. In mission-critical applications such as translating official documents, incorrect translation can have unpleasant or sometimes catastrophic consequences. This motivates recent research on the testing methodologies for machine translation systems. Existing methodologies mostly rely on metamorphic relations designed at the textual level (e.g., Levenshtein distance) or syntactic level (e.g., distance between grammar structures) to determine the correctness of translation results. However, these metamorphic relations do not consider whether the original and the translated sentences have the same meaning (i.e., semantic similarity). 
To address this problem, in this paper, we propose SemMT, an automatic testing approach for machine translation systems based on semantic similarity checking. 
SemMT applies round-trip translation and measures the semantic similarity between the original and the translated sentences. 
Our insight is that the semantics concerning logical relations and quantifiers
in sentences can be captured by regular expressions (or deterministic finite automata) where efficient semantic equivalence/similarity checking algorithms can be applied.
Leveraging the insight, we propose three semantic similarity metrics and implement them in SemMT. 
We compared SemMT with related state-of-the-art testing techniques, demonstrating the effectiveness of mistranslation detection.
The experiment results show that SemMT outperforms existing metrics, achieving an increase of 34.2\% and 15.4\% on accuracy and F-Score, respectively. 
We also study the possibility of further enhancing the performance by combining various metrics.
Finally, we discuss a solution to locate the suspicious trip in round-trip translation, which provides hints for bug diagnosis.
\end{abstract}


\maketitle

\section{Introduction}
\label{sec:1}
Machine translation systems, which provide automatic translation for text and speech from a \textit{source language} to another \textit{target language}, are widely used in daily lives~\cite{HazelwoodBBCDDF18,SIT}. However, machine translation systems can give incorrect or inappropriate translations, which could lead to harmful consequences such as embarrassment, misunderstanding, financial loss, or political conflicts~\cite{Attack17,Mistranslations15, LittelTranslation16,fbApologize17}. 
This motivates the research on testing methodologies to assure the quality of machine translation software. 

\begin{figure}[t!]
    \centering
    \includegraphics[width=1.0\linewidth]{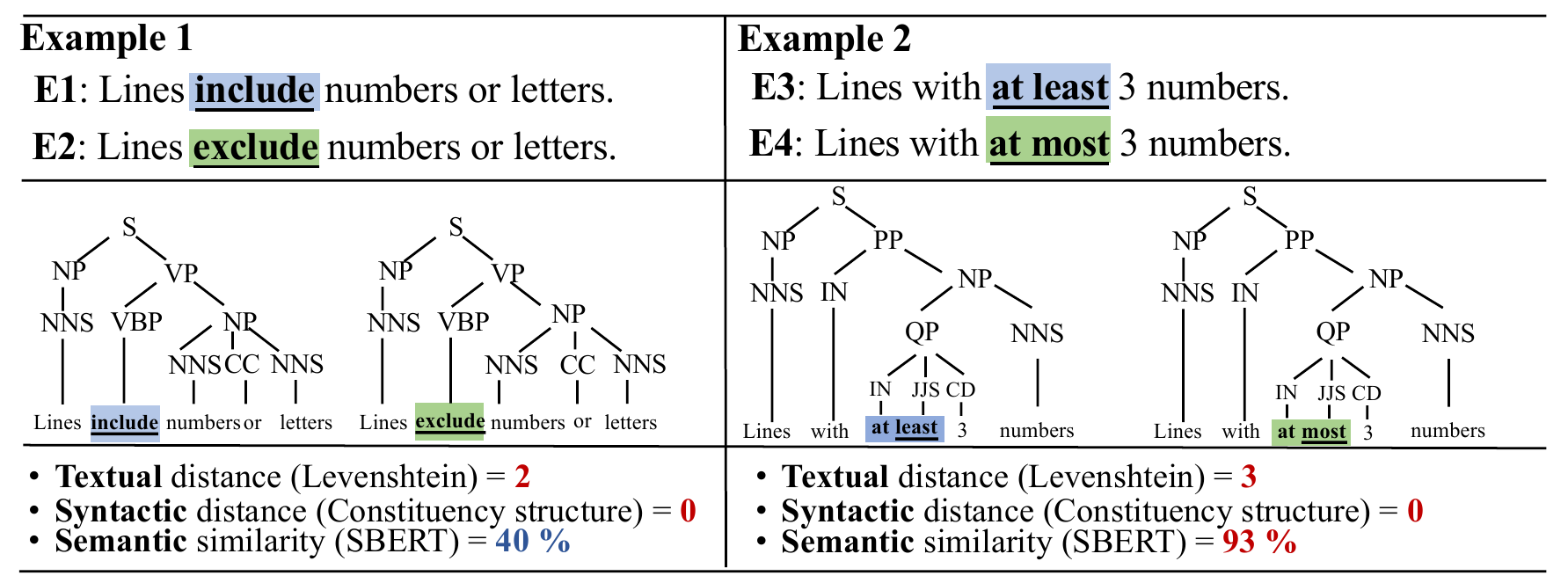}
    \caption{\textbf{Motivating Examples.} Examples of semantic differences that cannot be captured by existing textual, syntactic or semantic similarity metrics.}
    \label{fig:example}
\end{figure}

Recent works~\cite{monte18,MT4MT,SIT,transrepair20,pathology,DBLP:conf/dsn/WangZLZZDYHX19} on machine translation testing mostly adopt the metamorphic testing approach~\cite{metamorphic,chen2018metamorphic}.
The intuition is that similar sentences should be similarly translated~\cite{transrepair20,MT4MT,SIT}, while sentences with different meanings should not have the same translations~\cite{pathology}. 
Mistranslations are detected by examining the textual (e.g., Levenshtein distance) or syntactic
similarities between the original sentence and the translated sentence.
However, a close textual or syntactic distance between two sentences does not necessarily imply close semantic meaning, and thus cannot guarantee the correctness of translation. For example, in Fig.~\ref{fig:example} (Example 1), the verbs ``include'' and ``exclude'' are logically opposite, though their textual/syntactic distances are the same or close. 
Consequently, while the textual and syntactic distances between the two sentences are close, 
they deliver opposite semantic meanings.
Using semantic-based metrics such as the state-of-the-art SBERT~\cite{SBERT19}, the similarity measured over the two sentences in Example 1 is low (i.e., only 40\%), which indicates that the differences can be more precisely captured in this case. However, SBERT does not necessarily perform well in all cases.
In the Example 2 of Fig.~\ref{fig:example}, SBERT gives a high semantic similarity (93\%) although the phrase ``at least'' and ``at most'' in E3 and E4 are two opposite quantifiers. 
The two examples suggest that a better metric to measure sentence semantics is needed.

However, the semantic meaning of sentences is hard to be precisely captured, which exacerbates the challenges of measuring translation correctness.
It is non-trivial to judge the semantic equivalence/similarity between two sentences even for human~\cite{ZhouXC16}. The judgement made by human can be subjective~\cite{monte18,MT4MT}, making the decision varies across different individuals. 
Besides, the flexibility of natural language makes this problem even more challenging~\cite{MT4MT}. For example, a token or phrase may have multiple correct translations, in such cases, modern translation software does not perform well~\cite{SIT}. 

\begin{figure}[t!]
    \centering
    \includegraphics[width=1.0\linewidth]{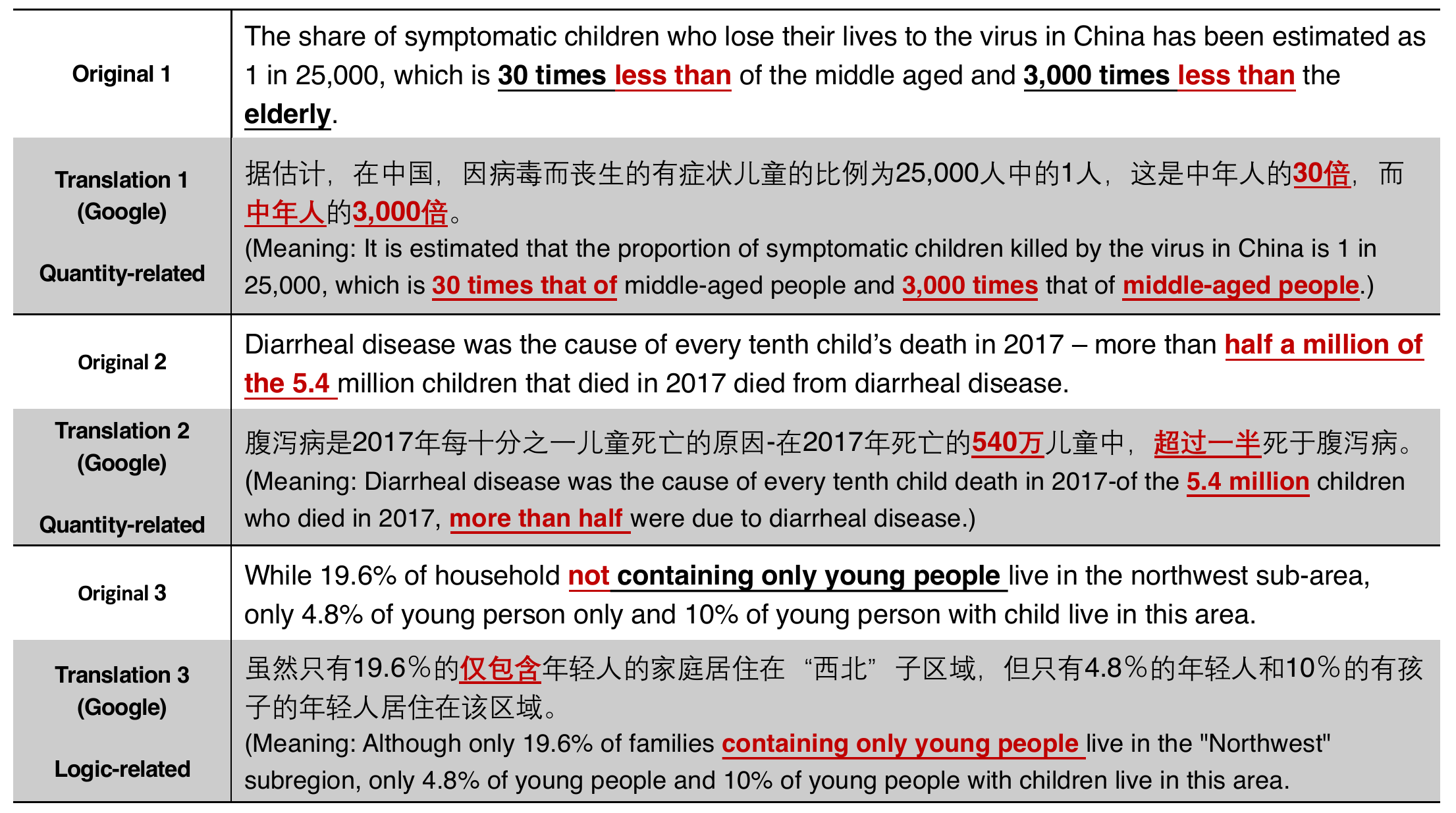}
    \caption{Examples of Mistranslation of Sentences with Quantifiers and Logical Relations.
    }
    \label{fig:quantity-example}
\end{figure}

In view of these challenges, we then approach the problem by
confining the scope to the semantics that concern quantifiers and logical relations (like examples in Fig.~\ref{fig:example}), and then test translation on sentences with such semantics.
For what it may worth to mention, sentences with quantifiers~\cite{ACL18some,ACL18quantification} and logical relations-\cite{semanticRelation2010,ACL20logic,emnlp20causual} are pervasive in the daily life and of central importance in linguistic semantics \cite{ACL18some,barwise1981generalized,peters2006quantifiers,Szymanik16quantifiers}. According to our investigation (see Section \ref{sec:2}), in every six sentences, there is one sentence involving quantifier{s} or logical relations, reflecting the prevalent use of such sentences in daily life. Since they mainly specify the quantity and logical meaning of the objects, and are commonly found in legal contracts, financial statements, healthcare reports, product instructions and so on, the mistranslation regarding them can cause misunderstanding or severe consequences. For example, in Fig.~\ref{fig:quantity-example}~\footnote{The translation results were collected on 
December 16, 2020 on Google Translator.}, the quantifier
``30 times less than'' in the first sentence (i.e., Original 1) is mistranslated to ``30 times'' by Google translator (i.e., Translation 1). The translation mistakenly reports the children's morbidity and mortality. It can result in unnecessary public panic. Note that the original sentence is taken from a brief policy launched recently\cite{policy20} on the impact of COVID-19 on children. Similarly, in the second example excerpted from a report on Diseases and Mortality~\cite{owiddiarrhealdiseases},
the number of children who suffered from disease is exaggerated after translation. Such mistranslations can cause severe consequences for the public. Apart from mistranslation of quantifiers, mistranslation of logical relations are also common. The third example (i.e., Original 3) taken from ~\cite{housingneeds} in Fig.~\ref{fig:quantity-example} illustrates the mistranslation of {a} logical relation, conveying opposite meaning (i.e., exclusion and inclusion) before and after translation, resulting in a wrong cognition of the household situation in northwest. Earlier work~\cite{pathology} has proposed to detect mistranslations based on an intuition that semantically different sentences should not have the same translation results. However, such detection technique may not be effective because many mistranslated sentences and their semantically mutated ones do have different translation results. For example, the sentence given by Original 3 in Fig~\ref{fig:quantity-example} is mistranslated to Chinese by Google. Yet, Google gives a different Chinese translation when the word ``not'' is removed from the sentence, escaping such detection of mistranslation based on the same translation results. The same situation occurs in the translation of the other two sentences in Fig~\ref{fig:quantity-example} (i.e., replacing ``less'' with ``more'' in Original 1, and replacing reversely in Original 2).
As such, it is important to design an effective test methodology for the translation of sentences that contain quantifiers and logical relations. Such testing methodologies have not been studied in prior work. 

In this paper, we propose SemMT, an automatic testing approach for machine translation systems based on semantic similarity checking of the concerned quantifiers and logical relations. The insight of SemMT is that the semantics concerning logical relations and quantifiers in sentences can be captured by regular expressions (or deterministic finite automata) where efficient equivalence/similarity checking algorithms are available. To be more specific, SemMT addresses the difficulty of capturing semantics similarities precisely and detecting semantic mistranslation on translation systems using the following three strategies:

\textbf{Transformation to regular expressions.}
Since the semantics regarding quantifier{s} and logical relations in sentences can be captured using regular expressions, the core step of SemMT is to transform the sentences to the semantic-equivalent regular expressions. This strategy enables the maximized precise semantic similarity capturing. 
As a well-explored, -tested and widely-used context-free grammar~\cite{wild18,largescale19,regexPython16,arewethere18,Rex10}, 
the semantics over regular expression (abbrev. regex) can be evaluated under a context-free paradigm, enabling us to capture and quantify semantic similarities precisely.

\textbf{Precise semantic similarity capturing.} 
Based on the above strategy, the semantic similarities over regular expressions then can be captured and quantified via well-established algorithms in formal language~\cite{DBLP:books/daglib/0016921,semregex18,softregex19}. 
If the semantic similarity quantified by these algorithms between the original
and translated sentences falls beyond the similarity threshold, such translations will be considered as suspicious mistranslations.

\textbf{Semantic checking in the same language.} 
Semantic equivalence can hardly be measured automatically across different languages~\cite{DBLP:conf/cscw/YamashitaI06}. 
Therefore, SemMT performs testing on round-trip translation, which translates a original sentence to another language, and then translates it back. In this way, the back-translated sentence and the original sentence are in the same language, allowing their semantics to be uniformly measured and compared.

On top of that, we implemented SemMT and compared with the state-of-the-art testing techniques. The experiment results show that SemMT achieves an increase of 23\%
in terms of F-Score as compared with other similarity measures. SemMT outperforms the state-of-the-art techniques, achieving an improvement of 34.2\% in accuracy with comparable amount of mistranlations detected. 
Besides, SemMT can detect 213 bugs in Google Translator as compared with 173 detected by other similarity metrics.
We also study the possibilities of improving accuracy and F-Score by combining different similarity metrics. Note that a mistranlation detected by SemMT may occur at the forward translation or the backward translation~\cite{monte18,MT4MT,goodfor05}. We discuss a method to locate the translation in which the bug resides. In addition, we discuss the types of bugs detected by SemMT. 

To sum up, this paper addresses the oracle problem in testing machine translation systems with respect to their semantics. Specifically, it makes the following four main contributions: 

\begin{itemize}
    \item We propose SemMT, a semantic-based machine translation testing framework. Specifically, it captures the semantics of quantifiers and logical relations during translation by semantic-equivalent regular language and detects mistranslation accordingly. To best of our knowledge, it is the first testing methodology proposed for machine translation systems based on the semantic similarity.
    \item We introduce an approach to determining and quantifying semantic differences. Specifically, we transform natural language to semantic-equivalent regular expressions, and then propose a metric to measure semantic similarities in formal ways.
    \item The experiment results show that our proposed metrics are more effective than existing similarity metrics in capturing the semantics of sentences that concern quantifiers and logical relations. Using the proposed metric, SemMT outperforms state-of-the-art techniques, achieving an increase of 34.2\% in accuracy and 15.4\% in F-Score. 
    \item We perform the first study to improve test effectiveness for machine translations via combining multiple similarity metrics. 
    Experiment results show that improvements can be achieved when proper metrics are combined. The experiment data and tool are publicly available~\cite{SemMT}. 
\end{itemize}

\section{Motivation}\label{sec:2}
Measuring the semantic difference/similarity of natural languages is still an open problem due to their 
subjective~\cite{monte18,MT4MT}, flexible~\cite{MT4MT} and context-aware~\cite{context-aware-acl19,DBLP:conf/emnlp/LaubliS018} features. We, therefore, confine our measurement to
the semantics of quantifiers and logical relations, which can be represented formally by regular expressions.
A \textbf{quantifier} is a word/phrase that usually goes before a noun to express the quantity of the object. It is traditionally defined using set-theoretic terms in linguistic theories~\cite{lobner1987quantification,bach2013quantification,kurtzman1993resolution,ACL18some,ACL18quantification}. Commonly used quantifiers include proportional quantifiers (e.g., ``some'', ``a few'', ``many'' and ``more than half''), logic quantifiers (e.g., ``none''), and quantifiable quantifiers (e.g., ``more than half'' and ``more than 3 times''). We refer to those sentences that contain quantifiers as {\it quantifier-related sentences}. For \textbf{logical relations}, we follow existing works~\cite{potts2007logic,logic_in_linguistics,partee1990mathematical,sharvy1980more,mcdaniel2004modal,mellor2006wholes} and focus on those semantics expressed in first-order logic such as conjunction, disjunction, negation, inclusive and exclusive relations, such as ``X {contains} Y''. We refer to those sentences that contain such logical semantics as {\it logic-related sentences}. 

Since we confine our scope to quantifier- and logic-related sentences, a legitimate question is: {\it Are quantifier- and logic-related sentences common?} To answer this question, we collected five large-scale corpora (i.e., Europarl~\cite{koehn2005europarl}, CommonCrawl and News available from the Workshop on Machine Translation (WMT)~\cite{WMT13}, News Commentary Parallel Corpus~\cite{Newscommentary}, Financial News from Reuters~\cite{Reuters14}) which are commonly-used for machine translation and other natural language processing tasks, and analyzed the proportion of sentences that involve quantifiers or logic relations. The corpora that we analyzed cover a broad range from policy, finance to daily news, which indicates that the findings made based on them can be generalized. We followed the methodology of earlier work~\cite{ACL18some, ahmad1992semantic} to identify quantifier-related sentences and logic-related sentences. 
The list of patterns that we used for the analysis is publicly available~\cite{SemMT}. 

Table~\ref{tab:prevalence} shows the statistics of the investigation. Initially, over 4 million sentences were collected. After filtering out those containing less than 10 words (e.g., the sentence ``I understand''),\footnote{{In the natural language processing area, a common practice~\cite{DBLP:conf/wmt/KurfaliO19,DBLP:conf/acl/AulamoVT20,vazquez2021helsinki} is to set such filtering condition as 5 to 10 according to the requirements of different tasks and corpus. We set the threshold as 10, which accounts for only 0.87\% sentences.}} the statistics were conducted over 3.7 million sentences. 
According to Table~\ref{tab:prevalence}, there are 639,179 (16.96\%) sentences that are quantifier- or logic-related. It reflects the popular use of such sentences in daily life. For a further breakdown, 12.52\% sentences are quantifier-related, and 5.22\% are logic-related, respectively. 

While not all of these quantifiers and logical relations can be precisely captured by regexes (e.g., ``a few'' and ``many'' cannot be quantified by an exact number), we further measured the proportion of quantifiers and logical relations that can be precisely captured by regexes.
As revealed by Table~\ref{tab:prevalence}, $89.6\%$ (=15.19/16.96) of the quantifier- and logic-related sentences in the selected dataset can be captured precisely. A closer examination reveals that $90.8\%$ of the quantifier- and $83.9\%$ of the logic-related sentences can be precisely captured. 
Such high ratios reflect the fact that quantifier- or logic-related sentences that can be captured precisely are also pervasive, and the focus on such sentences will not largely narrow the scope of application of our approach to detect mistranslation bugs. Besides, we also discuss the situation where the semantics can be captured approximately in \S\ref{sec:approx}.

\begin{table}[thbp]
\centering
\caption{\textbf{Prevalence of Quantifier and Logic-related Sentences in Real-World Corpora.} The table shows the total number of sentences before (\#Total) and after (\# Filtered) filtering, the number and ratio of sentences with quantifiers (\# Qty, \% Qty), with logical relations (\# Lgc, \% Lgc), and with either one (\# Related, \% Related). Those quantifier or logic-related sentences that can be quantified are labeled with subscript ${PRS}$.}
\label{tab:prevalence}
\renewcommand\arraystretch{1.5}
\resizebox{1.0\linewidth}{!}{%
\begin{tabular}{l|r|r|rr|ll|rr|ll|rr|ll}
\hline
\multirow{2}{*}{Corpora} & \multicolumn{1}{l|}{\multirow{2}{*}{\# Total}} & \multicolumn{1}{l|}{\multirow{2}{*}{\# Filtered}} & \multicolumn{4}{c|}{Quantity-related}                             & \multicolumn{4}{c|}{Logic-related}                               & \multicolumn{4}{c}{Related}                                                                      \\ \cline{4-15} 
                         & \multicolumn{1}{l|}{}                                & \multicolumn{1}{l|}{}                                                                                      & \# Qty             & \# Qty$_{PRS}$       & \% Qty            & \% Qty$_{PRS}$     & \# Lgc            & \# Lgc$_{PRS}$       & \% Lgc           & \% Lgc$_{PRS}$    & \# Related & \# Related$_{PRS}$       & \% Related            & \% Related$_{PRS}$     \\ \hline
ReutersNews              &     285,964                                           & 243,664                                                                                                    & 53,453           & 51,627           & 21.94\%          & 21.19\%          & 12,061           & 10,394           & 4.95\%          & 4.27\%          & 62,119                   & \multicolumn{1}{r|}{59,107}           & 25.49\%          & 24.26\%          \\
WMT-News                 & 14,522                                            & 12,219                                                                                                     & 2,331            & 2,189            & 19.08\%          & 17.91\%          & 507              & 443              & 4.15\%          & 3.63\%          & 2,711                    & \multicolumn{1}{r|}{2,526}            & 22.19\%          & 20.67\%          \\
CommonCrawl              & 1,845,286                                              & 1,640,632                                                                                                  & 199,945          & 175,989          & 12.19\%          & 10.73\%          & 93,572           & 77,653           & 5.70\%          & 4.73\%          & 279,598                  & \multicolumn{1}{r|}{243,909}          & 17.04\%          & 14.87\%          \\
Commentary               & 174,441                                            & 153,258                                                                                                    & 21,892           & 19,111           & 14.28\%          & 12.47\%          & 6,876            & 5,867            & 4.49\%          & 3.83\%          & 27,633                   & \multicolumn{1}{r|}{24,210}           & 18.03\%          & 15.80\%          \\
Europarl                 & 1,965,734                                              & 1,718,060                                                                                                  & 194,110          & 179,657          & 11.30\%          & 10.46\%          & 83,730           & 70,744           & 4.87\%          & 4.12\%          & 267,118                  & \multicolumn{1}{r|}{242,436}          & 15.55\%          & 14.11\%          \\ \hline
\textbf{Total}           & 4,285,947                                            & 3,767,833                                                                                                  & \textbf{471,731} & \textbf{428,573} & \textbf{12.52\%} & \textbf{11.37\%} & \textbf{196,746} & \textbf{165,101} & \textbf{5.22\%} & \textbf{4.38\%} & \textbf{639,179}         & \multicolumn{1}{r|}{\textbf{572,188}} & \textbf{16.96\%} & \textbf{15.19\%} \\ \hline
\end{tabular}
}
\end{table}

\section{Approach}
In this section, we first give an overview of SemMT, followed by the explanation of its methodology.

\subsection{Overview}
\label{sec:overview}

\begin{figure*}[t!]
    \centering
    \includegraphics[width=1.0\textwidth]{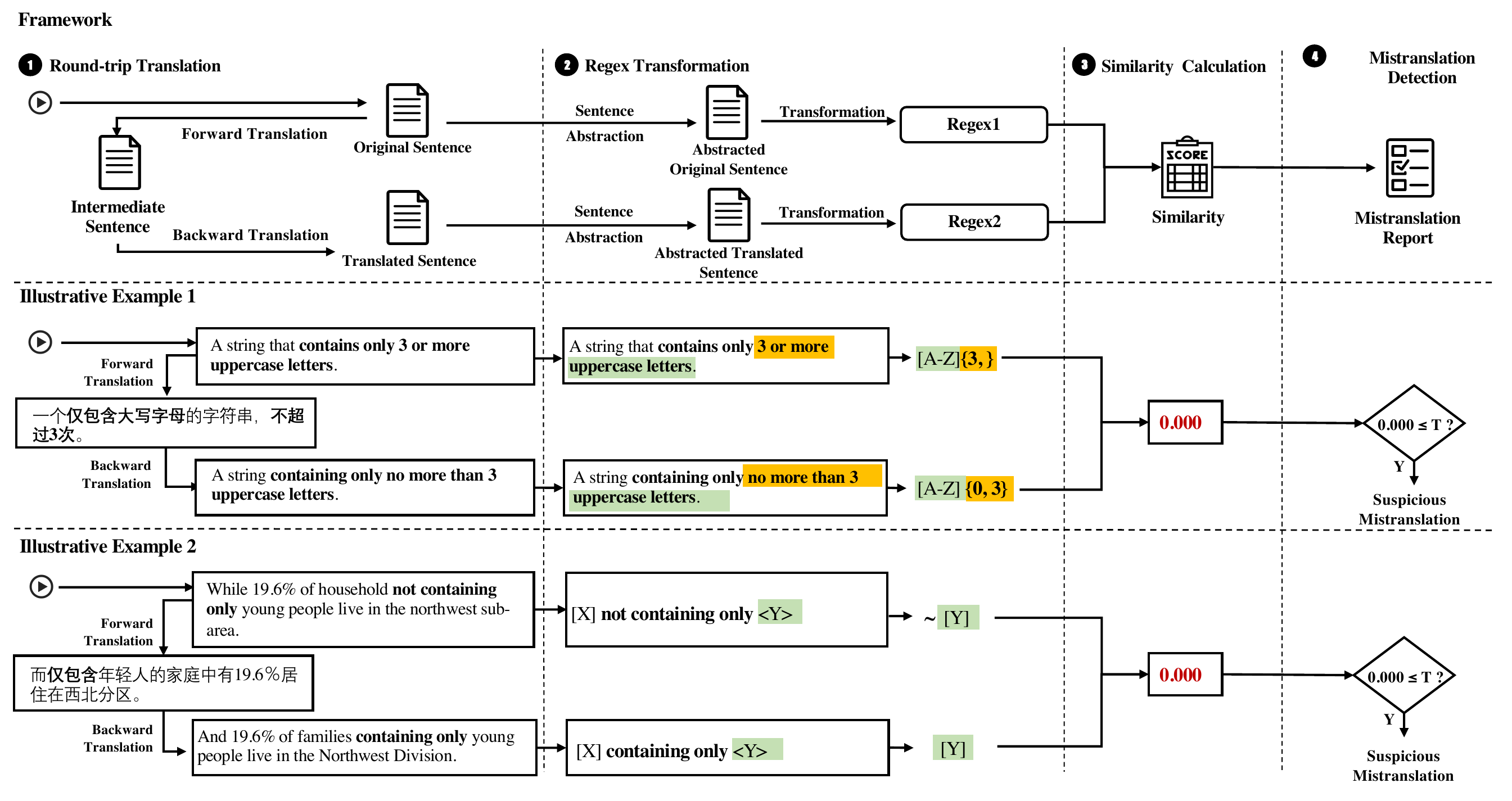}
    \caption{\textbf{Framework and Two Illustrative Examples.} The words highlighted in yellow are quantifiers that modify the object words highlighted in green.}
    \label{fig:framework}
\end{figure*}

Fig.~\ref{fig:framework} illustrates the framework of SemMT with two illustrative examples. The upper diagram in Fig.~\ref{fig:framework} shows the general workflow of SemMT. Given the original sentences in one language, SemMT detects mistranslation in four steps: (1) conducting the round-trip translation to collect the intermediate and translated sentences, (2) abstracting and transforming the {original} and translated sentences into regular expressions using existing tools, (3) calculating the semantic similarity between the regular expressions based on a set of regex-related metrics and (4) detecting mistranslation according to customized thresholds. Finally, the detected mistranslations are reported.

The middle and lower diagrams demonstrate two mistranslations that can be detected by SemMT~\footnote{The translation results were collected on December 17, 2020 on Google Translator.}. The first original sentences is taken from the NL-RX-Synth dataset~\cite{deepregex16}, and the second one is taken from an online housing needs document~\cite{housingneeds} (same as the third example in Fig.~\ref{fig:example}). Given the original sentences, the round-trip translation is firstly conducted by the forward translation from the source language (take English as example) to the intermediate language (take Chinese as example), then backward translation to the source language. Secondly, the semantics regarding the quantifiers and logical relations are identified from the original and translated sentences and transformed into the corresponding regular expressions. The purpose of sentence abstraction is to capture the quantifiers and logical relations as precise as possible, and the details of sentence abstraction and regular expression transformation are explained in \S\ref{sec:abstraction} and \S\ref{sec:transformation}. After the second step, the semantics of the original and translated sentences can be captured by regular expressions. Take Illustrative Example 1 as an example, the meaning of two English sentences can be described by the corresponding two regular expressions \verb|[A-Z]{3,}| and \verb|[A-Z]{0,3}|, where \verb|[A-Z]| represents an uppercase letter, and quantifiers \verb|{3,}| and \verb|{0,3}| serve as quantifiers ``3 or more'' and ``no more than 3'', respectively.
Next, semantic similarity is calculated over the two regular expressions by regex-based semantic similarity metrics. In particular, SemMT utilized three semantic similarity metrics to differentiate the semantic meanings. If the similarity is higher than the predefined threshold, the semantic meaning of the original sentence is well-preserved after translation; otherwise, SemMT reports it as a candidate round-trip mistranslation. As shown in the illustrative examples, take DFA-based similarity (details can be found in \S~\ref{sec:dfa})
as example, the semantic similarities of two examples are relatively low (0.000) for both translations, SemMT, therefore, reports the original, intermediate and and translated sentences as a potential round-trip mistranslation. 

In the following of this section, we will explain the details of the four steps using two illustrative examples, showing how SemMT is able to capture semantic difference and detect mistranslations in round-trip translations. Challenges arise in the measurement of similarity between regexes and the determination of thresholds. We will discuss them in \S\ref{sec:transformation}, \S\ref{sec:similarity} and \S\ref{sec:detection}.

\subsection{Round-Trip Translation}
Round trip translation (RTT) is also known as back-and-forth translation. It translates a given text or sentence into an intermediate language (the forward translation), and then translates the result back into the source language (the
back translation)~\cite{goodfor05}. 
It reflects the general quality of a translation system over longer texts~\cite{goodfor05} and is a cost-effective choice~\cite{monte18,MT4MT,revisiting20} to derive reference translation automatically.

The benefit of adopting RTT in our methodology is that the semantics of the original and back-translated sentences can be uniformly measured and compared in the same language. The selection of source
and intermediate languages are not restricted by our methodology. Yet, the selection of language pairs will affect the effectiveness of methodology in two aspects. First, in some languages (e.g., Chinese and Japanese), nouns are the same in both singular and plural forms, while they are in different forms in other languages like English and Russian. The RTT between these two kinds of languages may lose/switch the singularity or plurality information, making the semantics changed.
Second, the availability of automated transformation approaches used in the following steps (i.e., the second step in Fig.~\ref{fig:framework})
also needs to be considered. If the automatic approaches of deriving regex from natural language sentences in the source language is unavailable, the workflow is unlikely to proceed automatically. 
Hence the selection of language pairs under test should take these two aspects into account. Besides, since SemMT adopts a black-box testing manner, so the translators under test can be either open- or close-sourced. 

Nevertheless, one may concern that RTT involves testing two systems instead of one~\cite{monte18,MT4MT} and it is unclear which trip is buggy when a sentence is mistranslated. It subsequently motivates us to explore the possibility to locate the buggy trip automatically, and we discuss one possible solution in Section \ref{sec:localization}.

\subsection{Regular Expression and Deterministic Finite Automaton Transformation}
\label{sec:transformation}
SemMT is empowered by the recent advances made on the synthesis of regular expressions from natural languages using rule-based~\cite{ranta1998multilingual,KushmanB13} and learning-based~\cite{deepregex16,semregex18,softregex19} approaches. In this subsection, we present the key ideas of how these techniques can be applied in the methodology of SemMT.
In the following, we adapt two illustrative examples in Fig.~\ref{fig:framework} as running examples to show how the resulting transformed regexes and deterministic finite automata are derived together with the introduction of the related formalism.

\begin{itemize}
\item \textbf{Running Example 1 (Original Sentences)}
\begin{itemize}
    \item S1: A string that contains only \textbf{3 or more} uppercase letters.
    \item S2: A string containing only \textbf{no more than 3} uppercase letters.
\end{itemize}

\item \textbf{Running Example 2 (Original Sentences)}
\begin{itemize}
    \item S3: While 19.6\% of household \textbf{not containing only} young people live in the northwest sub-area.
    \item S4: And 19.6\% of families \textbf{containing only} young people live in the Northwest Division.
\end{itemize}
\end{itemize}

\subsubsection{\textbf{Sentence Abstraction.}}
\label{sec:abstraction}
Sentence abstraction is a preprocessing step conducted on the original
and translated sentences before converting them into regular expressions. It helps to focus on semantics that relate to quantifiers and logical relations. In general, the sentence abstraction works in two steps: (1) identify the non-terminal words/tokens that describe quantifiers and logical relations in the sentence, and (2) abstract the terminals or non-terminals that are grouped by the identified non-terminals as abstracted objects, denoted by symbols such as \verb|X| and \verb|Y|. Specifically, the non-terminals and terminals we used are specified and used in existing works~\cite{deepregex16,softregex19}, and are associated verbalization for both regular expressions and language descriptions.
For the first running example, the terminals (e.g., ``uppercase letters'') and non-terminals (e.g., ``contains only'', ``3 or more'', ``no more than 3'') are specified in ~\cite{deepregex16}, and therefore no abstraction is needed. While for the second running example, the terminal nouns/phrases grouped by non-terminals (i.e., ``not containing only'' and ``containing only'') are not specified, so they are abstracted and denoted as abstract symbols \verb|X| and \verb|Y|.
Hence, after abstraction, the sentences in two running examples are as follows:

\begin{itemize}
\item \textbf{Running Example 1 (Abstracted Sentences)}
\begin{itemize}
    \item S1': A string that contains only \textbf{3 or more} uppercase letters.
    \item S2': A string containing only \textbf{no more than 3} uppercase letters.
\end{itemize}

\item \textbf{Running Example 2 (Abstracted Sentences)}
\begin{itemize}
    \item S3': \verb|X| \textbf{not containing only} \verb|Y|.
    \item S4': \verb|X| \textbf{containing only} \verb|Y|.
\end{itemize}
\end{itemize}

Note that after abstraction, the meaning of sentences in the first example are almost preserved while in the second one, the most semantics are abstracted, remaining only the logical related meaning.

\subsubsection{\textbf{Regular Expression (regex).} }\label{sec:regex}
Regular expressions are widely used in practice. Let $\Sigma$ be a finite alphabet of symbols, a \textit{word} is a finite sequence of symbols chosen from this alphabet, and the set of all words over $\Sigma$ is denoted by $\Sigma^*$. For example, \verb|01101| is a \textit{word} from the binary alphabet $\Sigma=\{\verb|0|,\verb|1|\}$.
The empty word and the empty set are denoted by $\varepsilon$ and $\varnothing$,  respectively. 
Regexes over $\Sigma$ are defined inductively as follows: $\varepsilon$, $\varnothing$, $a \in \Sigma$, and [$C$] where $C \subseteq \Sigma$ are regular expressions; for regular expressions $r_1$ and $r_2$, the disjunction $r_1 | r_2$, the concatenation $r_1r_2$, the intersection $r_1\&r_2$, the negation $\neg r_1$, and the quantifier $r_1\{m,n\}$ where $m\in\mathbb{N}$, $n\in\mathbb{N}\cup\{{\infty\}}$, and $m\leq n$ are also regular expressions. Besides, $r?$, $r^*$, $r^+$ and $r\{i\}$ where $i\in\mathbb{N}$ are abbreviations of $r\{0,1\}$, $r\{0,\infty\}$, $r\{1,\infty\}$ and $r\{i,i\}$, respectively. $r_1\{m,\infty\}$ is often simplified as $r_1\{m,\}$.
For the running examples, the sentences are tokenized and fed into the state-of-the-art model ~\cite{softregex19}, resulting in the following regexes:

\begin{itemize}
\item \textbf{Running Example 1 (Transformed Regexes)}
\begin{itemize}
    \item R1:  \verb|[A-Z]{3, }|.
    \item R2:  \verb|[A-Z]{0,3}|.
\end{itemize}

\item \textbf{Running Example 2 (Transformed Regexes)}
\begin{itemize}
    \item R3: \verb|~[Y]|
    \item R4: \verb|[Y]|
\end{itemize}
\end{itemize}

\noindent where $[A-Z]$ denotes an uppercase letter, the quantifier $\{0,3\}$ represents the object before it (i.e., [A-Z] in the example) could appear zero to three times. 
While the quantifier $\{3, \}$ represents the object could appear three or more times.  
One may notice that 
after transformation, the abstract symbol \verb|X| in the second running example is not appear in the regex R4. It is because for the non-terminals which describe part-whole relations~\cite{sharvy1980more,mcdaniel2004modal,mellor2006wholes} such as inclusion (e.g., ``contain'') or exclusion (e.g., ``exclude''), the preceding terminal is  regarded as the collection while the latter one represents the part/component inside. For example, the sentence ``\verb|X| containing only \verb|Y|'' (S4') is transformed to \verb|[Y]| where \verb|X| is the collection which includes only the abstract symbol \verb|Y|. Similarly, the terminal ``a string'' in the first example is also regarded as the collection which is absent in R1 and R2.

\subsubsection{\textbf{Language of regular expression.}}
\label{sec:language}
The term \textit{language} denotes a set of strings chosen from the set of words $\Sigma^*$. Let $L(r)$ be the language of regular expression $r$. $L(r)$ can be defined inductively as follows: 
    $L(\varnothing)=\varnothing$;
    $L(\varepsilon)=\{\varepsilon\}$;
    $L(a)=\{a\}$;
    $L([C])=C$;
    $L(r_1|r_2)=L(r_1)\cup L(r_2)$;
    $L(r_1 r_2)= \{vw~|~v\in L(r_1), w\in L(r_2)\}$;
    $L(r_1\&r_2)=L(r_1)\cap L(r_2)$;
    $L({\footnotesize \neg} r_1)= \{v~|~v\notin L(r_1)\}$;
    $L(r\{m,n\})=\bigcup_{m\leqslant i \leqslant n} L(r)^{i}$.

For the running examples, the languages of the regular expressions in the running examples are:

\begin{itemize}
\item \textbf{Running Example 1 (Language of Regexes)}
\begin{itemize}
    \item L(R1) =  $L({[A-Z]\{3,~\}}) = \{AAA, AAB,\ldots, ZZZ, AAAA,\ldots\}$
    \item L(R2) =  $L({[A-Z]\{0,3\}}) = \{\varepsilon, A, B, C,\ldots, ZZZ\}$
\end{itemize}

\item \textbf{Running Example 2 (Language of Regexes)}
\begin{itemize}
    \item L(R3) = $L(\sim [Y]) = L(\Sigma^*)\setminus\{Y\}$~\footnote{We set the alphabet ($\Sigma$) of symbols is set to be letters and numbers.}
    \item L(R4) = $L([Y]) = \{Y\}$
\end{itemize}
\end{itemize}

In addition, two regular expressions are \textbf{equivalent} if they describe the same language. Also, we denote $|L(r)|$ as the \textit{language size} of regex $r$, which is the number of words described by $r$. In the running examples, $|L(R2)|$ and $|L(R4)|$ are $1 + 26 + 26^2 + 26^3 = 18279$ and $1$, respectively,  while $|L(R1)|$ and $|L(R3)|$ are infinite.

\subsubsection{\textbf{Deterministic Finite Automaton (DFA)}}
\begin{figure}[thbp]
    \centering
    \includegraphics[width=1.0\linewidth]{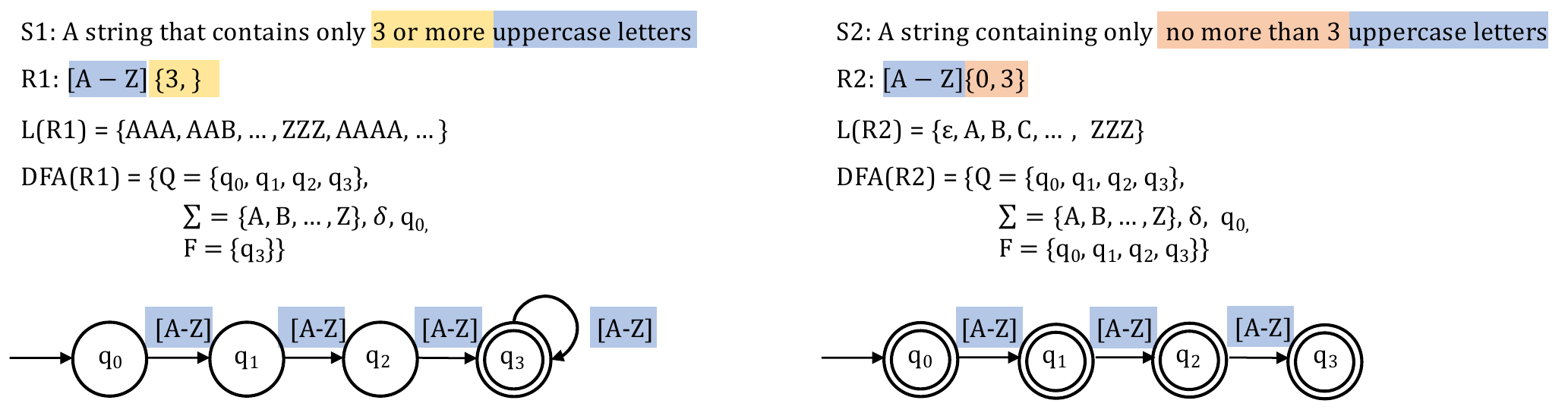}
    \caption{{A Running Example of the Transformation from Natural Language to Regular Expressions, Language of Regular Expressions and Deterministic Finite Automata.}
    }
    \label{fig:dfa-example}
\end{figure}

A Deterministic Finite Automaton (DFA) is a finite-state machine that accepts or rejects a given string of symbols, by running through a state sequence uniquely determined by the string~\cite{DBLP:books/daglib/0016921}. It has equivalent expressive power as the corresponding regular expression, which means any regular expression can be converted into a DFA that recognizes the language described, and vice versa. 
A DFA~\cite{DBLP:books/daglib/0016921} can be formally defined as a 5-tuple $(Q,\Sigma,\delta,q_{0},F)$, where $Q$ is a finite set of  \textit{states}, $\Sigma$ is an \textit{alphabet}, $\delta: Q\times \Sigma \longrightarrow Q$ is the \textit{transition function}, $q_0 \in Q$ is the \textit{start state}, and $F\subseteq Q$ is the \textit{set of accept states}. 
In Fig.~\ref{fig:dfa-example}, we visualized the DFAs of two regular expressions ($R_1$ and $R_2$) for better illustration.

In addition, according to \S\ref{sec:language}, each regular language can be expressed by a regular expression. There exists a unique minimal automaton that accepts the given regular language in a minimum number of states. This minimal automaton is known as a minimal DFA. A minimal DFA is guaranteed to have a regular expression which is semantically equivalent to the minimal DFA~\cite{semregex18}. For example, the illustration of the DFAs of the regular expressions $R1$ and $R2$ in \S\ref{sec:regex} are shown in Fig.~\ref{fig:dfa-example}.

\subsubsection{\textbf{Approximated Semantic Transformation}}\label{sec:approx-trans}
In the two running examples above, the meaning of quantifiers such as ``3 or more'' and ``no more than 3'' can be precisely captured with the quantifiers in regex \verb|{3,}| and \verb|{0,3}|, respectively.
However, as mentioned in \S\ref{sec:2}, some quantifiers and logical relations such as ``a few'' and ``many'' can only be approximately captured by regexes or DFAs. Hence, we give an example to illustrate how SemMT can be adapted to handle this situation. 

Take the following sentence as an example:

\begin{itemize}
   \item S5: {The U.S. contains \textbf{a few} states which choose to have the judges in the state courts serve for life terms.}~\footnote{This sentence is adapted from \url{https://en.wikipedia.org/wiki/U.S._state}}.
\end{itemize}

\noindent The vague quantifier ``a few'' is used with plural countable nouns. It emphasizes a small number of objects. To transform this sentence into regex, we first process it via sentence abstraction. 
The resulting abstracted sentence is :
    \begin{itemize}
       \item S5': {[X] contains \textbf{a few} [Y]}.
    \end{itemize}

\noindent where the abstract symbol \verb|[Y]| represents the ``states which choose to have the judges on the state's courts serve for life terms'', and \verb|X|, abstracted from the words ``The U.S.'', represents the collection which includes only the abstract symbol \verb|Y|.

Then, to approximate the semantics of S5 by regex, either over- and under-approximation can be applied. Specifically, the over-approximation enlarges the number of countable objects compared with the original number, while under-approximation underestimates the amount. For the example S5, two possible approximated sentences are as follows~\footnote{Since ``a few'' is a vague quantifier, there is no specific number that everyone would agree what it actually means. In this example, we quantify ``a few'' as ``at least 3'' following a common usage of ``a few''{~\cite{afew,afew-diff}}.}:

\begin{itemize}

        \item S5$_{O}$: {[X] contains \textbf{at least three} [Y]}.
        
        \item S5$_{U}$: {[X] contains \textbf{at most three} [Y]}.
\end{itemize}

\noindent which can be transformed into the following regexes:

\begin{itemize}
    \item  R5$_{O}$: \verb| Y{3, }|
    
    \item  R5$_{U}$: \verb| Y{1,3}|
\end{itemize}

\noindent where the quantifiers \verb|{3,}| and \verb|{1,3}| in R5$_{O}$ and R5$_{U}$ prescribe the symbol \verb|Y| could appears more than three times and once to three times, respectively.

By doing so, the semantics of sentences like S5 can be approximately captured by regexes, then the semantic similarities can be calculated on top of them, hence the mistranslation can be calculated accordingly. 
However, the approximated semantics may affect the effectiveness of mistranslation detection to some extent, so we also discussed such influence in \S\ref{sec:approx}.
Note that the conversion of sentences to regexes is a major research problem in natural language processing. We do not make contribution to such conversion in SemMT but leverage existing conversion techniques. 
Since the effectiveness of mistranslation detection can be affected by the quality of the approximated regexes, we evaluate SemMT based on the sentences with quantifiers or logical relations that can be precisely quantified due to the availability of automated tools for transforming such relations. The finding based on the five large-scale natural language corpora in \S\ref{sec:2} indicates that a large majority (89.6\%) of quantifiers and logical relations can be precisely quantified.

\subsection{Semantic Similarity Calculation}\label{sec:similarity}
The semantic differences of sentences involving quantifiers and logical relations cannot be adequately captured by SBERT~\cite{SBERT19}, which is the state-of-the-art metric proposed to measure semantic similarities. As shown in the second example of Fig.~\ref{fig:example}, SBERT gives nearly 100\% semantic similarity between ``at least'' and ``at most''. Such weakness is commonly found when SBERT is applied to the NL-RX-Synth~\cite{deepregex16} dataset used by the natural language processing community for sentences that contain quantifiers and logical relations. We will discuss it in more detail in our experiment.
This motivates us to develop new metrics based on regexes and DFAs to better measure the semantic similarities of sentences involving quantifiers and logical relations.

\subsubsection{{\textbf{Regex-based Similarity (SemMT-R)}}}
Considering the transformation rules of regular expression, the similarity between two regexes can be calculated by their Levenshtein distance. Specifically:
\begin{equation}
    S_{REG} = 1 - \frac{D_{L}(r1, r2)}{Max(len(r1), len(r2))}
\end{equation}
where $D_{L}(r1, r2)$ is a function that computes the Levenshtein distance between regular expressions $r1$ and $r2$. 
Note that in Levenshtein distance calculation, we followed the convention to count terminals (e.g., $[a-z]$ which denotes an arbitrary lower-case letter, and $[0-9]$ which represents an arbitrary number from $0$ to $9$) specified in ~\cite{deepregex16} as distance one. 
Hence, for the running examples, the regex-based similarities are:

\begin{itemize}
\item \textbf{Running Example 1 (SemMT-R Similarity)}
$$S_{REG}(R1, ~R2) = 1 -3/6 = 0.500$$

\item \textbf{Running Example 2 (SemMT-R Similarity)}
$$S_{REG}(R3, ~R4) = 1 - 1/4 = 0.750$$

\end{itemize}

Though efficient, this regex-based similarity may fail to capture the semantic difference in some situations~\cite{similarity13,regexPython16}. For example, R3 and R4
have semantically opposite meaning, while the Levenshtein distance between them is only 1. 
Hence, we further explore the measurement of semantic similarity using DFAs, which can capture the languages described by the regexes.

\subsubsection{{\textbf{DFA-based Similarity (SemMT-D)}}}\label{sec:dfa} 
An alternative to measure similarity between two regexes is to evaluate the Jaccard similarity between their regular languages. This approach can be effective in finding similarity between two finite sets of data, drawn from various application domains~\cite{similarity13,garcia2008database,DBLP:conf/vldb/ArasuGK06}. Specifically, given two regexes, one can construct two corresponding semantic equivalent minimal DFAs, then calculate the Jaccard similarity using the following equation~\cite{DBLP:books/daglib/0016921,similarity13}:
\begin{equation}
    S_{DFA}(r1, r2) = \frac{|L(r1)\cap L(r2)|}{|L(r1) \cup L(r2)|}
\end{equation}
whether $L(r1)$ and $L(r2)$ are languages of regexes $r1$ and $r2$, respectively.

However, an issue with such similarity is that regular languages can be infinite. To address this issue, we adapt ideas from existing works~\cite{shannon1948mathematical,chomsky1958finite,DFAinf10,bex2010inference} to improve the efficiency. 
We first define the function $S_{DFA}^{'}(r1, r2,\lambda)$ as follows:
\begin{equation}
    S_{DFA}^{'}(r1, r2,\lambda)=\frac{|L(r1)^{\leq \lambda}\cap L(r2)^{\leq \lambda}|}{|L(r1)^{\leq \lambda} \cup L(r2)^{\leq \lambda}|}
\end{equation}
\noindent where $\lambda \in \mathbb{N}$. Formally, for a language $L(r)$, let $|L(r)^{\leq \lambda}|$ denote the number of words in $L(r)$ of length at most $\lambda$. Then, we reduce the calculating function $S_{DFA}(r1, r2)$ to calculate the limit of function $S_{DFA}^{'}(r1, r2,\lambda)$, as shown below:
\begin{equation}
    S_{DFA}(r1, r2) = \lim\limits_{\lambda\to+\infty} S_{DFA}^{'}(r1, r2,\lambda)
\end{equation}

\noindent Finally, the calculation proceeds iteratively until the solution converges to a preset threshold. Note that the threshold is customized to balance the efficiency and effectiveness. In our evaluation, we set the threshold as 0.001.

So for the running examples, the DFA similarities between two groups of regexes are as follows:

\begin{itemize}
\item \textbf{Running Example 1 (SemMT-D Similarity)}
\begin{align*}
S_{DFA}(R1, R2) &= \lim\limits_{\lambda\to+\infty} S_{DFA}^{'}(R1, R2,\lambda)\approx S_{DFA}^{'}(R1, R2, 6)=\frac{|L(R1)^{\leq 6}\cap L(R2)^{\leq 6}|}{|L(R1)^{\leq 6} \cup L(R2)^{\leq 6}|}\\
&=\frac{26^{3}}{1+26 + 26^2 + \cdots + 26^{6}}=\frac{17576}{321272407}\approx0.000\\
\end{align*}

\item \textbf{Running Example 2 (SemMT-D Similarity)}
$$S_{DFA}(R3, R4) = \lim\limits_{\lambda\to+\infty} S_{DFA}^{'}(R3, R4,\lambda)\approx S_{DFA}^{'}(R3, R4, 1)=\frac{|L(R3)^{\leq 1}\cap L(R4)^{\leq 1}|}{|L(R3)^{\leq 1} \cup L(R4)^{\leq 1}|}=0.000$$

\end{itemize}

\noindent where 6 and 1 are the resulting $\lambda$s given that the limit of the function $S_{DFA}^{'}(r1, r2,\lambda)$ can approximate the DFA similarities over the infinite regular languages. 

\subsubsection{{\textbf{Hybrid Similarity (SemMT-H)}}}\label{sec:hyb}
The above described SemMT-R and SemMT-D methods have their own advantages in the measurement of semantic similarities. Regex-based similarity measures the semantics similarity between two regexes in the textual level, while SemMT-D similarity measures it from the perspective of language set.
Hence, we propose a hybrid metric to enjoy both advantages by combining SemMT-R and SemMT-D with customized weights.
The hybrid similarity is calculated by the following equation:
\begin{equation}
    S_{HYB}(r1, r2) = K \cdot S_{REG}(r1, r2) + (1-K) \cdot S_{DFA}(r1, r2) 
\end{equation}

\noindent where $K$ is a customized parameter which adjusts the balance of REG- and DFA-based similarity metrics. 
For example, if $K$ is set to be $0.5$, the hybrid similarity of the running examples are calculated as follow:
\begin{itemize}
    \item \textbf{Running Example 1 (SemMT-H Similarity)}
    $$ S_{HYB}(R1, R2)_{K=0.5} = K \cdot S_{REG}(R1, R2) + (1-K) \cdot S_{DFA}(R1, R2) = 0.250 $$
    
    \item \textbf{Running Example 2 (SemMT-H Similarity)}
    $$ S_{HYB}(R3, R4)_{K=0.5} = K \cdot S_{REG}(R3, R4) + (1-K) \cdot S_{DFA}(R3, R4) = 0.375$$
\end{itemize}

Note that the selection of different values for parameter K can influence the effectiveness of SemMT-H similarity.
The larger the parameter K is, the SemMT-H similarity $S_{HYB}$ depends more on SemMT-R similarity ($S_{REG}$), or vice versa. 
To further demonstrate the influence of K on the effectiveness of HYB-based similarity, we conducted an experiment to discuss this issue in \S\ref{sec:k-selection}.

\subsection{Mistranslation Detection}\label{sec:detection}
Using one of the above regex-based metrics, the semantic similarity between the original
and back-translated sentences can be calculated. We then decide whether a given translated sentence is far enough from the original
to indicate the presence of a mistranslation. To do so, following an existing study TransRepair~\cite{transrepair20}, we first conduct the threshold selection through fine-level granularity enumeration guided by the target evaluation metrics (i.e., F-Score), and then detect mistranslations based on the similarity threshold which achieves the optimum performance. 
Finally, we identify those sentences whose semantics are less similar to the original sentence than the threshold as suspicious mistranslated sentences. The intuition of guiding by F-Score is that it can better balance precision and recall.
Besides, the threshold can be customized since the user may prioritize minimizing false positives or maximizing recall depending on their goals. In \S\ref{sec:experiments}, we show the trade-offs for different threshold values. 

If the similarity between the original and back-translated 
sentences is less than the threshold, SemMT will report it as a suspicious mistranslation. Considering the fact that a forward mistranslation can hardly lead to a reasonable backward translation~\cite{goodfor05}, some may believe it is enough to only report the forward trip translation (i.e., the pair of original
and intermediate sentences). However, according to our analysis and investigation, there are 26\% mistranslations detected by SemMT are introduced in the backward trip (\S\ref{sec:localization}). Therefore, when the similarity falls beyond the threshold, SemMT will report the original, intermediate and back-translated sentences together.

\section{Experiments}
\label{sec:experiments}
This section reports the effectiveness of SemMT by studying the following four research questions (RQs):
\begin{itemize}
    \item \textbf{RQ1: How effective is SemMT in finding buggy translations?} We evaluated the effectiveness of SemMT in terms of accuracy, F-Score, precision and recall compared with other semantic and non-semantic similarity metrics. 
    We also quantified their capabilities in distinguishing buggy and correct translations.
    \item \textbf{RQ2: Can SemMT outperform the state-of-the-art works?} We evaluated the performance using the number of issues detected, precision, recall and F-score for each work. 
    \item \textbf{RQ3: Can SemMT's effectiveness be improved by combining different similarity metrics?} Different metrics tend to evaluate similarity from diverse aspects, so we explored combinations of similarities to see whether {their effectiveness can be mutually-improve.}
    \item {\textbf{RQ4: What is the applicability of SemMT?
    } 
    The applicability of a testing framework is also critical to measure its practical usefulness. 
    So we further investigated the applicability of our framework, 
    i.e., whether it can achieve similar effectiveness on different translators, language pairs and datasets.
    } 
\end{itemize}

\subsection{Experiment Setup}
\label{sec:setup}
We implemented SemMT in Python, and conducted experiments on a machine powered by one Intel i7-8700K@3.7GHz CPU that supports 6 cores, Nvidia GeForce Titan V 12GB VRAM, 64GB memory and Dual Tesla M60 8GB VRAM. 
The parameter $K$ for SemMT-H is set to be $0.5$ on the purpose of striking the balance between them. All experiment results have been released for validation~\cite{SemMT}.

\textbf{Dataset.} 
Since we mainly focus on the semantics of quantifier and logical relations,
the dataset selected should be consistent of sentences conveying such semantics, which is non-trivial for the evaluation.
In this study, we adopted two benchmarks, NL-RX-Synth~\cite{deepregex16} 
{and KB13~\cite{deepregex16,softregex19} as the source} of test inputs for SemMT. The two benchmarks are frequently-used~\cite{deepregex16,softregex19,pldi20} in generating regular expressions for natural language description tasks. 
In these datasets, the words/tokens can be represented as {the} semantic-equivalent alphabet of symbols, quantities, logic relations or quantitative modifiers. 
For example, the word ``numbers'' can be represented by the set of digits (\verb|[0-9]|), and the semantic relations ``A or B'' can be represented as the disjunction between them (\verb|A| $|$ \verb|B|).
In total, NL-RX-Synth {comprises} 10,000 pairs of English sentences and the corresponding regular expressions. 
The average character number of sentences in NL-RX-Synth dataset is 66.36, and 40.61 in KB13. 
For the first three RQs, the test inputs were randomly selected from NL-RX-Synth, while for RQ4, the test inputs were randomly selected from KB13.

\textbf{Labeling.} 
The output of SemMT is a list of suspicious issues, each of which consists of the original sentence, intermediate translation and target translation. 
Two of the authors inspected all the results separately and discussed all the inconsistent labels until convergence. Besides labeling whether the RTT is correct or not, we also labeled whether the forward and backward translations are correct. 
In addition, to compare with existing the state-of-the-art approaches, we also labeled the issues reported by each of them. The output issues of existing approaches are 
a list of suspicious issues including {the original}, intermediate and back-translated sentences. 

\textbf{Model Training.}
We adopted the state-of-the-art regex synthesis work~\cite{softregex19} to transform natural language to regex,
which {uses} reinforcement learning to train a sequence-to-sequence model. 
In addition, to make it better fit into our work, we proceeded {with} the grammar checking and data augmentation to the original NL-RX-Synth dataset in order to improve accuracy and enlarge the vocabulary size. 
Specifically, to augment data, we selected one fourth of the sentences and then replace with synonyms. 
Note that a manual checking of the replaced synonyms is necessary because we need to ensure the semantic meaning is maintained after the replacement. 
We also confirmed with a linguist if the semantics are preserved after the synonym substitution.
After augmentation, 13,588 sentences were obtained with an average vocabulary size of 123.
We then split the training, validation and testing set by 80\%, 5\%, 15\% respectively.
The model was trained for 30 epochs in total, achieving a 90.93\% accuracy on the test set. 
To answer RQ4, we also perform training on the KB13 dataset using a  process similar to that on the NL-RX-Synth dataset. The resulting model can achieve the accuracy of 77.67\% on the test set.~\footnote{Since the accuracy on the KB13 dataset is lower than that on the NL-RX-Synth dataset, which may affect the comparison with the effectiveness between two datasets, two authors manually checked and refined the resulting regexes until convergence for better comparison.}

\textbf{Similarity Metrics.}
Various similarities have been used by recent works to detect translation errors~\cite{SIT,transrepair20,pathology} and estimate translation quality~\cite{revisiting20,acl20-honor-metrics,context-aware-acl19,BLEU02,chrF15,EMSI17,EMSI19,YISI19,acl20-honor-metrics}. 
We selected three syntactic-based similarities that are recently used by related works~\cite{transrepair20,SIT} and one state-of-the-art semantic-based similarity~\cite{SBERT19} as baselines to compare the accuracy, precision and recall of bug detection with the three similarities supported by SemMT.

\begin{itemize}
    \item \textbf{Levenshtein-based similarity (LEVEN).}  It is a way of quantifying how dissimilar two strings are by calculating the Levenshtein distance (a.k.a., Edit distance), i.e., counting the minimum number of operations required to transform one string into the other~\cite{leven97} and normalizing it in the same way as~\cite{transrepair20,DBLP:conf/naacl/ZhangUSNN18,DBLP:conf/aaai/GuWCL18}

    \item \textbf{Dependency-based similarity (DEP).}
    Dependency relations describe the direct relationships~\cite{dependency} of strings. 
    We evaluated the distance between two sets of dependency relations by summing up the absolute difference in the numbers of each type of dependency relations as described in a recent work~\cite{SIT}. 

    \item \textbf{BLEU-based similarity (BLEU).}
    The BLEU (BiLingual Evaluation Under study)~\cite{BLEU02} 
    aims to automatically evaluate machine translation quality by checking the correspondence between the output of machines and that of humans~\cite{BLEU02,coughlin2003correlating}. For one target sentence, the score of BLEU is calculated by comparing it to a set of good quality reference translations. The details of BLEU can be found in \cite{BLEU02}. 

    \item \textbf{SBERT similarity (SBERT).}
    We also used the state-of-the-art sentence-level semantic approach, SBERT~\cite{SBERT19}, a refined version of BERT~\cite{BERT19}, as a baseline for semantic similarity metric in our evaluation, because it performs the best in RTT quality estimation as \cite{revisiting20} suggests. 

\end{itemize}

\textbf{Comparisons.}
In the experiment, we compared SemMT with SIT~\cite{SIT}, TransRepair~\cite{transrepair20} and PatInv~\cite{pathology}. 
These works utilize different methods to generate mutants {of the sentences under test}, then detect inconsistencies according to different metrics. 
In particular, SIT~\cite{SIT} generated syntactically equivalent mutants using BERT by replacing noun and adjective words in sentences. Similarly, TransRepair~\cite{transrepair20} generated mutants by replacing nouns, adjective words and numbers with their synonyms and further construct the word-mutant pair. 
For comparison, we implemented these works to generate mutants either using their release source code~\cite{SIT} or by carefully following the explanations in their paper~\cite{transrepair20,pathology}. We adjusted the parameters for these works following the original strategies published in their papers based on our dataset. 
TransRepair~\cite{transrepair20} used $0.9$ as the minimum cosine similarity of word embeddings to generate word pairs. In our experiment, we lowered the threshold to $0.8$ in order to generate sufficient number of pairs.
For SemMT, since it is not originally designed for mutant generation, we adapted similar mutant generation processes in the baselines by replacing the nouns, numbers and relational adverbs with their synonyms. We also manually validate the generated mutants to preserve the semantics after synonym substitution. The manual checking process was under the guidance of a linguist to reduce the threat to the reliability of experimental results that may be introduce.
and removed semantically-changed mutants by manual check. 
For each {sentence under test}, SemMT generated up to two mutants.

Note that since SemMT adopts the RTT paradigm, the reported issues may be caused by either the forward and backward trips. Therefore, we labeled both translation trips, and for fairness, the comparison only focused on the correctness of the forward translation while removing those whose backward translations are incorrect. 

\textbf{Evaluation metrics.}
To evaluate the effectiveness, we adopted accuracy, precision, recall and F1-Score (abbrev. F-Score). 
Given the number of 
true positives (TPs, a TP refers to a mistranslated sentence that is reported to be a mistranslation), 
false positives (FPs, a FP refers to a correctly translated sentence that is reported to be a mistranslation), 
true negatives (TNs, a TN refers to a correctly translated sentence that is not reported to be a mistranslation) and 
false negatives (FNs, a FN refers to a mistranslated sentence that is not reported to be a mistranslation), the metrics are defined as follow:
\begin{itemize}
    \item \textbf{Accuracy}: the proportion of correctly reported (whether the translation is correct or not) sentences.
    \begin{equation}
        Accuracy = \frac{TP + TN}{TP + TN + FP + FN}
    \end{equation}
    \item \textbf{Precision}: the proportion of real mistranslations over the reported mistranslations.
    \begin{equation}
        Precision = \frac{TP}{TP + FP}
    \end{equation}
    \item \textbf{Recall}: the ratio of reported mistranslations over all the real mistranslations.
    \begin{equation}
        Recall = \frac{TP}{TP + FN}
    \end{equation}
    \item \textbf{F-Score}: twice the multiplication of precision and recall divided by the sum of them. 
    \begin{equation}
        F-Score = 2 * \frac{Precision * Recall}{Precision + Recall}
    \end{equation}
\end{itemize}

\subsection{RQ1: Effectiveness of SemMT}
\label{sec:RQ1}

To evaluate the effectiveness of our SemMT, we randomly sampled 500 sentences from the NL-RX-Synth dataset, applied the round-trip translation and collected the translation results.\footnote{The translation results  were collected on July 27, 2020 on Google translator.}
We then transformed both the original sentences and the round-trip translation results into regular expressions by the trained transformation model. This is described earlier in the experiment setup, and the three SemMT similarities (SemMT-R, SemMT-D, SemMT-H) were then calculated based on regular expressions. Other metrics were calculated on the {original} and target sentences. Worth mentioning is that in order to evaluate all four evaluation metrics, we labeled all 500 translations following the previously mentioned labeling process.

\begin{table}[htbp]
\centering
\caption{\textbf{Effectiveness of Different Similarity Metrics For Bug Detection.} Each entry reports the average/highest score among threshold values ranging from 0 to 1 at step 0.01.}
\label{tab:ave}
\renewcommand\arraystretch{1.5}
\resizebox{0.6\linewidth}{!}{%
\begin{tabular}{r|rrrr}
\hline
& Accuracy & Precision & Recall & F-Score\\ \hline
LEVEN & 0.52 / 0.56 & 0.53 / 0.75 & 0.42 / 1.00 & 0.40 / 0.70 \\
DEP & 0.47 / 0.55 & 0.41 / 0.54 & 0.34 / 1.00 & 0.31 / 0.70 \\
SBERT & 0.47 / 0.56 & 0.34 / 0.55 & 0.22 / 1.00 & 0.23 / 0.70 \\
BLEU & 0.48 / 0.53 & 0.47 / 0.72 & 0.40 / 1.00 & 0.37 / 0.69 \\ \hline
SemMT-R & 0.54 / \textbf{0.60} & \textbf{0.67 / 1.00} & 0.42 / 1.00 & 0.41 / 0.70 \\
SemMT-D & \textbf{0.56} / 0.59 & 0.56 / 0.59 & \textbf{0.71} / 1.00 & \textbf{0.63} / 0.69 \\
SemMT-H & {0.55 / 0.58} & {0.61 / 1.00} & 0.57 / 1.00 & {0.52} / 0.70 \\ \hline
\end{tabular}%
}
\end{table}

Table~\ref{tab:ave} shows the comparison of the average and highest accuracy, recall, precision and F-Score of different similarity metrics against thresholds (from $0.1$ to $1.0$ with step $0.01$). 
Generally, the three similarity metrics in SemMT perform better than the other similarity metrics regarding all four measurements. Moreover,
SemMT-D achieves the highest average recall (0.71) and F-Score (0.63), while the highest value of other similarity metrics (i.e., LEVEN, DEP, SBERT, BLEU) are 29\% and 23\% lower. 
Particularly, SemMT-D performs the best among our three similarities and baseline similarities with respect to most of the evaluation metrics.

\begin{figure}[thbp]
\centering
\subfigure[Accuracy]{
\includegraphics[width=0.75\linewidth]{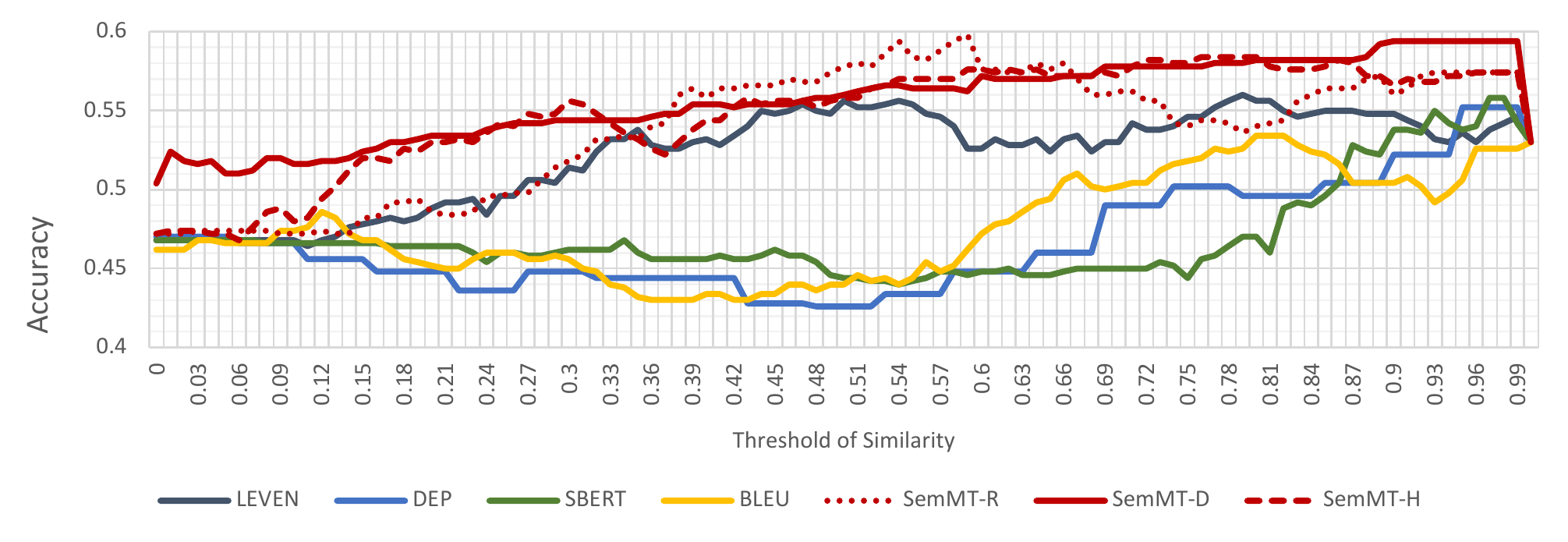}
}
\subfigure[Precision]{
\includegraphics[width=0.75\linewidth]{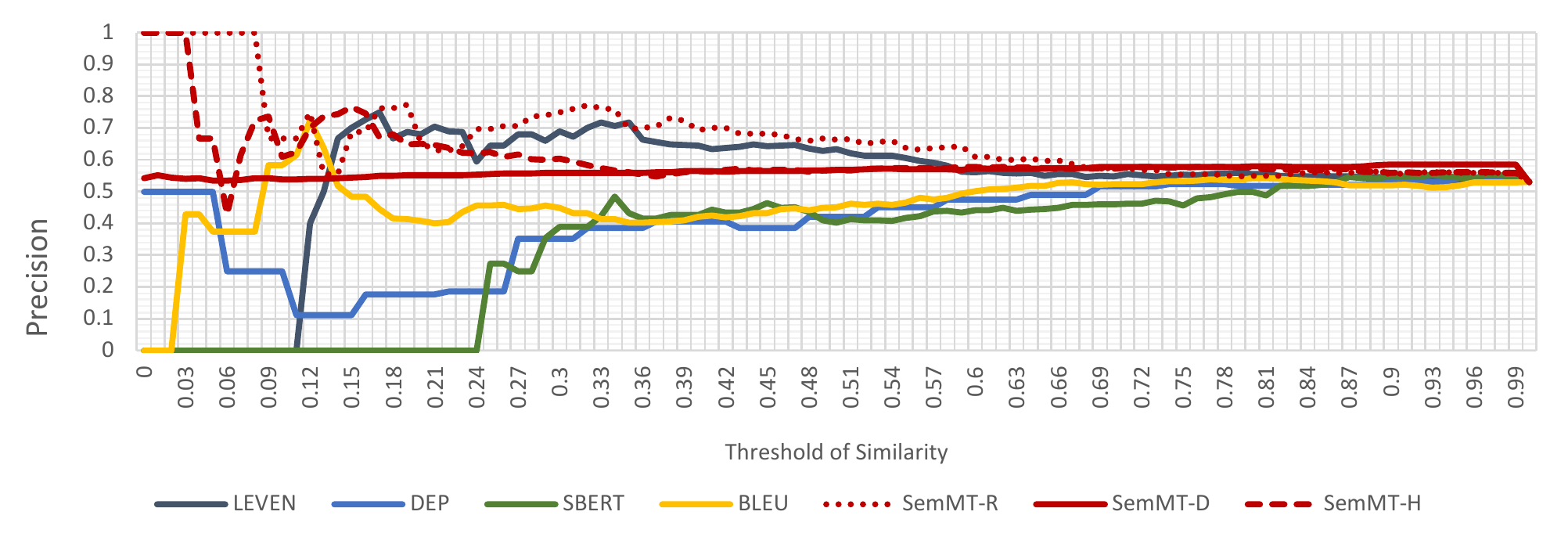}
}
\subfigure[Recall]{
\includegraphics[width=0.75\linewidth]{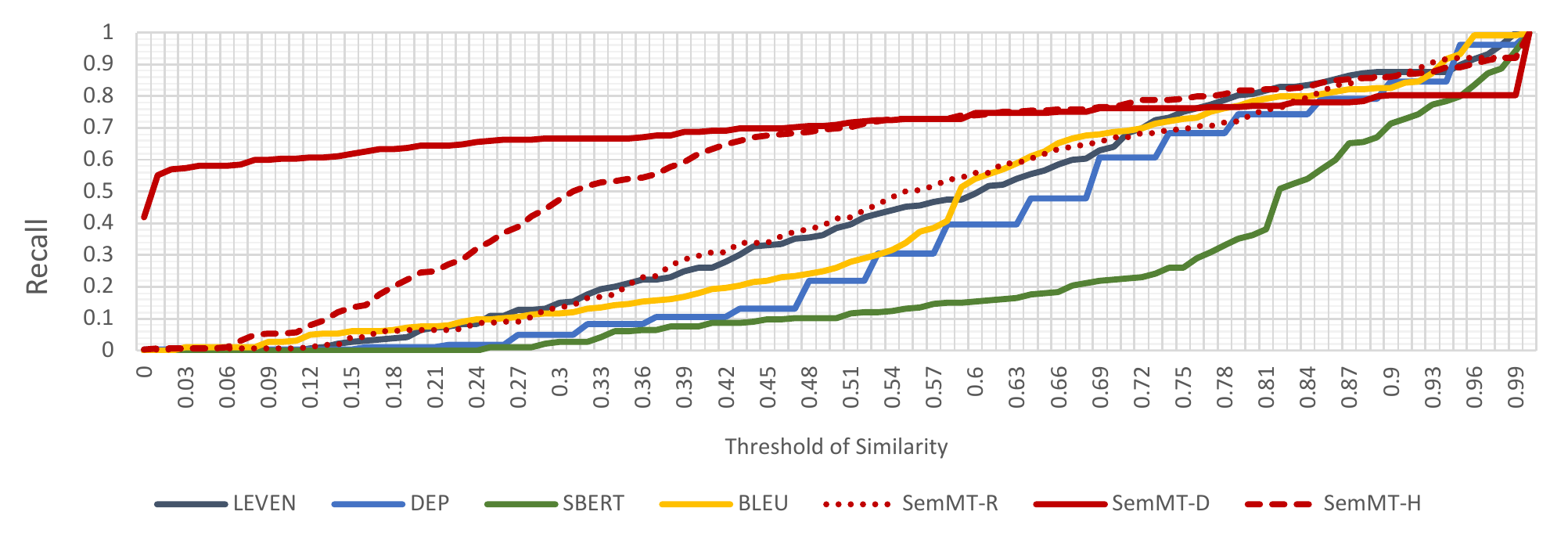}
}
\subfigure[F-Score]{
\includegraphics[width=0.75\linewidth]{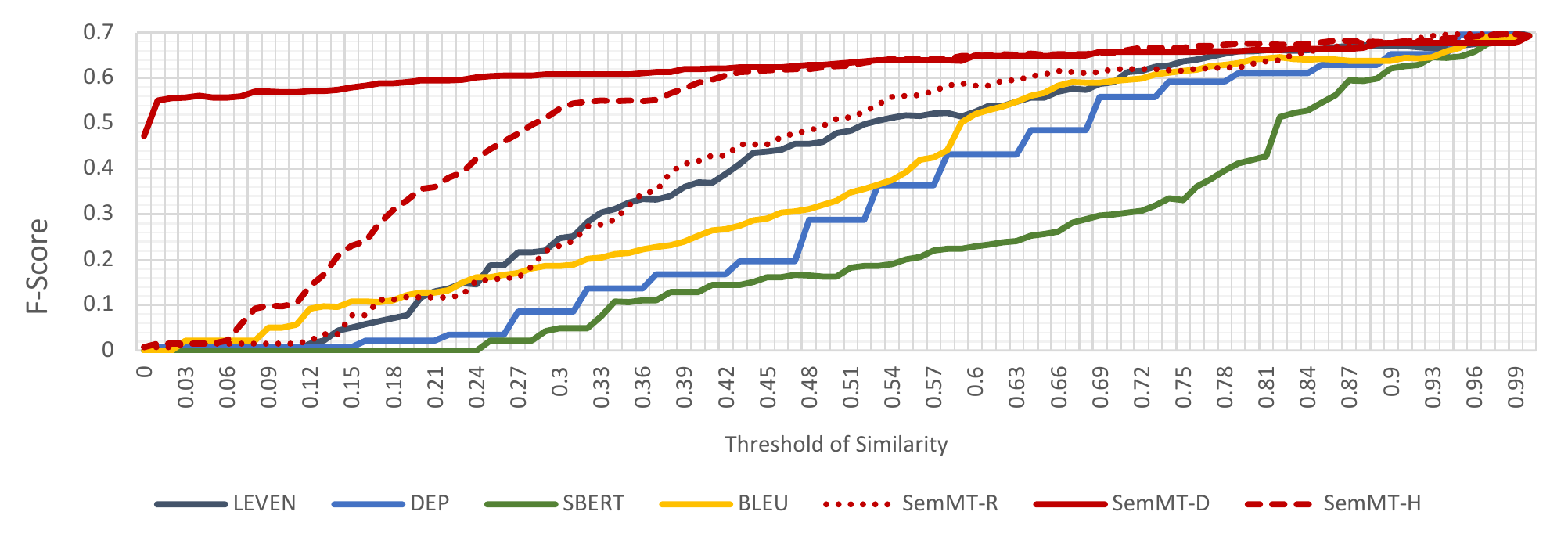}
}
\caption{Effectiveness of Mistranslation Detection Over Different Similarity Thresholds}
\label{fig:exp-effectiveness}
\end{figure}

Since the effectiveness of these metrics are affected by customized threshold, we also illustrated how the performance of these evaluation metrics vary against different thresholds. 
Specifically, we normalized all similarity values, and set the threshold from $0.0$ to $1.0$, with step $0.01$. 
In Fig~\ref{fig:exp-effectiveness}, we presented the trends on accuracy, precision, recall and F-Score.
Overall, all three of our similarity metrics (drawn in red) outperform the others in terms of accuracy, precision, recall and F-Score for most of the threshold values. 
Apart from our three similarities, most times LEVEN outperforms other similarities. Although SBERT measures semantic similarity, its performance is not as good as expected.

\begin{mdframed}[style=MyFrame]
\textbf{Summary of Findings Related to RQ1:} The three similarity metrics used in SemMT are effective for mistranslation detection. They outperform the other metrics against almost all threshold values. Specifically, in terms of precision, recall and F-Score, our metrics achieve an increase of 13\%, 30\% and 23\% compared with the highest value achieved by other metrics. 
\end{mdframed}

\subsection{RQ2: Comparison with Existing Works}

In this section, we compared SemMT with SIT~\cite{SIT}, TransRepair~\cite{transrepair20} and PatInv~\cite{pathology} on mutant generation and the effectiveness of bug detection in terms of accuracy, precision, recall and F-Score under the En-Zh language setting on the Google translator. We randomly selected 200 sentences from the dataset
and used them as {sentences to be tested.}

We first generated and filtered out mutants in the way as described in the original paper of baselines \cite{SIT,transrepair20,pathology}.
The numbers of generated mutants are listed in Table~\ref{tab:mutants}. These mutants are generated using their own generation approaches, and 
after filtering, there are 223 to 452 mutants left for each work. 
Note that the number of filtered mutants does not necessarily indicate a better capability on bug detection, it is depended on different strategy of mutant generation, the effectiveness is continued to be evaluated.
Specifically, we evaluated the effectiveness of each method in terms of the four evaluation metrics. The results are presented in Table~\ref{tab:mutants}. 
The threshold that achieves the optimum performance with respect to F-Score is chosen for each work, as listed in Table~\ref{tab:mutants}.
Note that the threshold of SIT is the distance between the dependency parse trees, while PatInv is not tuned by the threshold. 
We choose two most well-performing (with the highest F-Score at 0.82 and 0.82 with thresholds 0.963 and 0.906, respectively) metric values out of the four in TransRepair, which are the Levenshtein- (denoted as ``ED'' in ~\cite{transrepair20}) and BLEU-based method, written as TransRepair(L) and TransRepair(B). For TransRepair(L) and (B), a buggy translation issue is reported when the metric value is smaller than or equal to the selected thresholds, For SIT, a buggy translation issue is reported when the metric value is larger than or equal to the selected threshold.

\begin{table}[t!]
\centering
\caption{\textbf{Precision and the Number of Mistranslation Using Different Threshold Values.} The symbol `-' denotes the values are inapplicable.}

\label{tab:mutants}
\renewcommand\arraystretch{1.3}
\resizebox{0.75\linewidth}{!}{%
\begin{tabular}{r|cc|rrrrr}
\hline
& \begin{tabular}[c]{@{}c@{}}\# Mutants \\ Filter (Gen) \end{tabular} & 
\begin{tabular}[c]{@{}l@{}}Thrsh\end{tabular} & 
\begin{tabular}[c]{@{}l@{}}\# Issues\end{tabular} &
\begin{tabular}[c]{@{}l@{}}Acc\end{tabular} & 
\begin{tabular}[c]{@{}l@{}}Recall\end{tabular} & 
\begin{tabular}[c]{@{}l@{}}Prec\end{tabular} & 
\begin{tabular}[c]{@{}l@{}}F-Score\end{tabular}\\ \hline
SIT & 452 (635) 
& 5 & {124} & 0.497 & \textbf{0.932} & 0.331 & 0.488 \\
TransRepair(L) & 223 (869) 
& 0.96 & 108 & 0.520 & 0.506 & 0.398 & 0.446 \\
TransRepair(B) & 223 (869) 
& 0.88 & 93 & 0.552 & 0.459 & 0.419 & 0.438 \\
PatInv & 266 (358) 
& --- & 16 & --- & --- & \textbf{0.565} & --- \\ \hline
{SemMT-R} & 328 (384)
& 0.62 & 105 & {0.730} & 0.710 & 0.466 & \textbf{0.563} \\
{SemMT-D} & 328 (384)
& 0.42 & \textbf{166} & 0.585 & {0.855} & 0.355 & 0.502 \\
{SemMT-H} & 328 (384)
& 0.32 & 86 & \textbf{0.741} & 0.594 & {0.476} & {0.529} \\ \hline
\end{tabular}%
}
\end{table}

According to Table~\ref{tab:mutants}, SemMT achieves the highest accuracy and F-Score compared with existing works, with a similar number of issues detected. 
In particular, the highest accuracy achieved by SemMT (74.1\% by {SemMT-H}) is 34.2\% higher than the highest accuracy achieved by TransRepair(B) (55.2\%), 
and the highest F-Score (56.3\%) achieved by SemMT (SemMT-R) is 15.4\% larger than the highest value achieved by SIT (48.8\%). 
In addition, although SIT achieves the best recall (93.2\%), its precision is relatively low (33.1\%). 
The highest precision is achieved by PatInv (56.5\%), yet the number of issues identified is small (16).
In constrast, SemMT ({SemMT-D}) achieves a comparatively high recall (85.5\%) with 150 (166 - 16 = 150) more issues reported. 

\begin{figure}[t!]
    \centering
    \includegraphics[angle=0,width=0.80\linewidth]{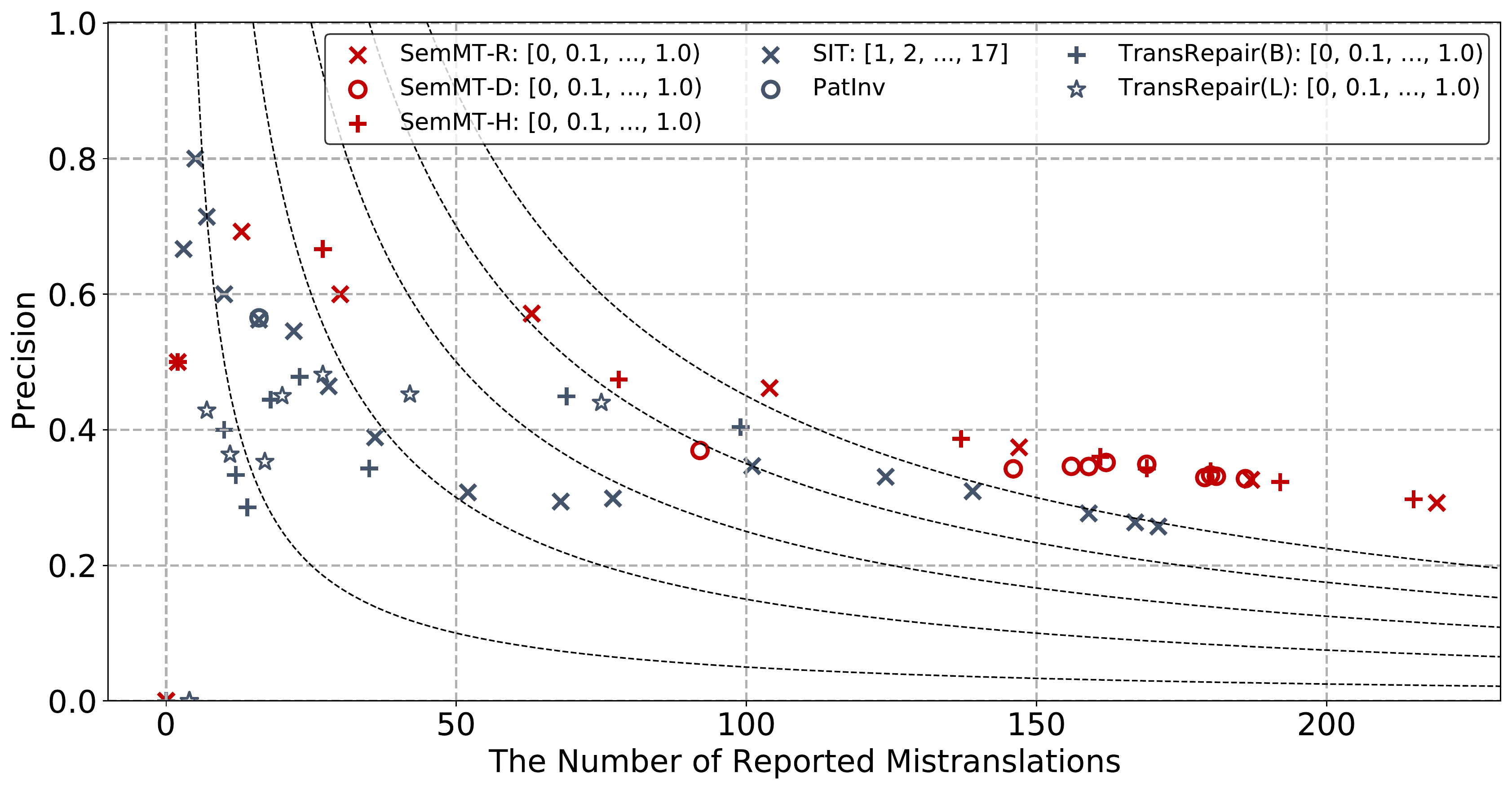}
    \caption{\textbf{Precision and the Number of Mistranslation Reported by SemMT Compared With Other Works.}}
    \label{fig:compare-sit}
\end{figure}

Moreover, we plotted the correlation between precision and the number of mistranslation detected in Fig~\ref{fig:compare-sit} as threshold varies. A translation is regarded as a candidate mistranslation if its similarity is smaller than (for the thresholds of distance, it should be larger than) or equal to the given threshold. The X and Y axes represent the number of issues detected and precision, respectively. The threshold setup for each method is proceeded as follow: For SemMT and TransRepair, we normalized the similarity values to the range of [0,1], and set the threshold from $0.0$ to $1.0$, with a step of $0.1$. 
The threshold of distance for SIT ranges from 1 to 17. No threshold is required for PatInv. 
Therefore, there are 11 dots for SemMT-R, SemMT-D, SemMT-H, TransRepair(B) and TransRepair(L), 17 dots for SIT and 1 dot for PathInv. In the figure, we did not denote every threshold for each dots, while the threshold values can be implied - the closer the dot is to the y-axis, the larger the similarity threshold (the smaller the distance threshold).

According to Fig~\ref{fig:compare-sit}, we can see that there is a trade-off between the precision and the number of mistranslation issues reported, i.e., with the similarity threshold increases, the more translations are regarded as mistranslation (i.e., the larger the number of issues reported), while the more false positives may be involved, resulting in a lower precision. 
To better illustrate such balance, we set precision and the number of mistranslation reported as axes, and the more the dot is closed to the top-right corner, the more the result strikes the balance. 
Overall, three variants of SemMT (as shown in red) outperform other baseline works for most of the cases with the change of thresholds, followed by SIT and TransRepair(B). On the contrary, PathInv and TransRepair(B) perform less satisfying, with less number of reported mistranslations and lower accuracy.

\begin{mdframed}[style=MyFrame]
\textbf{Summary of Findings Related to RQ2:} SemMT outperforms the state-of-the-art approaches. 
In particular, compared with the optimum performance achieved by other works, SemMT achieves an improvement of 34.2\% and 15.4\% on accuracy and F-Score with similar numbers of issues detected.
Further experiment shows that in general, SemMT can report more mistranslations with higher precision over the change in thresholds.
\end{mdframed}

\subsection{RQ3: Can Metrics of SemMT Find Mistranslations That Cannot Be Detected By Other Metrics?}
\label{sec:RQ3}

To answer RQ3, we analyzed whether the mistranslations reported by different metrics overlap. 
We also explored whether the combination of metrics can improve the performance with respect to accuracy, F-Score and the number of issues detected. 
As previously mentioned, a metric's performance varies with threshold values.
Therefore, we chose a threshold value that maximizes its performance based on the largest product of true positives and false negatives for each metric in order for fair comparison.

\begin{figure}[htbp]
    \centering
    \includegraphics[angle=0,width=1\linewidth]{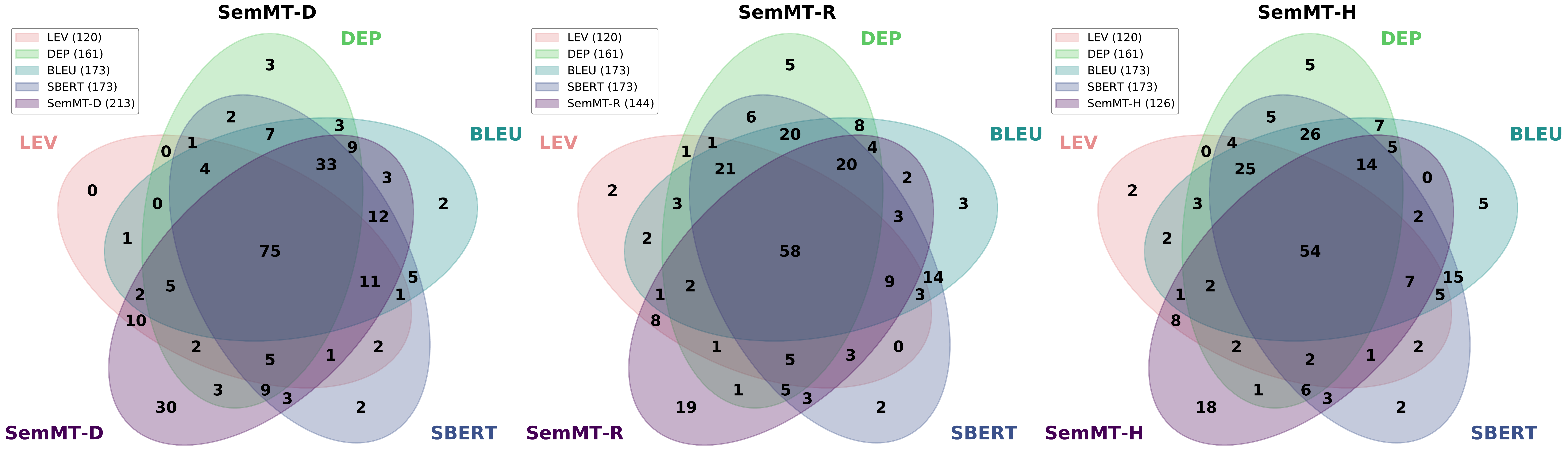}
    \caption{\textbf{Mistranslations Reported By Different Similarity Metrics.}}
    \label{fig:overlap}
\end{figure}

Fig.~\ref{fig:overlap} compares the number of bugs uniquely and commonly detected by three semantic-based metrics (i.e., REG, DFA, HYB) with other metrics. 
As shown in the figure, 
DFA detects the most mistranslations (213) compared with other metrics, with 40 (i.e., 213 - 173) more than the second most. 
Besides, though the total number of mistranslations reported by REG and HYB are not the most, the number of unique mistranslations (i.e., mistranslations that can only be detected by one metric) are high, with 19 and 18, respectively. 
Such results also reveal that existing metrics mildly complement each other since they can detect different mistranslations, indicating that their combination may lead to improvement.
Motivated by this, we studied if the performance of {SemMT} can be boosted by combining it with other existing techniques.
Specifically, we adopted a simple strategy which assumes a translation to be buggy if either of the two combined metrics report it as a mistranslation.

\begin{figure}[htbp]
    \centering
    \includegraphics[angle=0,width=1.0\linewidth]{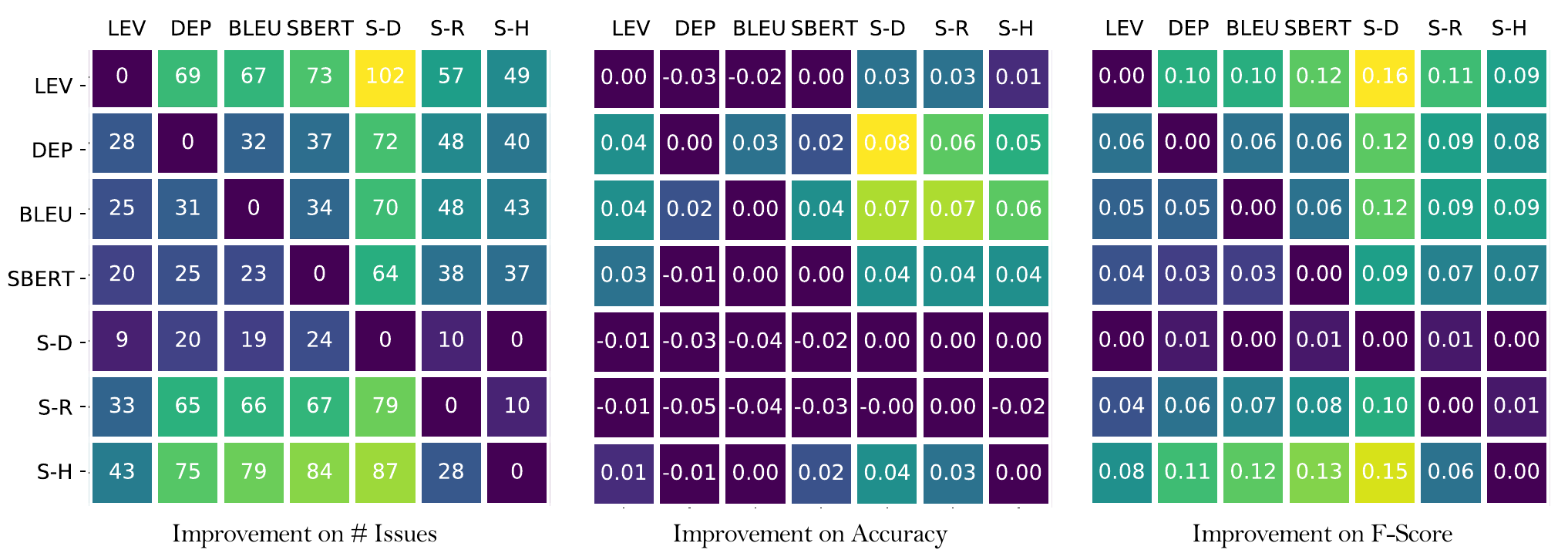}
    \caption{\textbf{Improvement on Number/Accuracy/F-Score of Mistranslations Detection Reported By Different Similarity Metrics.} 
    (S-D, S-R and S-H are abbreviations for SemMT-D, SemMT-R and SemMT-H, respectively.)}
    \label{fig:heatmap}
\end{figure}

The experiment result is illustrated in Fig.~\ref{fig:heatmap}. The heatmaps show the increase of the number of issues, ratio of accuracy and F-Score accordingly achieved by different combinations. 
The value in each grid (e.g., v[i][j] in i-th row, j-th column) represents the improvement to the i-th similarity metric by combing with the j-th metric. 
The greatest improvement is achieved when LEVEN is combined with DFA, detecting 102 more issues and achieving 16\% improvement in F-Score. 
In particular, a combined use of any of our three metrics with an existing one can detect 38 to 102 more issues. 
Our metrics also detect 9 to 79 more mistranslations when combined with another existing metric.
Even for DFA which has already reported 213 issues, a combination with SBERT can help to detect 24 more bugs. 
The F-Score of all similarity metrics are mostly improved when a metric is combined with another one. Finally, we combined all 7 metrics and found that 246 bugs can be found, with a 1\% increase in F-score.

\begin{mdframed}[style=MyFrame]
\textbf{Summary of Findings Related to RQ3:} 
The combination of different metrics can improve the effectiveness mutually to a large degree in terms of number of mistranslations reported, accuracy and F-Score. 
In particular, DFA boosts the performance of other similarity metrics the most, with 102 more mistranslations detected and 16\% higher F-Scores achieved. 
Besides, DFA can, in turn, be boosted by SBERT with 24 more mistranslations. 
\end{mdframed}

\subsection{RQ4: Applicability of SemMT } 
In the above RQs, we evaluate the effectiveness of SemMT by testing Google translator under the EN-ZH (i.e., English-Chinese) language setting.
In the following, we present three experiments to further investigate whether similar effectiveness can be observed on other popular translators, language pairs and test sets, respectively.

\subsubsection{\textbf{Impact of Translator Under Test.}}
Our first experiment is to repeat the experiment in RQ1 by replacing the Google translator with Microsoft Bing Translator. 
Specifically, we randomly selected 100 sentences as test inputs from the NL-RX-Synth dataset, 
applied the round-trip translation on the Bing translator and collected the translation results.\footnote{The translation results were collected on March 11, 2021 on Microsoft Bing translator.} 
We repeated the labeling procedure as described in \S~\ref{sec:setup} and calculated similarities across thresholds 0.0 to 1.0 at the step of 0.01, 
making sure the experiment setup is the same as that of RQ1 apart from the translator under test.

\begin{figure}[htbp]
    \centering
    \subfigure[Accuracy]{
    \includegraphics[width=0.75\linewidth]{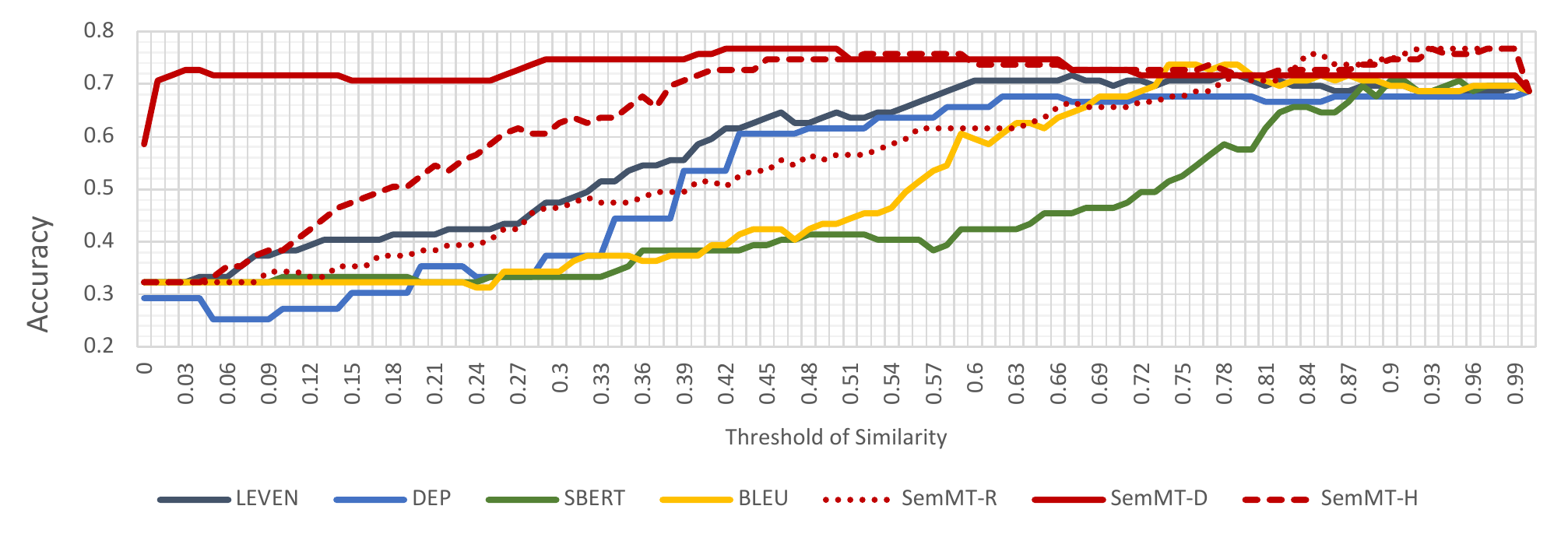}
    }
    \subfigure[F-Score]{
    \includegraphics[width=0.75\linewidth]{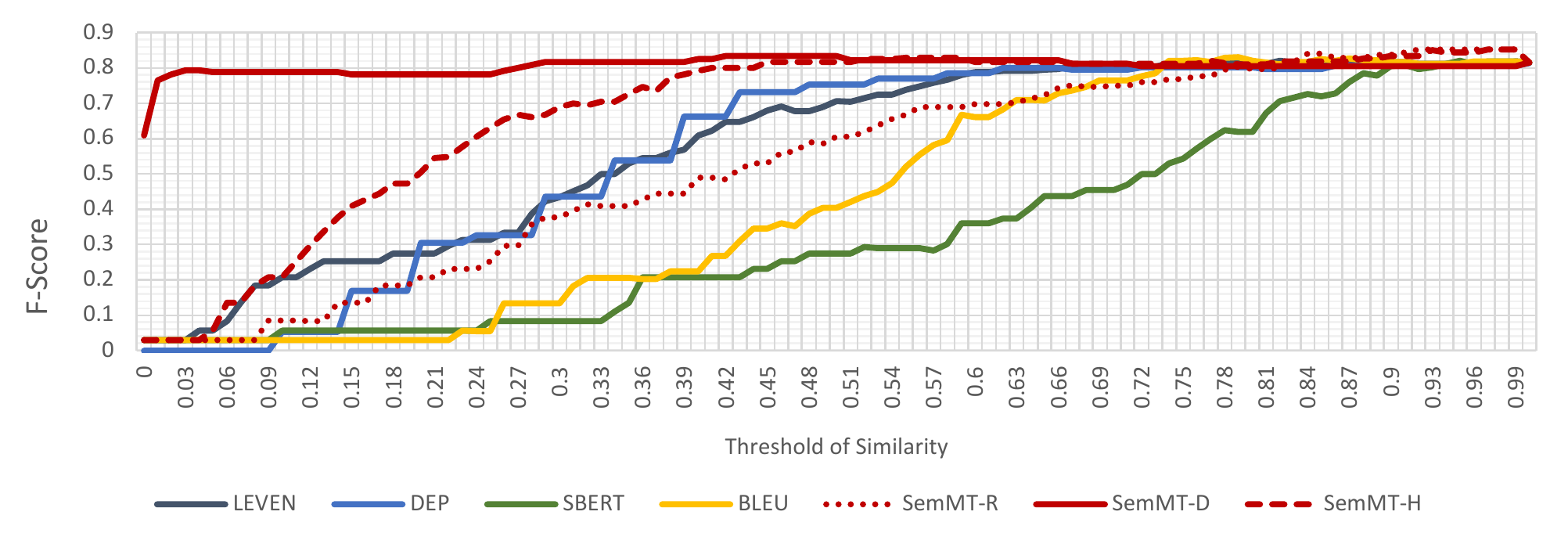}
    }
    \caption{Effectiveness of Mistranslation Detection On the Microsoft Bing Translator.}
    \label{fig:bing}
\end{figure}

The experiment result is shown in Fig.~\ref{fig:bing}. 
We can see that SemMT-D and SemMT-H outperform other metrics on both accuracy and F-Score for most threshold values. 
Specifically, SemMT-R achieves the highest accuracy (76\%) and F-Score (84\%) when the threshold is above 0.9. 
{Among the existing metrics, LEVEN and DEP achieve the highest accuracy and F-Score than others in general, while SBERT is the least effective one for most threshold values.} 
The result also shows that changing the translator under test cast little impact on the effectiveness of the three SemMT metrics. 
Specifically, different variants of SemMT still outperform existing baselines significantly while SemMT-D performs the best over most threshold values.

\subsubsection{\textbf{Impact of Language Pair.}}
\begin{figure}[htbp]
    \centering
    \subfigure[Accuracy]{
    \includegraphics[width=0.75\linewidth]{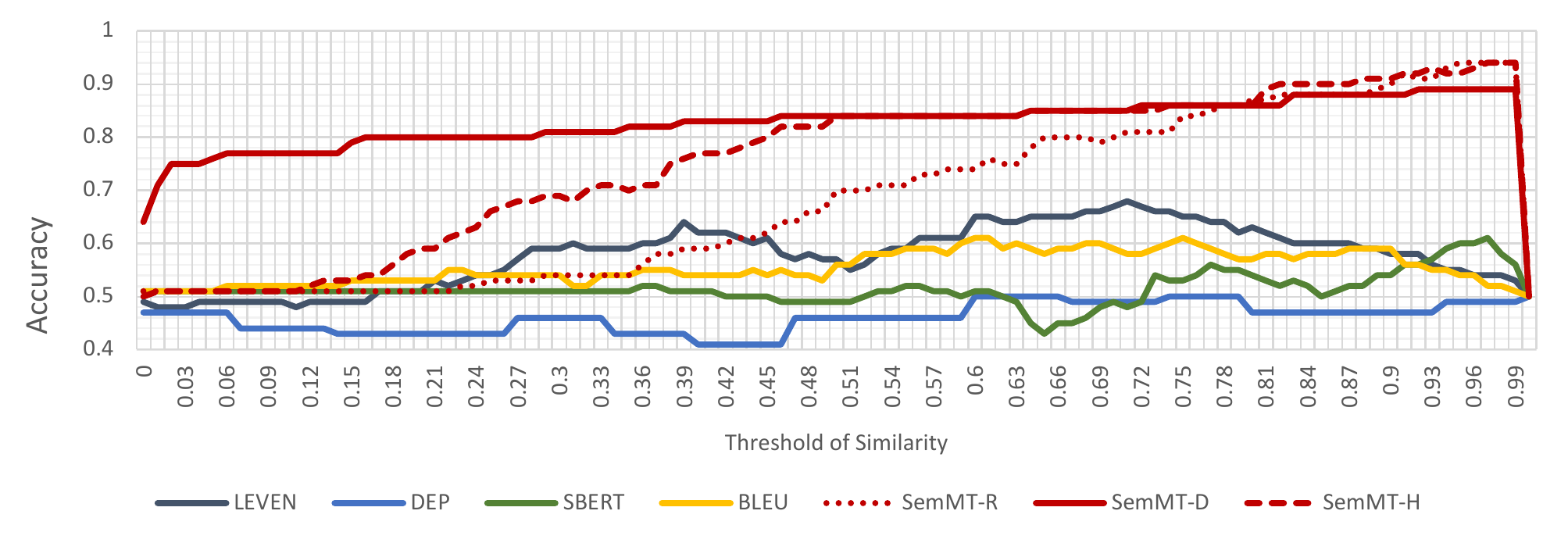}
    }
    \subfigure[F-Score]{
    \includegraphics[width=0.75\linewidth]{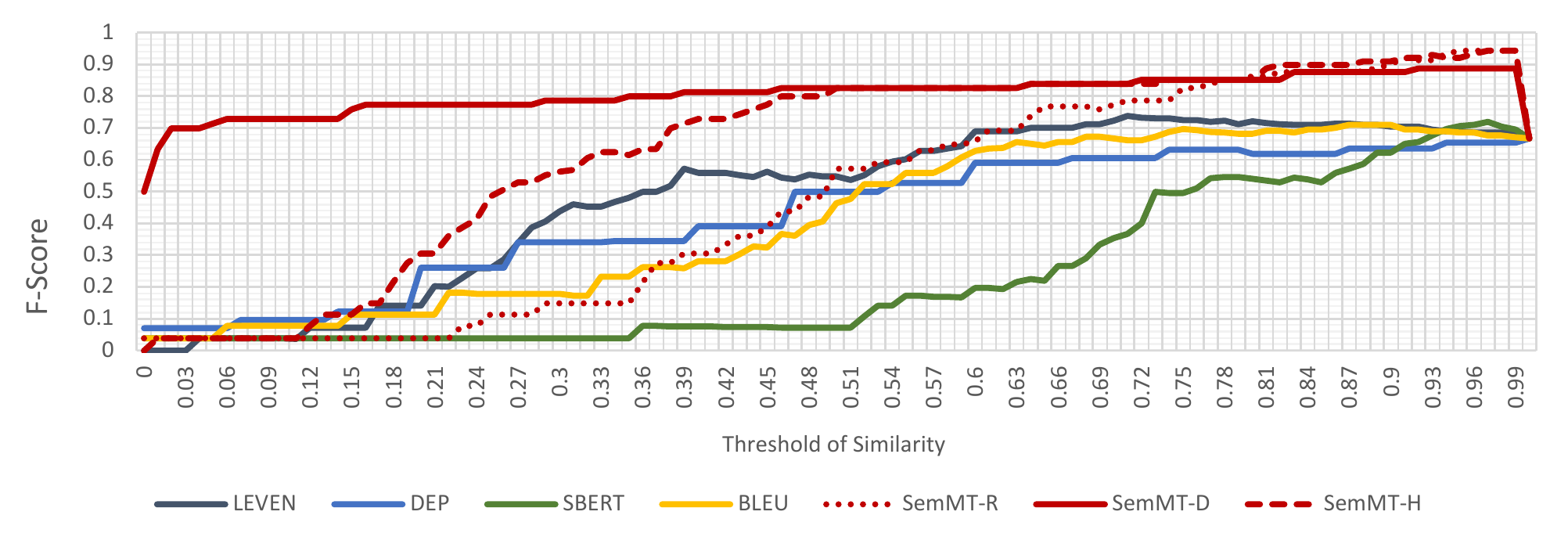}
    }
    \caption{Effectiveness of Mistranslation Detection on the English-Japanese Language Pair.}
    \label{fig:en-jp}
\end{figure}

We also examined whether the use of language pair would affect the effectiveness of SemMT. 
Specifically, besides the translation between English and Chinese, we conducted another round-trip translation between English and Japanese using the Google translator on 100 randomly selected sentences from the NL-RX-Synth dataset.\footnote{The translation results were collected on March 11, 2021 on Google translator.}
The experiment results are illustrated in Fig.~\ref{fig:en-jp}. 
The result shows that our three metrics outperform others for most of the thresholds. 
And among our three similarity metrics, SemMT-D outperforms SemMT-R and SemMT-H at most times.  
Among other existing metrics, LEVEN outperforms other existing metrics in terms of accuracy and F-Score, 
while DEP and SBERT reach the lowest accuracy and F-Score, respectively.
The result also echos that in Fig.~\ref{fig:exp-effectiveness}, 
indicating that the change of language pair has little impact on the effectiveness of our three SemMT metrics, i.e., they outperform existing baselines significantly while SemMT-D achieves the best overall effectiveness.
Furthermore, as compared with the result of RQ1 (as shown in Fig.~\ref{fig:exp-effectiveness}), 
SemMT shows similar effectiveness over the thresholds when changing the language pair, which indicates the effectiveness of SemMT may hold under the change of language pairs.

\subsubsection{\textbf{Impact of Test Dataset.}}
\begin{figure}[htbp]
    \centering
    \subfigure[Accuracy]{
    \includegraphics[width=0.75\linewidth]{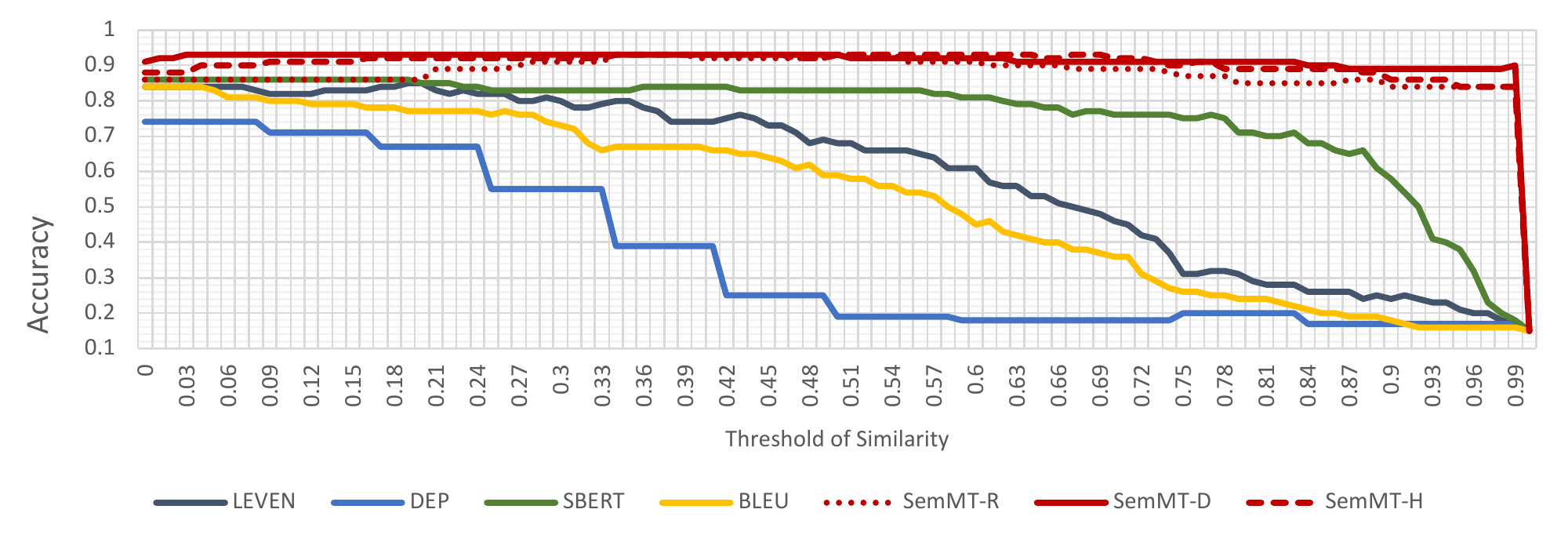}
    }
    \subfigure[F-Score]{
    \includegraphics[width=0.75\linewidth]{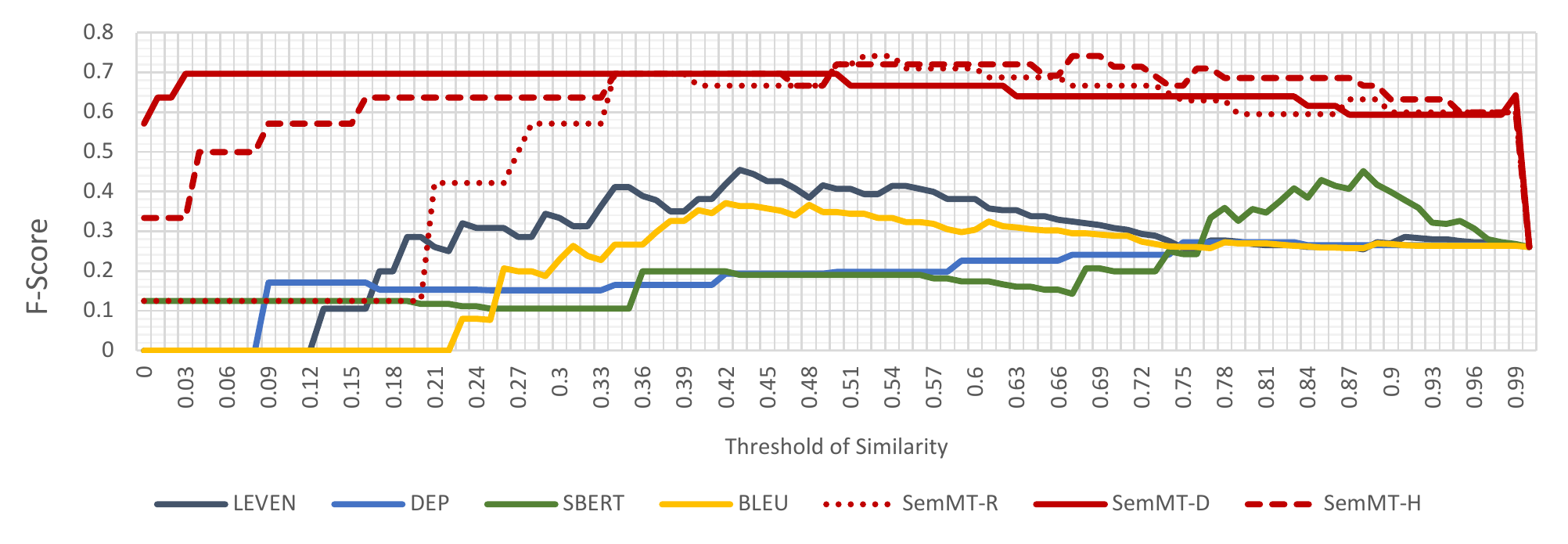}
    }
    \caption{Effectiveness of Mistranslation Detection Over KB13 Dataset.}
    \label{fig:kb13}
\end{figure}

Finally, we explored whether similar effectiveness can be achieved given test inputs extracted from other datasets. We thus randomly selected 100 sentences from the KB13 dataset 
to proceed the evaluation. 
The round-trip translation was conducted on the Google translator between English and Chinese.\footnote{The translation results were collected on March 11, 2021 on Google translator.}
The result is illustrated in Fig.~\ref{fig:kb13}. 
As shown in the red lines, we can see that three metrics of SemMT outperform other metrics as the threshold changes. 
Specifically, SemMT-D achieves the highest effectiveness on average, while SBERT and BLEU are less effective among these similarity metrics. 
In addition, compared with the results displayed in Fig.~\ref{fig:exp-effectiveness}, 
our three metrics follow similar trends on both datasets, indicating the potential applicability of applying SemMT on various datasets. 
While other existing metrics (such as LEVEN, SBERT and BLEU) show apparently less effectiveness in terms of F-Score than that on NL-RX-Synth dataset.

\begin{mdframed}[style=MyFrame]
    \textbf{Summary of Findings Related to RQ4:} 
    The applicability of SemMT has been evaluated by changing the translator (i.e., Bing translator), language pair (English-Japanese) and test set (KB13~\cite{deepregex16,softregex19}). 
    The results obtained are similar with those in RQ1, which  indicate that SemMT is applicable to performing testing under different settings, and three similarity metrics of SemMT outperform the existing ones for most thresholds.
\end{mdframed}

\subsection{Summary and Recommendation}
We made three observations on SemMT's performance from our experiments. 
First, the experiment results reveal that our proposed similarity metrics (i.e., SemMT-R, SemMT-D and SemMT-H) outperform the baselines on a wide range of threshold values in terms of accuracy and F-Score, etc. 
Second, compared with other state-of-the-art works, SemMT offers a better balance in precision and number of mistranslation detected. It also achieves an improvement of 34.2\% and 15.4\% on accuracy and F-Score, respectively, with a similar number of issues detected. 
Third, we investigate any potential improvement that can be achieved by combining different metrics, and find that SemMT-D can boost the performance of other similarity metrics with 102 more issues detected and 16\% higher F-Scores achieved. 

The three SemMT similarities have their own merits. For best precision (> 0.8), SemMT-R with small threshold (< 0.1) is a good choice. 
For best recall (> 0.8), SemMT-H with high threshold (> 0.8) and a K value of 0.5 is recommended. 
For best F-Score, SemMT-D outperforms the other two metrics in a wide threshold range. 
One may switch between the three similarity metrics and adjust the threshold value according to the needs of an application.

\section{Discussion}\label{sec:discussion}

\subsection{Characteristics of Mistranslations Detected by SemMT }

To follow up on RQ3, since each similarity metric tends to capture certain aspect of mistranslations, 
we then analyzed the types of mistranslations detected by each metric for further investigation.
We manually labeled the type of mistranslations according to the existing work~\cite{SIT}, which has concluded five types of mistranslations (i.e., Under-Translation, Over-Translation, Word/phrase Mistranslation, Incorrect Modification and Unclear Logic). 

The statistics are listed in Table~\ref{tab:stats}. We can see that for the first three types, the number of mistranslations detected by each metric type are relatively similar, while for ``Unclear logic'' and ``Modification'', the number varies a lot. 
As such, we further subdivided these two mistranslation types to better characterize these two categories of mistranslations detected by SemMT. 
These new subcategories are motivated by the mistranslations detected in sentences that are 
mainly logic- or quantifier-related. 

We subdivide ``Unclear Logic'' (i.e., all the tokens/phrases are correctly translated but the sentence logic is incorrect~\cite{SIT}) into three subcategories.
\begin{itemize}
    \item[1] \textbf{Incorrect Order.} If all tokens and phrases are translated correctly, yet the order of tokens/phrases are organized in different orders after translation, it is an incorrect order mistranslation. 
    For example, as shown in Table~\ref{tab:qualitative-quantitative-modification} (Example 1), in the original sentence, the string \textit{``dog''} is arranged before the string \textit{``truck''} or a \textit{letter} 
    , while after translation, the \textit{letter} may be arranged before string \textit{``truck''}.
    
    \item[2] \textbf{Incorrect Affiliation.} If all tokens and phrases are translated correctly, yet the affiliation relation is incorrect, it is an incorrect affiliation mistranslation. As shown in Table~\ref{tab:qualitative-quantitative-modification} (Example 2), the string contains letters and lowercase letters in the original sentence, while after translation, the affiliation is missed. 
    
    \item[3] \textbf{Incorrect Semantics.} If tokens/phrases are correctly translated, yet the semantic logic of the original sentence is not preserved after translation, it is an incorrect semantics mistranslation. An example is presented in Table ~\ref{tab:qualitative-quantitative-modification} (Example 3), the original semantic logic describes that the lines contains 5 or more numeric characters, while after translation, the semantic meaning is changed, describing the number of lines instead of the lines themselves.
\end{itemize}

For ``{Incorrect Modification}'' (i.e., if the modifier modify the wrong element), we subdivide it into two subcategories according to the type of modifiers. 

\begin{itemize}
    \item[1] \textbf{Incorrect Qualitative Modification}. If a qualitative modifier modifies the wrong element in a sentence, it is an incorrect qualitative modification mistranslation. An example is illustrated in Table~\ref{tab:qualitative-quantitative-modification} (Example 4). The attribute \textit{``string''} modifies \textit{``dog''} in the original sentence, while after translation, the modifier of \textit{``dog''} becomes \textit{``numeric character''}.
    \item[2] \textbf{Incorrect Quantitative Modification.} Similarly, if a qualitative modifier modifies the wrong elements, it is an incorrect quantitative modification mistranslation. An example is illustrated in Table~\ref{tab:qualitative-quantitative-modification} (Example 5). The quantitative modifier \textit{``at least once''} modifies different elements after translation.
\end{itemize}

\begin{table}[thbp]
\centering
\caption{Number of Quantifier- and Logic-Related Mistranslations}
\label{tab:stats}
\renewcommand\arraystretch{1.5}
\resizebox{0.8\linewidth}{!}{%
\begin{tabular}{l|cccc|ccc}
\hline
 & LEVEN & DEP & BLEU & SBERT & SemMT-D & SemMT-R & SemMT-H \\ \hline
 Under-Translation (35) & 27 & \textbf{30} & 32 & 28 & 28 & 24 & 20 \\
 Over-Translation (6) & 5 & 5 & 5 & 1 & 5 & 4 & 4 \\
 Mistranslation (1) & 1 & 1 & 1 & 1 & 1 & 1 & 1 \\  \hline \hline
 Unclear logic (60) & 36 & 34 & 37 & 30 & \textbf{59} & 47 & 43 \\ \hline
Affiliation (32) & 13 & 13 & 13 & 10 & \textbf{31} & 21 & 21 \\
Order (19) & 14 & 12 & 15 & 11 & \textbf{19} & 17 & 13 \\ 
Semantic (9) & 9 & 9 & 9 & 9 & 9 & 9 & 9\\ \hline \hline
Modification (183) & 65 & 107 & 114 & 121 & \textbf{136} & 79 & 65\\ \hline
Quality (8) & 4 & 6 & 6 & \textbf{7} & \textbf{7} & 6 & 5\\
Quantity (175) & 61 & 101 & 108 & 114 & \textbf{129} & 73 & 60 \\
 
\hline
\end{tabular}%
}
\end{table}

\begin{table}[thbp]
\centering
\tiny
\caption{Examples of Five Subcategories of Mistranslations Reported by SemMT.}
\label{tab:qualitative-quantitative-modification}
\renewcommand\arraystretch{1.3}
\resizebox{0.8\linewidth}{!}{%
\begin{tabular}{l|l}

\hline 
 & \begin{tabular}[c]{@{}l@{}}lines with the string ``dog'' \textbf{\underline{before}} the string ``truck'' or \textbf{\underline{a letter}}. \end{tabular} \\
 & \cellcolor[HTML]{E7E7E7}\begin{CJK}{UTF8}{gbsn}在字符串``truck''\textbf{\underline{之前}}的字符串``dog''或\textbf{\underline{字母}}\end{CJK} \\
\multirow{-3}{*}{\textbf{\begin{tabular}[c]{@{}l@{}}Example 1\\ (Incorrect\\ Order)\end{tabular}}} & \begin{tabular}[c]{@{}l@{}}the string ``dog'' or \textbf{\underline{letter}} \textbf{\underline{before}} the string ``truck''.\end{tabular} \\ 
\hline
 & \begin{tabular}[c]{@{}l@{}}strings \textbf{\underline{with a letter followed by}} a lower-case letter, zero or more times. \end{tabular} \\
 & \cellcolor[HTML]{E7E7E7}\begin{CJK}{UTF8}{gbsn}字符串，其\textbf{\underline{后跟}}一个小写字母的\textbf{\underline{字母}}，零次或多次
\end{CJK} \\
\multirow{-3}{*}{\textbf{\begin{tabular}[c]{@{}l@{}}Example 2\\ (Incorrect\\ Affiliation)\end{tabular}}} & \begin{tabular}[c]{@{}l@{}}string , \textbf{\underline{followed by a letter with}} a lowercase letter, zero or more times.\end{tabular}\\
\hline
 & \begin{tabular}[c]{@{}l@{}}\textbf{\underline{lines with a number}} , 5 or more times. \end{tabular} \\
 & \cellcolor[HTML]{E7E7E7}\begin{CJK}{UTF8}{gbsn}\textbf{\underline{行数}}大于等于5次\end{CJK} \\
\multirow{-3}{*}{\textbf{\begin{tabular}[c]{@{}l@{}}Example 3\\ (Incorrect\\ Semantic)\end{tabular}}} & \begin{tabular}[c]{@{}l@{}}\textbf{\underline{the number of rows}} is greater than or equal to 5.\end{tabular}\\
\hline
 & \begin{tabular}[c]{@{}l@{}}lines with the \textbf{\underline{string ``dog''}} before a vowel or a numeric character.\end{tabular} \\
 & \cellcolor[HTML]{E7E7E7}\begin{CJK}{UTF8}{gbsn}在元音或数字字符之前的\textbf{\underline{字符串``dog''}}\end{CJK} \\
\multirow{-3}{*}{\textbf{\begin{tabular}[c]{@{}l@{}}Example 4\\ (Qualitative \\Modification)\end{tabular}}} & string before vowel or \textbf{\underline{numeric character ``dog''}}. \\ 
\hline
 & \begin{tabular}[c]{@{}l@{}}lines starting with \textcolor[HTML]{1F4E79}{\textbf{a lower case letter}} \textbf{\underline{at least once}} or  \textcolor[HTML]{C00001}{\textbf{a capital letter}}. \end{tabular} \\
 & \cellcolor[HTML]{E7E7E7}\begin{CJK}{UTF8}{gbsn}以\textcolor[HTML]{1F4E79}{\textbf{小写字母}}开头的行或\textbf{\underline{至少一个}}\textcolor[HTML]{C00001}{\textbf{大写字母}}的行\end{CJK} \\ 
\multirow{-3}{*}{\textbf{\begin{tabular}[c]{@{}l@{}}Example 5\\ (Quantitative \\Modification)\end{tabular}}} & \begin{tabular}[c]{@{}l@{}}lines beginning with \textcolor[HTML]{1F4E79}{\textbf{lowercase letters}} or lines with \textbf{\underline{at least one}} \\ \textcolor[HTML]{C00001}{\textbf{capital letter}}\end{tabular}\\
\hline
\end{tabular}%
}
\end{table}


\subsection{False Positives of SemMT}
\noindent\textbf{False Positives Caused by Singularity and Plurality.}
When we analyzed the mistranslations reported by SemMT in dataset used in \S\ref{sec:RQ1}, there are 148 mistranslations among 500 are plural-related, i.e., nouns are the same in both singular and plural forms in Chinese, while they are in different forms in English.
As a result, the singularity/plurality is mistakenly missed or imposed during translation. Such plural-related mistranslations are commonly-seen across languages. For example, languages such as Slovenian, Russian, and Welsh have several plural forms, while languages such as Chinese and Japanese do not have counterparts to the forms of singular and plural in languages like English. Though minor and easy-to-neglect, in aid of regex and DFA, our SemMT is able to capture such subtle differences to some degree, nevertheless false positives may be caused due to this reason. 

\noindent\textbf{False Positives Caused by Inaccurate Transformation From Natural Language to Regular Expression.}
The inaccurate transformation from natural language sentences to the corresponding regexes may also lead to false positives of SemMT. The main reason is caused by the existence of out-of-vocabulary words in the round-trip translated sentences, leading to the inaccurate transformation from natural language to regex.
To alleviate such concerns, we enlarged the vocabulary of the training dataset in order to achieve accuracy as high as we can.
{Specifically, we collected the translation results from the Google translator, and obtained a list of parsed tokens. 
Then we augmented the training data by synonym substitution (i.e., replacing the tokens in the original training dataset by the tokens that are not in the original dataset but are derived in the returned dataset.). 
}

\subsection{Influence of Approximated Semantics Measurement on Mistranslation Detection}
\label{sec:approx}
In \S\ref{sec:approx-trans}, we explained how can the semantics be captured over- or under-approximately. However, if we apply the 
approximated semantics to detect mistranslations over original
and translated sentences, the derived results are likely to be unreliable due to the approximated semantics. 
Take the example (i.e., S5) and its back-translated sentence (i.e., T5) 
in \S\ref{sec:approx-trans} as examples~\footnote{The translation results were collected on December 24, 2020 on Google Translator, using Chinese as the intermediate language.}:
\begin{itemize}
    \item S5: The U.S. contains \textbf{a few} states which choose to have the judges on the state's courts serve for life terms.
    \item T5: The United States contains \textbf{several} states, which choose to let judges in state courts serve for life.
\end{itemize}

After sentence abstraction, we will obtain the following abstracted sentences:

\begin{itemize}
    \item S5': [X] contains \textbf{a few} [Y].
    \item T5': [X] contains \textbf{several} [Y].
\end{itemize}

\noindent Since ``a few'' and ``several'' are vague quantifiers for which the semantics are hard to be precisely quantified, we approximate their semantics using regexes as follows:

\begin{itemize}
    \item R\_{S5'}$_{O}$: \verb|[Y]{3,}|
    \item R\_{T5'}$_{O}$: \verb|[Y]{5,}|
\end{itemize}

After calculation, the SemMT-D semantic similarity is up to 0.957, meaning that there is little difference after translation. 
However, people tend to believe ``a few'' is less than ``several''~\cite{hintikka1994quantifier}, while such semantic difference between these two vague quantifiers is hard to be captured after approximation. 
As a result, it may be harder to capture semantic difference due to the wider range of quantification.

Therefore, we discuss one possible solution, i.e., to quantify vague quantifiers more precisely by considering the common practice and the context of the sentence. Plenty of studies have been performed on quantifying vague quantifiers~\cite{wright1994much, lappin2000intensional, hox2003response,Vague1979,solt2011vagueness} from logical, linguistic and psychological aspects. For example, ``a few'' and ``several'' are believed to be less than the quantifier ``a half''. And being aware of the fact that the number of states in the U.S. is at most 50, then the quantification-related semantics of the above sentences (S5' and T5') can be more precisely quantified by the following regexes:
\begin{itemize}
    \item RS5'$_{P}$: \verb|[Y]{3,25}|
    \item RT5'$_{P}$: \verb|[Y]{5,25}|
\end{itemize}

\noindent where the quantifiers ``a few'' and ``several'' are quantified to be \verb|{3,25}| and \verb|{5,25}|, respectively. By doing so, the SemMT-D similarity between them is $0.84$, where the subtle semantic difference can be detected.

\subsection{Buggy Trip Localization}
\label{sec:localization}
Despite the advantage of round-trip translation, it has been criticized for not testing one translation system but two~\cite{goodfor05,revisiting20}. Hence, in this section, we discuss a potential solution to localize buggy trip (the trip that produces wrong translation) using the idea of cross reference. 
The intuition is as follows: if the translation returned by one translator is different from the results of other translators, it is likely to be incorrect. The less similar with other translation results, the more likely the original sentence is error-prone. 

\begin{table}[t!]
\centering
\caption{\textbf{Statistics of Buggy Trip Localization.} The first three major columns denote the average similarity scores or distances over correctly/mistranslated sentences across different translators. The last major column shows the number and accuracy of correctly identified buggy trip using different similarity metrics. The values in bold represent the maximum number of correctly located buggy trip.}
\label{tab:mistranslation-trip}
\renewcommand\arraystretch{1.5}
\resizebox{1.0\linewidth}{!}{%
\begin{tabular}{r|cc|cc|cc|ccc}
\hline
\multirow{2}{*}{{}} & \multicolumn{2}{c|}{{AveSim\_Correct}} & \multicolumn{2}{c|}{{AveSim\_BuggyFW}} & \multicolumn{2}{c|}{{AveSim\_BuggyBW}} & \multirow{2}{*}{\# FW} & \multirow{2}{*}{\# BW} & \multirow{2}{*}{Accuracy} \\ \cline{2-7}
 & {Sim\_{{FW}}} & {Sim\_{{BW}}} & {Sim\_{FW}} & \multicolumn{1}{c|}{{Sim\_{BW}}} & {Sim\_{FW}} & \multicolumn{1}{c|}{{Sim\_{BW}}} &  &  &  \\ \hline
{LEVEN} & {0.50} & {0.64} & {0.39} & 0.66 & 0.42 & {0.54} & 57 & 19 & 0.65 \\
DEP & 3.15 & 6.34 & 3.07 & 5.70 & {3.28} & {8.87} & 36 & 8 & 0.38 \\
BLEU & 0.42 & 0.67 & {0.36} & 0.74 & 0.44 & {0.62} & 55 & \textbf{23} & 0.67 \\
SBERT & 0.94 & 0.90 & {0.91} & 0.92 & 0.95 & {0.89} & \textbf{74} & 15 & {0.76} \\ \hline
\end{tabular}%
}
\end{table}

For better understanding, we analyzed 500 pairs of round-trip translations used in \S\ref{sec:RQ1} and manually identified the buggy trip for all the 265 mistranslations. If both trips are mistranslated, we regarded it as a forward mistranslation because it is where the mistranslation was first introduced. Apart from the 148 plural-related mistranslations, there are 87 forward mistranslations and 30 backward mistranslations in total. For cross reference, we used the Microsoft Bing and Youdao translator. 
Then we conducted preliminary statistics, calculating the average similarity scores using four similarity metrics (i.e., LEVEN, DEP, BLEU and SBERT) across two translation trips (i.e., forward trip translating from English to Chinese, and backward trip from Chinese to English) for correctly and incorrectly translated sentences.

The result is shown in Table~\ref{tab:mistranslation-trip}. The values in the first three major columns (i.e., \texttt{AveSim\_Correct}, \texttt{AveSim\_BuggyFW)} and \texttt{AveSim\_BuggyBW} denote the average similarity scores of the forward translations (columns \texttt{Sim\_{FW}}) in Chinese and  the backward sentences (columns \texttt{Sim\_{BW}}) except for DEP which calculates the distance. The higher the similarity scores, the more similar the sentences that are translated by different translators. 
We can see that similarity scores are not identical in different translation trips, and on average, the similarity scores of the correct translated sentences (column \texttt{AveSim\_Correct}) are higher than that of mistranslated ones (column \texttt{AveSim\_BuggyFW} and \texttt{AveSim\_BuggyBW}). For example, for LEVEN, the average similarity scores in the forward and backward trips are $0.50$ and $0.64$, respectively. And the average similarity on correctly translated sentences is $0.50$, while if the forward translation on Google is mistranslated, the similarity between sentences translated by Google was lower (with $0.39$) than $0.50$. Similar patterns can be observed for other similarity metrics. 
{To sum up, we made three observations: (1) The similarity score for correct or incorrect translations vary from languages. (2) Within the same language, the similarity scores also vary from different similarity metrics. (3) The similarity scores of the correct translated sentences are higher than that of mistranslated ones on average.}

On top of these observations, we tried the following strategy: given an original sentence which has been mistranslated in either trip, if the difference between the average similarity scores (i.e., as shown in Table \ref{tab:mistranslation-trip})
for the correct translations and the forward similarity scores is larger than or equal to backward trip, then this mistranslation is considered as a forward-trip mistranslation, otherwise a backward-trip mistranslation. 
The result is listed in the last three columns in Table~\ref{tab:mistranslation-trip}. We can see that the capability of differentiating mistranslation trip differs for different similarity metrics. Specifically, SBERT achieved the highest accuracy (76\%) by correctly identifying 74 forward mistranslations over 87 and 15 backward mistranslations over 30, while the syntactic-based DEP has the lowest accuracy (38\%). In addition, the result indicates different similarity metrics tend to have diverse capabilities in identifying different translation trips.
For example, SBERT founds the most forward buggy trip (74) while BLEU performs better in identifying the buggy backward trip.

\subsection{Effect of the Parameter K on SemMT-H Similarity}\label{sec:k-selection}

\begin{figure}[t!]
\centering
\subfigure[Accuracy]{
\includegraphics[width=0.65\linewidth]{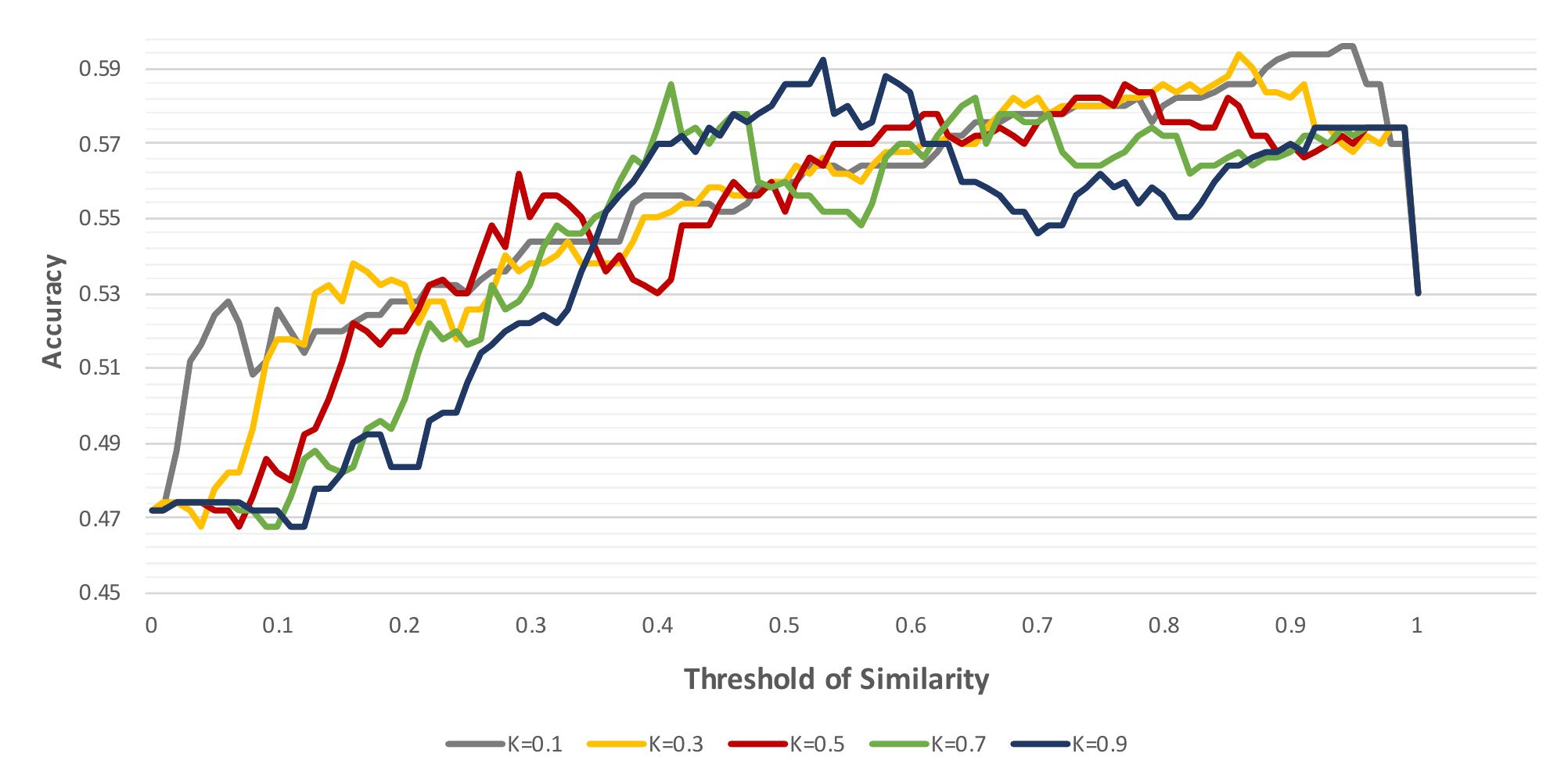}
}
\subfigure[F-Score]{
\includegraphics[width=0.65\linewidth]{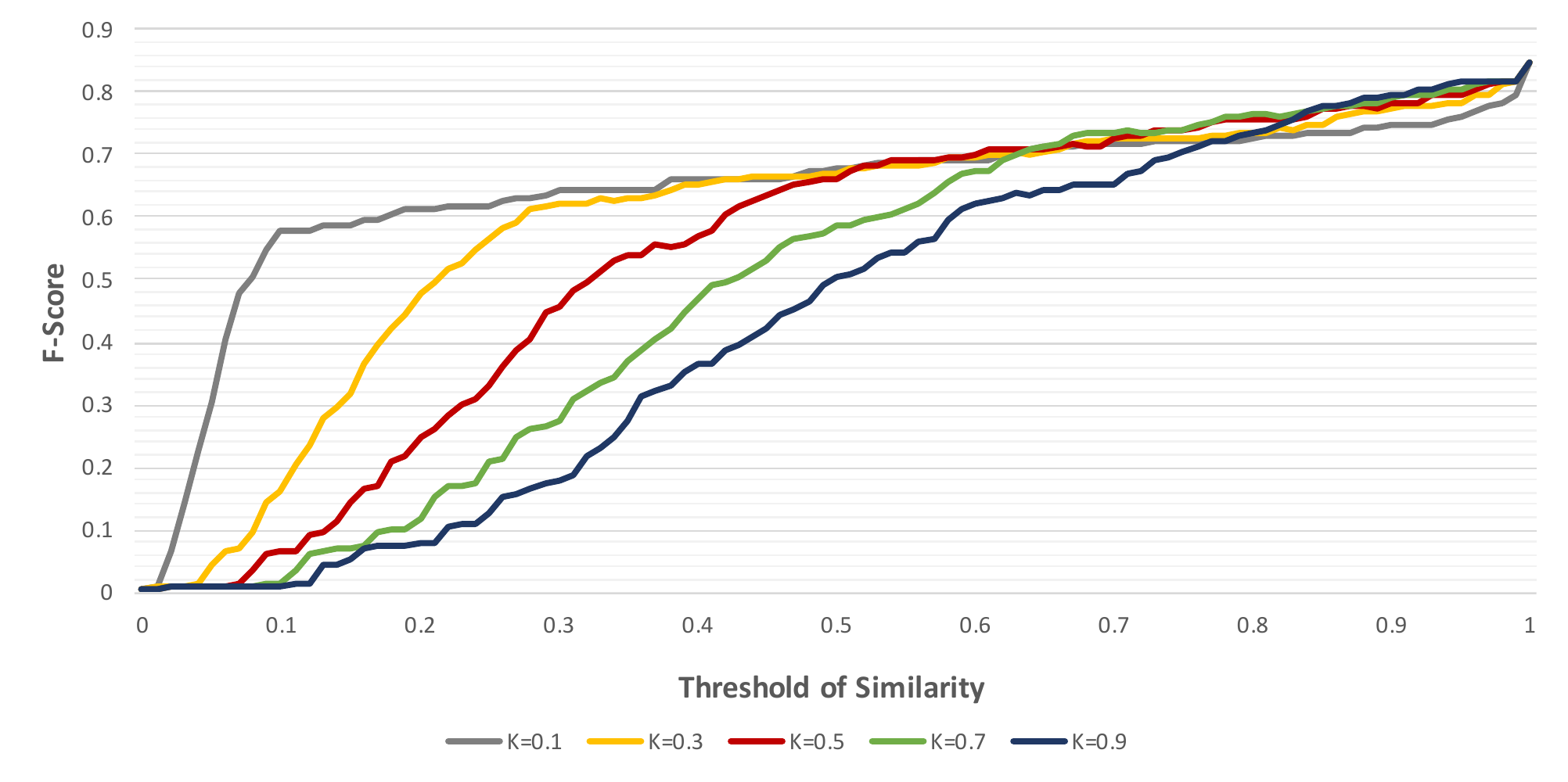}
}
\caption{Effectiveness of Mistranslation Detection Over Different Parameter K on Hybrid-based Similarity.}
\label{fig:k-sensitive}
\end{figure}

On top of the remarkable results achieved by SemMT, we explore a further question: \textit{whether the setting of K will affect the effectiveness of SemMT-H?}
To answer this question, we set K from 0.1 to 0.9 with a step of 0.2 (i.e., K is set to be $0.1$, $0.3$, $0.5$, $0.7$ and $0.9$), and examined the accuracy and F-Score on SemMT-H with different values of K against all the threshold values. 
We can see from Fig.~\ref{fig:k-sensitive} that the general trends of accuracy are similar under various fluctuations, while the trends of F-Score vary largely. Specifically, for accuracy, choosing a smaller K value in the hybrid metric would achieve better performance when the similarity threshold is less than $0.4$ or larger than $0.8$.
For F-Score, a small K value in the hybrid metric outperforms that using a large K value when threshold is less than 0.8, and the reverse situation is observed when using a threshold beyond (0.8 to 1.0). Furthermore, considering accuracy and F-Score together, SemMT-H similarity offers a better performance over a wide threshold range when K assumes a value between 0.3 and 0.5. And if SemMT-H similarity is to be used under a small threshold value where precision takes priority, a small K such as 0.1 is preferable.

\section{Related work}
\subsection{Machine Translation Testing}
Machine translation testing aims at finding sentences that trigger translation errors~\cite{pathology}. 
Pesu et al.~\cite{monte18} first applied metamorphic testing on machine translation systems. They proposed a Monte Carlo method to avoid round-trip translation by selecting eight target languages given the fixed source language, English. Under the factorial design and analysis, they evaluated the performance of translation over different combination of the source and target languages. After that, more metamorphic testings were developed. 
Sun et al.~\cite{MT4MT} designed straight-forward metamorphic relations focusing on short sentences in \textit{subject-verb-object} structure (e.g., ``Mike loves to eat KFC'' and ``Mouse loves to eat KFC''). They generated test inputs by replacing human names before ``likes'' or ``hates'', and brands after them. 
Wang et al.~\cite{wechat19} detected under- and over-translation in the absence of reference translation. By checking the frequency of occurrence and learning mappings between bilingual words and phrase, their work is able to detect these two kinds of mistranslations efficiently and scalably.
Later, He et al.~\cite{SIT} and Sun et al.~\cite{transrepair20} developed metamorphic testing techniques for general translation errors based on the assumption that similar sentences should have similar translation results. 
To be more specific, SIT~\cite{SIT} generated similar testing inputs by substituting one word in a given sentence, such that the generated inputs are semantically-similar and syntactically equivalent as the given sentence. Then they 
further reported the suspected issues if structures of translated sentences are different.
Similarly, TransRepair~\cite{transrepair20} conducted mutation testing via context-similar word replacement. The intuition is that the translations of both original sentence and mutants should be consistent except for the changed token.
While PatInv~\cite{pathology}, on the other hand, considered pathological invariance: sentences of {with} different meanings should not have the same translation. Following this intuition, they generated syntactically similar but semantically different sentences by either replacing one word with a non-synonymous word using masked language models or removing one word based on its constituency structure.  They further detected the potential mistranslations with closer textual similarity.
We can see that the existing techniques mainly focused on textual or syntactical level, while the preservation of semantics during translation has not gained enough attention.
Therefore, SemMT aims at filling this gap by evaluating whether the semantic meaning is preserved during translation, complementary to the existing works.

\subsection{Robustness of Neural Machine Translator}
Evaluating on adversarial examples has become a standard procedure to measure the robustness of deep learning models~\cite{Goodfellow15ICLR}.
Adversarial examples are inputs designed to slightly manipulate the real-world examples such that a well-trained machine learning model performs poorly against these adversarial examples~\cite{ebrahimi-etal-2018-adversarial}. 
In general, these works mainly fall into two categories: black-box and white-box methodologies. In particular, for black-box adversarial samples generation, they assume the implementation of translation system is agnostic. 
Belinkov et al.~\cite{DBLP:conf/iclr/BelinkovB18} showed that character-level machine translation systems are overly sensitive to random character manipulations, such as keyboard typos. They used black-box heuristics to generate character-level adversarial examples, without using the model parameters or gradients to generate adversarial examples. Zhao et al.~\cite{DBLP:conf/iclr/ZhaoDS18} searched for black-box adversarial examples in the space of encoded sentences and generate adversarial examples by perturbing the latent representation until the model is tricked.
On the other hand, the white-box methodologies are model-aware. Ebrahimi et al.~\cite{ebrahimi-etal-2018-adversarial} investigated adversarial examples of both untargeted and targeted attack for character-level neural machine translation in a white-box manner. They transferred this problem into an optimization problem, then generated the adversarial examples utilizing gradients of translation models to inflict more damaging manipulations for a larger decrease in the BLEU score or other target metrics.

\subsection{Quality Estimation of Machine Translation}
Unlike machine translation testing, quality estimation considers beyond correctness - it aims at deriving similar estimation results made by humans. 
Traditionally, it estimates the required amount of post-editing efforts for converting the given translation result to the reference translation~\cite{snover2006study}. In the recent decade, the trend is to find effective quality estimation metrics which can directly provide scores to the translation result without human-written reference~\cite{findingsWMT19}. To alleviate the manual effort in providing reference translation, round-trip translation (RTT) has been proposed~\cite{European06,BT07}. 
The general idea of RTT is to use the original sentence as reference, compared the translated sentence with it, and calculated estimation metrics such as BLEU score to show the correlation with human judgement.
In early 2010s, Aiken el al.~\cite{efficacy10} manually reassessed the correlation on input sentences and translated outputs and reassessed this correlation in a RTT manner. The result implied that if a suitable semantic-level metric is provided, RTT-based method can be reliably used for machine translation evaluation. They also pointed out that RTT quality might reflect the general quality of machine translation system used over the length of a longer text or multiple language pairs. Afterwards, with the emergence of BERT~\cite{BERT19} and SBERT~\cite{SBERT19}, semantic similarity on both word- and sentence-level can be better captured~\cite{DBLP:conf/semeval/CerDALS17}. Moon et al.~\cite{revisiting20} then revisited RTT for quality estimation and achieved the highest correlations with human judgments compared with the state-of-the-art works, indicating that RTT-based method can be used to evaluate machine translation systems when semantic similarity is considered. By observing the correlated results, they illustrated the robustness of the choice of backward translation system on RTT-based quality estimation. It motivates us to adopt RTT and develop three semantic similarity metrics.

\section{Threats to Validity}

The semantic metrics used in SemMT 
rely on the transformation of regexes from natural language. The precision of regex synthesis may affect SemMT's performance. 
If the derived regexes are imprecise, our proposed semantic similarity metrics might not accurately measure the real semantic relationship between source and translated sentence. To alleviate the influence of imprecise regex transformation, we adopted the state-of-the-art model and trained the model on augmented dataset to minimize the inaccurate prediction caused by out-of-vocabulary problem.

Our analysis on false positives reported by SemMT shows that the performance of SemMT can also be influenced by invalid synthesized regexes such as unpaired or mis-paired brackets (e.g., \verb|[a-z{0,3}| is an invalid regex due to lacking of a square bracket) or incorrect combination of operators (e.g., \verb|+{0,3}|). The invalid regex cannot be transformed into DFA, making the similarity computation of DFAs infeasible. 
To solve this problem, we performed post processing of transformed regexes to mitigate this problem, pairing/repairing the unpaired/mis-paired brackets. 

Moreover, the precision cannot be ensured if the original sentence presents unclear/ambiguous logic~\cite{softregex19,semregex18,deepregex16}. 
Even though the round-trip testing that we used is to compare two sentences in the same language, the ambiguity of translation between the two languages in the forward trip and backward trip can influence our test results. The consequences of ambiguity include mistranslation of sentences and incorrect regex transformation (i.e., the regex is mistakenly transformed, leading to the change in semantic meaning after transformation).
Besides, the reliance on published transformation tools from natural language to regex limits our evaluation to English datasets.

{Finally, the English proficiency of authors may cast impacts on the evaluation. 
To alleviate such impacts, 
we consulted a linguist to ensure the semantics are preserved during the synonym substitution 
(during dataset augmentation and mutation operator construction described in \S\ref{sec:setup}). 
In addition, for quantifying vague quantifiers when conducting semantics approximation (\S\ref{sec:approx-trans} and \S\ref{sec:approx}), 
we also discussed with the expert linguist for confirmation.
}

\section{Conclusion}
In this paper, we proposed SemMT, a semantic-based machine translation testing framework. It tests the semantic similarity during translation, taking the first step to semantic-aware testing approach for translation systems. 
Specifically, we focused on the semantics of quantifiers and logical relations, which take up a non-trivial ratio in the daily life, and the mistranslation of them may cause severe consequences. Via transforming such sentences into regular expressions, SemMT can capture the semantics of sentences during translation, and detect the suspicious mistranslation by semantic similarity measurements. 
The evaluation showed that SemMT can achieve higher effectiveness compared with state-of-the-art works, achieving an increase of 34.2\% on accuracy. 
Furthermore, considering the unique mistranslation detection, SemMT can cover the most of the bugs found by existing techniques and can locate 67 additional bugs that are ignored by existing techniques. 
Furthermore, our exploration also indicated the potential improvement may occur when proper combinations of various similarity metrics are adopted.

\section*{Acknowledgment}
The authors would like to thank the anonymous reviewers for their comments and suggestions. We also like to thank the communicator tutor, Mrs. Shauna Dalton who helped with proofreading and advice on the English usage from the linguistic perspective. This work was supported by the National Key Research and Development Program of China, No. 2019YFE0198100, National Natural Science Foundation of China (Grant Nos. 61932021, 61872339, 61472405, and 62002125), Hong Kong ITF (Grant No: MHP/055/19), Hong Kong RGC/RIF (Grant No. R5034-18), MSRA Collaborative Research Grant and Huawei PhD Fellowship.


\bibliographystyle{ACM-Reference-Format}
\bibliography{reference}


\begin{thebibliography}{99}


\ifx \showCODEN    \undefined \def \showCODEN     #1{\unskip}     \fi
\ifx \showDOI      \undefined \def \showDOI       #1{#1}\fi
\ifx \showISBNx    \undefined \def \showISBNx     #1{\unskip}     \fi
\ifx \showISBNxiii \undefined \def \showISBNxiii  #1{\unskip}     \fi
\ifx \showISSN     \undefined \def \showISSN      #1{\unskip}     \fi
\ifx \showLCCN     \undefined \def \showLCCN      #1{\unskip}     \fi
\ifx \shownote     \undefined \def \shownote      #1{#1}          \fi
\ifx \showarticletitle \undefined \def \showarticletitle #1{#1}   \fi
\ifx \showURL      \undefined \def \showURL       {\relax}        \fi
\providecommand\bibfield[2]{#2}
\providecommand\bibinfo[2]{#2}
\providecommand\natexlab[1]{#1}
\providecommand\showeprint[2][]{arXiv:#2}

\bibitem[\protect\citeauthoryear{??}{WMT}{2013}]%
        {WMT13}
 \bibinfo{year}{2013}\natexlab{}.
\newblock \bibinfo{title}{Workshop on Machine Translation}.
\newblock
\newblock
\urldef\tempurl%
\url{http://www.statmt.org/wmt13/}
\showURL{%
\tempurl}


\bibitem[\protect\citeauthoryear{??}{Sem}{2021}]%
        {SemMT}
 \bibinfo{year}{2021}\natexlab{}.
\newblock \bibinfo{title}{SemMT}.
\newblock
\newblock
\urldef\tempurl%
\url{https://github.com/ArabelaTso/SemMT}
\showURL{%
\tempurl}


\bibitem[\protect\citeauthoryear{Ahmad and Fulford}{Ahmad and Fulford}{1992}]%
        {ahmad1992semantic}
\bibfield{author}{\bibinfo{person}{Khurshid Ahmad} {and}
  \bibinfo{person}{Heather Fulford}.} \bibinfo{year}{1992}\natexlab{}.
\newblock \showarticletitle{Semantic relations and their use in elaborating
  terminology}.
\newblock \bibinfo{journal}{\emph{Computing Sciences Report CS-92-07.
  Guildford: University of Surrey}} (\bibinfo{year}{1992}).
\newblock


\bibitem[\protect\citeauthoryear{Aiken and Park}{Aiken and Park}{2010}]%
        {efficacy10}
\bibfield{author}{\bibinfo{person}{Milam Aiken} {and} \bibinfo{person}{Mina
  Park}.} \bibinfo{year}{2010}\natexlab{}.
\newblock \showarticletitle{The Efficacy of Round-trip Translation for MT
  Evaluation}.
\newblock \bibinfo{journal}{\emph{Translation Journal}} \bibinfo{volume}{14},
  \bibinfo{number}{1} (\bibinfo{year}{2010}).
\newblock


\bibitem[\protect\citeauthoryear{Allwood, Andersson, Andersson, and
  Dahl}{Allwood et~al\mbox{.}}{1977}]%
        {logic_in_linguistics}
\bibfield{author}{\bibinfo{person}{Jens Allwood},
  \bibinfo{person}{Gunnar-Gunnar Andersson}, \bibinfo{person}{Lars-Gunnar
  Andersson}, {and} \bibinfo{person}{Osten Dahl}.}
  \bibinfo{year}{1977}\natexlab{}.
\newblock \bibinfo{booktitle}{\emph{Logic in Linguistics}}.
\newblock \bibinfo{publisher}{Cambridge University Press}.
\newblock


\bibitem[\protect\citeauthoryear{Arasu, Ganti, and Kaushik}{Arasu
  et~al\mbox{.}}{2006}]%
        {DBLP:conf/vldb/ArasuGK06}
\bibfield{author}{\bibinfo{person}{Arvind Arasu}, \bibinfo{person}{Venkatesh
  Ganti}, {and} \bibinfo{person}{Raghav Kaushik}.}
  \bibinfo{year}{2006}\natexlab{}.
\newblock \showarticletitle{Efficient Exact Set-Similarity Joins}. In
  \bibinfo{booktitle}{\emph{Proceedings of the 32nd International Conference on
  Very Large Data Bases, Seoul, Korea, September 12-15, 2006}}.
  \bibinfo{publisher}{{ACM}}, \bibinfo{pages}{918--929}.
\newblock


\bibitem[\protect\citeauthoryear{Aulamo, Virpioja, and Tiedemann}{Aulamo
  et~al\mbox{.}}{2020}]%
        {DBLP:conf/acl/AulamoVT20}
\bibfield{author}{\bibinfo{person}{Mikko Aulamo}, \bibinfo{person}{Sami
  Virpioja}, {and} \bibinfo{person}{J{\"{o}}rg Tiedemann}.}
  \bibinfo{year}{2020}\natexlab{}.
\newblock \showarticletitle{OpusFilter: {A} Configurable Parallel Corpus
  Filtering Toolbox}. In \bibinfo{booktitle}{\emph{Proceedings of the 58th
  Annual Meeting of the Association for Computational Linguistics: System
  Demonstrations, {ACL} 2020, Online, July 5-10, 2020}},
  \bibfield{editor}{\bibinfo{person}{Asli Celikyilmaz} {and}
  \bibinfo{person}{Tsung{-}Hsien Wen}} (Eds.). \bibinfo{publisher}{Association
  for Computational Linguistics}, \bibinfo{pages}{150--156}.
\newblock
\urldef\tempurl%
\url{https://doi.org/10.18653/v1/2020.acl-demos.20}
\showDOI{\tempurl}


\bibitem[\protect\citeauthoryear{Bach, Jelinek, Kratzer, and Partee}{Bach
  et~al\mbox{.}}{2013}]%
        {bach2013quantification}
\bibfield{author}{\bibinfo{person}{Elke Bach}, \bibinfo{person}{Eloise
  Jelinek}, \bibinfo{person}{Angelika Kratzer}, {and}
  \bibinfo{person}{Barbara~BH Partee}.} \bibinfo{year}{2013}\natexlab{}.
\newblock \bibinfo{booktitle}{\emph{Quantification in natural languages}}.
  Vol.~\bibinfo{volume}{54}.
\newblock \bibinfo{publisher}{Springer Science \& Business Media}.
\newblock


\bibitem[\protect\citeauthoryear{Barwise and Cooper}{Barwise and
  Cooper}{1981}]%
        {barwise1981generalized}
\bibfield{author}{\bibinfo{person}{Jon Barwise} {and} \bibinfo{person}{Robin
  Cooper}.} \bibinfo{year}{1981}\natexlab{}.
\newblock \showarticletitle{Generalized quantifiers and natural language}.
\newblock In \bibinfo{booktitle}{\emph{Philosophy, language, and artificial
  intelligence}}. \bibinfo{publisher}{Springer}, \bibinfo{pages}{241--301}.
\newblock


\bibitem[\protect\citeauthoryear{Belinkov and Bisk}{Belinkov and Bisk}{2018}]%
        {DBLP:conf/iclr/BelinkovB18}
\bibfield{author}{\bibinfo{person}{Yonatan Belinkov} {and}
  \bibinfo{person}{Yonatan Bisk}.} \bibinfo{year}{2018}\natexlab{}.
\newblock \showarticletitle{Synthetic and Natural Noise Both Break Neural
  Machine Translation}. In \bibinfo{booktitle}{\emph{6th International
  Conference on Learning Representations, {ICLR} 2018, Vancouver, BC, Canada,
  April 30 - May 3, 2018, Conference Track Proceedings}}.
  \bibinfo{publisher}{OpenReview.net}.
\newblock
\urldef\tempurl%
\url{https://openreview.net/forum?id=BJ8vJebC-}
\showURL{%
\tempurl}


\bibitem[\protect\citeauthoryear{Bex, Gelade, Neven, and Vansummeren}{Bex
  et~al\mbox{.}}{2010a}]%
        {DFAinf10}
\bibfield{author}{\bibinfo{person}{Geert~Jan Bex}, \bibinfo{person}{Wouter
  Gelade}, \bibinfo{person}{Frank Neven}, {and} \bibinfo{person}{Stijn
  Vansummeren}.} \bibinfo{year}{2010}\natexlab{a}.
\newblock \showarticletitle{Learning Deterministic Regular Expressions for the
  Inference of Schemas from {XML} Data}.
\newblock \bibinfo{journal}{\emph{{ACM} Trans. Web}} \bibinfo{volume}{4},
  \bibinfo{number}{4} (\bibinfo{year}{2010}), \bibinfo{pages}{14:1--14:32}.
\newblock


\bibitem[\protect\citeauthoryear{Bex, Neven, Schwentick, and Vansummeren}{Bex
  et~al\mbox{.}}{2010b}]%
        {bex2010inference}
\bibfield{author}{\bibinfo{person}{Geert~Jan Bex}, \bibinfo{person}{Frank
  Neven}, \bibinfo{person}{Thomas Schwentick}, {and} \bibinfo{person}{Stijn
  Vansummeren}.} \bibinfo{year}{2010}\natexlab{b}.
\newblock \showarticletitle{Inference of concise regular expressions and DTDs}.
\newblock \bibinfo{journal}{\emph{ACM Transactions on Database Systems (TODS)}}
  \bibinfo{volume}{35}, \bibinfo{number}{2} (\bibinfo{year}{2010}),
  \bibinfo{pages}{1--47}.
\newblock


\bibitem[\protect\citeauthoryear{BRADBURN and MILES}{BRADBURN and
  MILES}{1979}]%
        {Vague1979}
\bibfield{author}{\bibinfo{person}{NORMAN~M. BRADBURN} {and}
  \bibinfo{person}{CARRIE MILES}.} \bibinfo{year}{1979}\natexlab{}.
\newblock \showarticletitle{{Vague Quantifiers}}.
\newblock \bibinfo{journal}{\emph{Public Opinion Quarterly}}
  \bibinfo{volume}{43}, \bibinfo{number}{1} (\bibinfo{year}{1979}),
  \bibinfo{pages}{92--101}.
\newblock


\bibitem[\protect\citeauthoryear{Cer, Diab, Agirre, Lopez{-}Gazpio, and
  Specia}{Cer et~al\mbox{.}}{2017}]%
        {DBLP:conf/semeval/CerDALS17}
\bibfield{author}{\bibinfo{person}{Daniel~M. Cer}, \bibinfo{person}{Mona~T.
  Diab}, \bibinfo{person}{Eneko Agirre}, \bibinfo{person}{I{\~{n}}igo
  Lopez{-}Gazpio}, {and} \bibinfo{person}{Lucia Specia}.}
  \bibinfo{year}{2017}\natexlab{}.
\newblock \showarticletitle{SemEval-2017 Task 1: Semantic Textual Similarity
  Multilingual and Crosslingual Focused Evaluation}. In
  \bibinfo{booktitle}{\emph{Proceedings of the 11th International Workshop on
  Semantic Evaluation, {SemEval-2017}}}. \bibinfo{publisher}{Association for
  Computational Linguistics}, \bibinfo{pages}{1--14}.
\newblock


\bibitem[\protect\citeauthoryear{Chapman and Stolee}{Chapman and
  Stolee}{2016}]%
        {regexPython16}
\bibfield{author}{\bibinfo{person}{Carl Chapman} {and}
  \bibinfo{person}{Kathryn~T. Stolee}.} \bibinfo{year}{2016}\natexlab{}.
\newblock \showarticletitle{Exploring regular expression usage and context in
  Python}. In \bibinfo{booktitle}{\emph{Proceedings of the 25th International
  Symposium on Software Testing and Analysis, {ISSTA}}}.
  \bibinfo{publisher}{{ACM}}, \bibinfo{pages}{282--293}.
\newblock


\bibitem[\protect\citeauthoryear{Chen and Manning}{Chen and Manning}{2014}]%
        {dependency}
\bibfield{author}{\bibinfo{person}{Danqi Chen} {and}
  \bibinfo{person}{Christopher~D. Manning}.} \bibinfo{year}{2014}\natexlab{}.
\newblock \showarticletitle{A Fast and Accurate Dependency Parser using Neural
  Networks}. In \bibinfo{booktitle}{\emph{Proceedings of the 2014 Conference on
  Empirical Methods in Natural Language Processing, {EMNLP} 2014, October
  25-29, 2014, Doha, Qatar, {A} meeting of SIGDAT, a Special Interest Group of
  the {ACL}}}. \bibinfo{publisher}{{ACL}}, \bibinfo{pages}{740--750}.
\newblock


\bibitem[\protect\citeauthoryear{Chen, Wang, Ye, Durrett, and Dillig}{Chen
  et~al\mbox{.}}{2020b}]%
        {pldi20}
\bibfield{author}{\bibinfo{person}{Qiaochu Chen}, \bibinfo{person}{Xinyu Wang},
  \bibinfo{person}{Xi Ye}, \bibinfo{person}{Greg Durrett}, {and}
  \bibinfo{person}{Isil Dillig}.} \bibinfo{year}{2020}\natexlab{b}.
\newblock \showarticletitle{Multi-modal synthesis of regular expressions}. In
  \bibinfo{booktitle}{\emph{Proceedings of the 41st {ACM} {SIGPLAN}
  International Conference on Programming Language Design and Implementation
  (PLDI)}}. \bibinfo{publisher}{{ACM}}, \bibinfo{pages}{487--502}.
\newblock


\bibitem[\protect\citeauthoryear{Chen, Zhu, Ling, Wei, Jiang, and Inkpen}{Chen
  et~al\mbox{.}}{2017}]%
        {EMSI17}
\bibfield{author}{\bibinfo{person}{Qian Chen}, \bibinfo{person}{Xiaodan Zhu},
  \bibinfo{person}{Zhen{-}Hua Ling}, \bibinfo{person}{Si Wei},
  \bibinfo{person}{Hui Jiang}, {and} \bibinfo{person}{Diana Inkpen}.}
  \bibinfo{year}{2017}\natexlab{}.
\newblock \showarticletitle{Enhanced {LSTM} for Natural Language Inference}. In
  \bibinfo{booktitle}{\emph{Proceedings of the 55th Annual Meeting of the
  Association for Computational Linguistics (ACL)}}.
  \bibinfo{publisher}{Association for Computational Linguistics},
  \bibinfo{pages}{1657--1668}.
\newblock


\bibitem[\protect\citeauthoryear{Chen, Cheung, and Yiu}{Chen
  et~al\mbox{.}}{1998}]%
        {metamorphic}
\bibfield{author}{\bibinfo{person}{Tsong~Yueh Chen}, \bibinfo{person}{S.~C.
  Cheung}, {and} \bibinfo{person}{Siu{-}Ming Yiu}.}
  \bibinfo{year}{1998}\natexlab{}.
\newblock \showarticletitle{Metamorphic Testing: {A} New Approach for
  Generating Next Test Cases}. Technical Report HKUST-CS98-01,
  \bibinfo{publisher}{Department of Computer Science, Hong Kong University of
  Science and Technology, Hong Kong.}
\newblock


\bibitem[\protect\citeauthoryear{Chen, Kuo, Liu, Poon, Towey, Tse, and
  Zhou}{Chen et~al\mbox{.}}{2018}]%
        {chen2018metamorphic}
\bibfield{author}{\bibinfo{person}{Tsong~Yueh Chen}, \bibinfo{person}{Fei-Ching
  Kuo}, \bibinfo{person}{Huai Liu}, \bibinfo{person}{Pak-Lok Poon},
  \bibinfo{person}{Dave Towey}, \bibinfo{person}{TH Tse}, {and}
  \bibinfo{person}{Zhi~Quan Zhou}.} \bibinfo{year}{2018}\natexlab{}.
\newblock \showarticletitle{Metamorphic testing: A review of challenges and
  opportunities}.
\newblock \bibinfo{journal}{\emph{Comput. Surveys}} \bibinfo{volume}{51},
  \bibinfo{number}{1} (\bibinfo{year}{2018}), \bibinfo{pages}{1--27}.
\newblock


\bibitem[\protect\citeauthoryear{Chen, Tian, Xiao, He, and Jin}{Chen
  et~al\mbox{.}}{2020a}]%
        {emnlp20causual}
\bibfield{author}{\bibinfo{person}{Wenqing Chen}, \bibinfo{person}{Jidong
  Tian}, \bibinfo{person}{Liqiang Xiao}, \bibinfo{person}{Hao He}, {and}
  \bibinfo{person}{Yaohui Jin}.} \bibinfo{year}{2020}\natexlab{a}.
\newblock \showarticletitle{Exploring Logically Dependent Multi-task Learning
  with Causal Inference}. In \bibinfo{booktitle}{\emph{Proceedings of the 2020
  Conference on Empirical Methods in Natural Language Processing, {EMNLP} 2020,
  Online, November 16-20, 2020}}. \bibinfo{publisher}{Association for
  Computational Linguistics}, \bibinfo{pages}{2213--2225}.
\newblock


\bibitem[\protect\citeauthoryear{Chomsky and Miller}{Chomsky and
  Miller}{1958}]%
        {chomsky1958finite}
\bibfield{author}{\bibinfo{person}{Noam Chomsky} {and}
  \bibinfo{person}{George~A Miller}.} \bibinfo{year}{1958}\natexlab{}.
\newblock \showarticletitle{Finite state languages}.
\newblock \bibinfo{journal}{\emph{Information and control}}
  \bibinfo{volume}{1}, \bibinfo{number}{2} (\bibinfo{year}{1958}),
  \bibinfo{pages}{91--112}.
\newblock


\bibitem[\protect\citeauthoryear{Coughlin}{Coughlin}{2003}]%
        {coughlin2003correlating}
\bibfield{author}{\bibinfo{person}{Deborah Coughlin}.}
  \bibinfo{year}{2003}\natexlab{}.
\newblock \showarticletitle{Correlating automated and human assessments of
  machine translation quality}. In \bibinfo{booktitle}{\emph{Proceedings of MT
  summit IX}}. Citeseer, \bibinfo{pages}{63--70}.
\newblock


\bibitem[\protect\citeauthoryear{Cui, Dang, Fischer, and Ibarra}{Cui
  et~al\mbox{.}}{2013}]%
        {similarity13}
\bibfield{author}{\bibinfo{person}{Cewei Cui}, \bibinfo{person}{Zhe Dang},
  \bibinfo{person}{Thomas~R. Fischer}, {and} \bibinfo{person}{Oscar~H.
  Ibarra}.} \bibinfo{year}{2013}\natexlab{}.
\newblock \showarticletitle{Similarity in languages and programs}.
\newblock \bibinfo{journal}{\emph{Theor. Comput. Sci.}}  \bibinfo{volume}{498}
  (\bibinfo{year}{2013}), \bibinfo{pages}{58--75}.
\newblock


\bibitem[\protect\citeauthoryear{Dadonaite and Ritchie}{Dadonaite and
  Ritchie}{2018}]%
        {owiddiarrhealdiseases}
\bibfield{author}{\bibinfo{person}{Bernadeta Dadonaite} {and}
  \bibinfo{person}{Hannah Ritchie}.} \bibinfo{year}{2018}\natexlab{}.
\newblock \showarticletitle{Diarrheal diseases}.
\newblock \bibinfo{journal}{\emph{Our World in Data}} (\bibinfo{year}{2018}).
\newblock
\newblock
\shownote{https://ourworldindata.org/diarrheal-diseases}.


\bibitem[\protect\citeauthoryear{Davies}{Davies}{2017}]%
        {Attack17}
\bibfield{author}{\bibinfo{person}{Gareth Davies}.}
  \bibinfo{year}{2017}\natexlab{}.
\newblock \showarticletitle{Palestinian man is arrested by police after posting
  `Good morning' in Arabic on Facebook which was wrongly translated as `attack
  them'}.
\newblock  (\bibinfo{year}{2017}).
\newblock
\urldef\tempurl%
\url{https://www.dailymail.co.uk/news/article-5005489/Good-morning-
  Facebook-post-leads-arrest-Palestinian.html}
\showURL{%
\tempurl}


\bibitem[\protect\citeauthoryear{Davis, Moyer, Kazerouni, and Lee}{Davis
  et~al\mbox{.}}{2019}]%
        {largescale19}
\bibfield{author}{\bibinfo{person}{James~C. Davis}, \bibinfo{person}{Daniel
  Moyer}, \bibinfo{person}{Ayaan~M. Kazerouni}, {and} \bibinfo{person}{Dongyoon
  Lee}.} \bibinfo{year}{2019}\natexlab{}.
\newblock \showarticletitle{Testing Regex Generalizability And Its
  Implications: {A} Large-Scale Many-Language Measurement Study}. In
  \bibinfo{booktitle}{\emph{34th {IEEE/ACM} International Conference on
  Automated Software Engineering, {ASE}}}. \bibinfo{publisher}{{IEEE}},
  \bibinfo{pages}{427--439}.
\newblock


\bibitem[\protect\citeauthoryear{Devlin, Chang, Lee, and Toutanova}{Devlin
  et~al\mbox{.}}{2019}]%
        {BERT19}
\bibfield{author}{\bibinfo{person}{Jacob Devlin}, \bibinfo{person}{Ming{-}Wei
  Chang}, \bibinfo{person}{Kenton Lee}, {and} \bibinfo{person}{Kristina
  Toutanova}.} \bibinfo{year}{2019}\natexlab{}.
\newblock \showarticletitle{{BERT:} Pre-training of Deep Bidirectional
  Transformers for Language Understanding}. In
  \bibinfo{booktitle}{\emph{Proceedings of the 2019 Conference of the North
  American Chapter of the Association for Computational Linguistics: Human
  Language Technologies, {NAACL-HLT}}}. \bibinfo{publisher}{Association for
  Computational Linguistics}, \bibinfo{pages}{4171--4186}.
\newblock


\bibitem[\protect\citeauthoryear{Ding, Zhang, Liu, and Duan}{Ding
  et~al\mbox{.}}{2014}]%
        {Reuters14}
\bibfield{author}{\bibinfo{person}{Xiao Ding}, \bibinfo{person}{Yue Zhang},
  \bibinfo{person}{Ting Liu}, {and} \bibinfo{person}{Junwen Duan}.}
  \bibinfo{year}{2014}\natexlab{}.
\newblock \showarticletitle{Using Structured Events to Predict Stock Price
  Movement: An Empirical Investigation}. In
  \bibinfo{booktitle}{\emph{Proceedings of the 2014 Conference on Empirical
  Methods in Natural Language Processing, {EMNLP} 2014, October 25-29, 2014,
  Doha, Qatar, {A} meeting of SIGDAT, a Special Interest Group of the {ACL}}}.
  \bibinfo{publisher}{{ACL}}, \bibinfo{pages}{1415--1425}.
\newblock


\bibitem[\protect\citeauthoryear{Ebrahimi, Lowd, and Dou}{Ebrahimi
  et~al\mbox{.}}{2018}]%
        {ebrahimi-etal-2018-adversarial}
\bibfield{author}{\bibinfo{person}{Javid Ebrahimi}, \bibinfo{person}{Daniel
  Lowd}, {and} \bibinfo{person}{Dejing Dou}.} \bibinfo{year}{2018}\natexlab{}.
\newblock \showarticletitle{On Adversarial Examples for Character-Level Neural
  Machine Translation}. In \bibinfo{booktitle}{\emph{Proceedings of the 27th
  International Conference on Computational Linguistics}}.
  \bibinfo{publisher}{Association for Computational Linguistics},
  \bibinfo{address}{Santa Fe, New Mexico, USA}, \bibinfo{pages}{653--663}.
\newblock
\urldef\tempurl%
\url{https://www.aclweb.org/anthology/C18-1055}
\showURL{%
\tempurl}


\bibitem[\protect\citeauthoryear{Fonseca, Yankovskaya, Martins, Fishel, and
  Federmann}{Fonseca et~al\mbox{.}}{2019}]%
        {findingsWMT19}
\bibfield{author}{\bibinfo{person}{Erick~R. Fonseca}, \bibinfo{person}{Lisa
  Yankovskaya}, \bibinfo{person}{Andr{\'{e}} F.~T. Martins},
  \bibinfo{person}{Mark Fishel}, {and} \bibinfo{person}{Christian Federmann}.}
  \bibinfo{year}{2019}\natexlab{}.
\newblock \showarticletitle{Findings of the {WMT} 2019 Shared Tasks on Quality
  Estimation}. In \bibinfo{booktitle}{\emph{Proceedings of the Fourth
  Conference on Machine Translation, (WMT)}}. \bibinfo{publisher}{Association
  for Computational Linguistics}, \bibinfo{pages}{1--10}.
\newblock


\bibitem[\protect\citeauthoryear{Garcia-Molina}{Garcia-Molina}{2008}]%
        {garcia2008database}
\bibfield{author}{\bibinfo{person}{Hector Garcia-Molina}.}
  \bibinfo{year}{2008}\natexlab{}.
\newblock \bibinfo{booktitle}{\emph{Database systems: the complete book}}.
\newblock \bibinfo{publisher}{Pearson Education India}.
\newblock


\bibitem[\protect\citeauthoryear{Goodfellow, Shlens, and Szegedy}{Goodfellow
  et~al\mbox{.}}{2015}]%
        {Goodfellow15ICLR}
\bibfield{author}{\bibinfo{person}{Ian~J. Goodfellow},
  \bibinfo{person}{Jonathon Shlens}, {and} \bibinfo{person}{Christian
  Szegedy}.} \bibinfo{year}{2015}\natexlab{}.
\newblock \showarticletitle{Explaining and Harnessing Adversarial Examples}. In
  \bibinfo{booktitle}{\emph{3rd International Conference on Learning
  Representations, {ICLR} 2015, San Diego, CA, USA, May 7-9, 2015, Conference
  Track Proceedings}}, \bibfield{editor}{\bibinfo{person}{Yoshua Bengio} {and}
  \bibinfo{person}{Yann LeCun}} (Eds.).
\newblock


\bibitem[\protect\citeauthoryear{Grammarly}{Grammarly}{[n.\,d.]}]%
        {afew-diff}
\bibfield{author}{\bibinfo{person}{Grammarly}.}
  \bibinfo{year}{[n.\,d.]}\natexlab{}.
\newblock \bibinfo{title}{Few, a Few—What’s the Difference?}
\newblock
  \bibinfo{howpublished}{\url{https://www.grammarly.com/blog/few-a-few/}}.
\newblock


\bibitem[\protect\citeauthoryear{Gu, Wang, Cho, and Li}{Gu
  et~al\mbox{.}}{2018}]%
        {DBLP:conf/aaai/GuWCL18}
\bibfield{author}{\bibinfo{person}{Jiatao Gu}, \bibinfo{person}{Yong Wang},
  \bibinfo{person}{Kyunghyun Cho}, {and} \bibinfo{person}{Victor O.~K. Li}.}
  \bibinfo{year}{2018}\natexlab{}.
\newblock \showarticletitle{Search Engine Guided Neural Machine Translation}.
  In \bibinfo{booktitle}{\emph{Proceedings of the Thirty-Second {AAAI}
  Conference on Artificial Intelligence, (AAAI-18), the 30th innovative
  Applications of Artificial Intelligence (IAAI-18), and the 8th {AAAI}
  Symposium on Educational Advances in Artificial Intelligence (EAAI-18)}}.
  \bibinfo{publisher}{{AAAI} Press}, \bibinfo{pages}{5133--5140}.
\newblock


\bibitem[\protect\citeauthoryear{Gupta, He, Meister, and Su}{Gupta
  et~al\mbox{.}}{2020}]%
        {pathology}
\bibfield{author}{\bibinfo{person}{Shashij Gupta}, \bibinfo{person}{Pinjia He},
  \bibinfo{person}{Clara Meister}, {and} \bibinfo{person}{Zhendong Su}.}
  \bibinfo{year}{2020}\natexlab{}.
\newblock \showarticletitle{Machine Translation Testing via Pathological
  Invariance}. In \bibinfo{booktitle}{\emph{Proceedings of the ACM Joint
  Meeting on European Software Engineering Conference and Symposium on the
  Foundations of Software Engineering {ESEC/FSE}}}. \bibinfo{pages}{863–875}.
\newblock
\showISBNx{9781450399999}


\bibitem[\protect\citeauthoryear{Haruta, Mineshima, and Bekki}{Haruta
  et~al\mbox{.}}{2020}]%
        {ACL20logic}
\bibfield{author}{\bibinfo{person}{Izumi Haruta}, \bibinfo{person}{Koji
  Mineshima}, {and} \bibinfo{person}{Daisuke Bekki}.}
  \bibinfo{year}{2020}\natexlab{}.
\newblock \showarticletitle{Logical Inferences with Comparatives and
  Generalized Quantifiers}. In \bibinfo{booktitle}{\emph{Proceedings of the
  58th Annual Meeting of the Association for Computational Linguistics: Student
  Research Workshop, {ACL} 2020, Online, July 5-10, 2020}}.
  \bibinfo{publisher}{Association for Computational Linguistics},
  \bibinfo{pages}{263--270}.
\newblock


\bibitem[\protect\citeauthoryear{Hazelwood, Bird, Brooks, Chintala, Diril,
  Dzhulgakov, Fawzy, Jia, Jia, Kalro, Law, Lee, Lu, Noordhuis, Smelyanskiy,
  Xiong, and Wang}{Hazelwood et~al\mbox{.}}{2018}]%
        {HazelwoodBBCDDF18}
\bibfield{author}{\bibinfo{person}{Kim~M. Hazelwood}, \bibinfo{person}{Sarah
  Bird}, \bibinfo{person}{David~M. Brooks}, \bibinfo{person}{Soumith Chintala},
  \bibinfo{person}{Utku Diril}, \bibinfo{person}{Dmytro Dzhulgakov},
  \bibinfo{person}{Mohamed Fawzy}, \bibinfo{person}{Bill Jia},
  \bibinfo{person}{Yangqing Jia}, \bibinfo{person}{Aditya Kalro},
  \bibinfo{person}{James Law}, \bibinfo{person}{Kevin Lee},
  \bibinfo{person}{Jason Lu}, \bibinfo{person}{Pieter Noordhuis},
  \bibinfo{person}{Misha Smelyanskiy}, \bibinfo{person}{Liang Xiong}, {and}
  \bibinfo{person}{Xiaodong Wang}.} \bibinfo{year}{2018}\natexlab{}.
\newblock \showarticletitle{Applied Machine Learning at Facebook: {A}
  Datacenter Infrastructure Perspective}. In \bibinfo{booktitle}{\emph{{IEEE}
  International Symposium on High Performance Computer Architecture (HPCA)}}.
  \bibinfo{publisher}{{IEEE} Computer Society}, \bibinfo{pages}{620--629}.
\newblock


\bibitem[\protect\citeauthoryear{He, Meister, and Su}{He et~al\mbox{.}}{2020}]%
        {SIT}
\bibfield{author}{\bibinfo{person}{Pinjia He}, \bibinfo{person}{Clara Meister},
  {and} \bibinfo{person}{Zhendong Su}.} \bibinfo{year}{2020}\natexlab{}.
\newblock \showarticletitle{Structure-Invariant Testing for Machine
  Translation}. In \bibinfo{booktitle}{\emph{Proceedings of the 42nd
  International Conference on Software Engineering}}
  \emph{(\bibinfo{series}{ICSE '20})}. \bibinfo{pages}{961–973}.
\newblock


\bibitem[\protect\citeauthoryear{Hintikka and Sandu}{Hintikka and
  Sandu}{1994}]%
        {hintikka1994quantifier}
\bibfield{author}{\bibinfo{person}{Jaakko Hintikka} {and}
  \bibinfo{person}{Gabriel Sandu}.} \bibinfo{year}{1994}\natexlab{}.
\newblock \showarticletitle{What is a Quantifier?}
\newblock \bibinfo{journal}{\emph{Synthese}} (\bibinfo{year}{1994}),
  \bibinfo{pages}{113--129}.
\newblock


\bibitem[\protect\citeauthoryear{Hopcroft, Motwani, and Ullman}{Hopcroft
  et~al\mbox{.}}{2007}]%
        {DBLP:books/daglib/0016921}
\bibfield{author}{\bibinfo{person}{John~E. Hopcroft}, \bibinfo{person}{Rajeev
  Motwani}, {and} \bibinfo{person}{Jeffrey~D. Ullman}.}
  \bibinfo{year}{2007}\natexlab{}.
\newblock \bibinfo{booktitle}{\emph{Introduction to automata theory, languages,
  and computation, 3rd Edition}}.
\newblock \bibinfo{publisher}{Addison-Wesley}.
\newblock
\showISBNx{978-0-321-47617-3}


\bibitem[\protect\citeauthoryear{Hox, Borgers, and Sikkel}{Hox
  et~al\mbox{.}}{2003}]%
        {hox2003response}
\bibfield{author}{\bibinfo{person}{JJ Hox}, \bibinfo{person}{N Borgers}, {and}
  \bibinfo{person}{Dirk Sikkel}.} \bibinfo{year}{2003}\natexlab{}.
\newblock \showarticletitle{Response quality in survey research with children
  and adolescents: the effect of labeled response options and vague
  quantifiers}.
\newblock \bibinfo{journal}{\emph{International Journal of Public Opinion
  Research}} \bibinfo{volume}{15}, \bibinfo{number}{1} (\bibinfo{year}{2003}),
  \bibinfo{pages}{83--94}.
\newblock


\bibitem[\protect\citeauthoryear{Insights}{Insights}{2019}]%
        {afew}
\bibfield{author}{\bibinfo{person}{Editage Insights}.}
  \bibinfo{year}{2019}\natexlab{}.
\newblock \bibinfo{title}{What is the difference between ``few'' and ``a
  few''?}
\newblock
  \bibinfo{howpublished}{\url{https://www.editage.com/insights/few-and-a-few-what-is-the-difference}}.
\newblock


\bibitem[\protect\citeauthoryear{Koehn}{Koehn}{2005}]%
        {koehn2005europarl}
\bibfield{author}{\bibinfo{person}{Philipp Koehn}.}
  \bibinfo{year}{2005}\natexlab{}.
\newblock \showarticletitle{Europarl: A parallel corpus for statistical machine
  translation}. In \bibinfo{booktitle}{\emph{MT summit}},
  Vol.~\bibinfo{volume}{5}. Citeseer, \bibinfo{pages}{79--86}.
\newblock


\bibitem[\protect\citeauthoryear{Koehn and Monz}{Koehn and Monz}{2006}]%
        {European06}
\bibfield{author}{\bibinfo{person}{Philipp Koehn} {and}
  \bibinfo{person}{Christof Monz}.} \bibinfo{year}{2006}\natexlab{}.
\newblock \showarticletitle{Manual and Automatic Evaluation of Machine
  Translation between European Languages}. In
  \bibinfo{booktitle}{\emph{Proceedings on the Workshop on Statistical Machine
  Translation, {(StatMT)}}}. \bibinfo{publisher}{Association for Computational
  Linguistics}, \bibinfo{pages}{102--121}.
\newblock


\bibitem[\protect\citeauthoryear{Kurfali and {\"{O}}stling}{Kurfali and
  {\"{O}}stling}{2019}]%
        {DBLP:conf/wmt/KurfaliO19}
\bibfield{author}{\bibinfo{person}{Murathan Kurfali} {and}
  \bibinfo{person}{Robert {\"{O}}stling}.} \bibinfo{year}{2019}\natexlab{}.
\newblock \showarticletitle{Noisy Parallel Corpus Filtering through Projected
  Word Embeddings}. In \bibinfo{booktitle}{\emph{Proceedings of the Fourth
  Conference on Machine Translation, {WMT} 2019, Florence, Italy, August 1-2,
  2019 - Volume 3: Shared Task Papers, Day 2}},
  \bibfield{editor}{\bibinfo{person}{Ondrej Bojar}, \bibinfo{person}{Rajen
  Chatterjee}, \bibinfo{person}{Christian Federmann}, \bibinfo{person}{Mark
  Fishel}, \bibinfo{person}{Yvette Graham}, \bibinfo{person}{Barry Haddow},
  \bibinfo{person}{Matthias Huck}, \bibinfo{person}{Antonio Jimeno{-}Yepes},
  \bibinfo{person}{Philipp Koehn}, \bibinfo{person}{Andr{\'{e}} Martins},
  \bibinfo{person}{Christof Monz}, \bibinfo{person}{Matteo Negri},
  \bibinfo{person}{Aur{\'{e}}lie N{\'{e}}v{\'{e}}ol},
  \bibinfo{person}{Mariana~L. Neves}, \bibinfo{person}{Matt Post},
  \bibinfo{person}{Marco Turchi}, {and} \bibinfo{person}{Karin Verspoor}}
  (Eds.). \bibinfo{publisher}{Association for Computational Linguistics},
  \bibinfo{pages}{277--281}.
\newblock
\urldef\tempurl%
\url{https://doi.org/10.18653/v1/w19-5438}
\showDOI{\tempurl}


\bibitem[\protect\citeauthoryear{Kurtzman and MacDonald}{Kurtzman and
  MacDonald}{1993}]%
        {kurtzman1993resolution}
\bibfield{author}{\bibinfo{person}{Howard~S Kurtzman} {and}
  \bibinfo{person}{Maryellen~C MacDonald}.} \bibinfo{year}{1993}\natexlab{}.
\newblock \showarticletitle{Resolution of quantifier scope ambiguities}.
\newblock \bibinfo{journal}{\emph{Cognition}} \bibinfo{volume}{48},
  \bibinfo{number}{3} (\bibinfo{year}{1993}), \bibinfo{pages}{243--279}.
\newblock


\bibitem[\protect\citeauthoryear{Kushman and Barzilay}{Kushman and
  Barzilay}{2013}]%
        {KushmanB13}
\bibfield{author}{\bibinfo{person}{Nate Kushman} {and} \bibinfo{person}{Regina
  Barzilay}.} \bibinfo{year}{2013}\natexlab{}.
\newblock \showarticletitle{Using Semantic Unification to Generate Regular
  Expressions from Natural Language}. In \bibinfo{booktitle}{\emph{Proceedings
  of the 2013 Conference of the North American Chapter of the Association for
  Computational Linguistics: Human Language Technologies}}.
  \bibinfo{publisher}{The Association for Computational Linguistics},
  \bibinfo{pages}{826--836}.
\newblock


\bibitem[\protect\citeauthoryear{Lappin}{Lappin}{2000}]%
        {lappin2000intensional}
\bibfield{author}{\bibinfo{person}{Shalom Lappin}.}
  \bibinfo{year}{2000}\natexlab{}.
\newblock \showarticletitle{An intensional parametric semantics for vague
  quantifiers}.
\newblock \bibinfo{journal}{\emph{Linguistics and philosophy}}
  (\bibinfo{year}{2000}), \bibinfo{pages}{599--620}.
\newblock


\bibitem[\protect\citeauthoryear{L{\"{a}}ubli, Sennrich, and Volk}{L{\"{a}}ubli
  et~al\mbox{.}}{2018}]%
        {DBLP:conf/emnlp/LaubliS018}
\bibfield{author}{\bibinfo{person}{Samuel L{\"{a}}ubli}, \bibinfo{person}{Rico
  Sennrich}, {and} \bibinfo{person}{Martin Volk}.}
  \bibinfo{year}{2018}\natexlab{}.
\newblock \showarticletitle{Has Machine Translation Achieved Human Parity? {A}
  Case for Document-level Evaluation}. In \bibinfo{booktitle}{\emph{Proceedings
  of the 2018 Conference on Empirical Methods in Natural Language Processing,
  (EMNLP)}}. \bibinfo{publisher}{Association for Computational Linguistics},
  \bibinfo{pages}{4791--4796}.
\newblock


\bibitem[\protect\citeauthoryear{Lo}{Lo}{2019}]%
        {YISI19}
\bibfield{author}{\bibinfo{person}{Chi{-}kiu Lo}.}
  \bibinfo{year}{2019}\natexlab{}.
\newblock \showarticletitle{YiSi - a Unified Semantic {MT} Quality Evaluation
  and Estimation Metric for Languages with Different Levels of Available
  Resources}. In \bibinfo{booktitle}{\emph{Proceedings of the Fourth Conference
  on Machine Translation, (WMT)}}. \bibinfo{publisher}{Association for
  Computational Linguistics}, \bibinfo{pages}{507--513}.
\newblock


\bibitem[\protect\citeauthoryear{L{\"o}bner}{L{\"o}bner}{1987}]%
        {lobner1987quantification}
\bibfield{author}{\bibinfo{person}{Sebastian L{\"o}bner}.}
  \bibinfo{year}{1987}\natexlab{}.
\newblock \showarticletitle{Quantification as a major module of natural
  language semantics}.
\newblock \bibinfo{journal}{\emph{Studies in discourse representation theory
  and the theory of generalized quantifiers}}  \bibinfo{volume}{8}
  (\bibinfo{year}{1987}), \bibinfo{pages}{53}.
\newblock


\bibitem[\protect\citeauthoryear{Locascio, Narasimhan, DeLeon, Kushman, and
  Barzilay}{Locascio et~al\mbox{.}}{2016}]%
        {deepregex16}
\bibfield{author}{\bibinfo{person}{Nicholas Locascio}, \bibinfo{person}{Karthik
  Narasimhan}, \bibinfo{person}{Eduardo DeLeon}, \bibinfo{person}{Nate
  Kushman}, {and} \bibinfo{person}{Regina Barzilay}.}
  \bibinfo{year}{2016}\natexlab{}.
\newblock \showarticletitle{Neural Generation of Regular Expressions from
  Natural Language with Minimal Domain Knowledge}. In
  \bibinfo{booktitle}{\emph{Proceedings of the 2016 Conference on Empirical
  Methods in Natural Language Processing (EMNLP)}}. \bibinfo{publisher}{The
  Association for Computational Linguistics}, \bibinfo{pages}{1918--1923}.
\newblock


\bibitem[\protect\citeauthoryear{Ltd}{Ltd}{[n.\,d.]}]%
        {housingneeds}
\bibfield{author}{\bibinfo{person}{Fordham~Research Ltd}.}
  \bibinfo{year}{[n.\,d.]}\natexlab{}.
\newblock \bibinfo{title}{Housing Needs Study}.
\newblock
  \bibinfo{howpublished}{\url{https://www.ipswich.gov.uk/sites/default/files/Housing\_Needs\_Study.pdf}}.
\newblock


\bibitem[\protect\citeauthoryear{Macdonald}{Macdonald}{2015}]%
        {Mistranslations15}
\bibfield{author}{\bibinfo{person}{Fiona Macdonald}.}
  \bibinfo{year}{2015}\natexlab{}.
\newblock \showarticletitle{The Greatest Mistranslations Ever}.
\newblock  (\bibinfo{year}{2015}).
\newblock
\urldef\tempurl%
\url{https://www.bbc.com/culture/article/20150202-the-greatest-mistranslations-ever}
\showURL{%
\tempurl}


\bibitem[\protect\citeauthoryear{Mathur, Baldwin, and Cohn}{Mathur
  et~al\mbox{.}}{2019}]%
        {EMSI19}
\bibfield{author}{\bibinfo{person}{Nitika Mathur}, \bibinfo{person}{Timothy
  Baldwin}, {and} \bibinfo{person}{Trevor Cohn}.}
  \bibinfo{year}{2019}\natexlab{}.
\newblock \showarticletitle{Putting Evaluation in Context: Contextual
  Embeddings Improve Machine Translation Evaluation}. In
  \bibinfo{booktitle}{\emph{Proceedings of the 57th Conference of the
  Association for Computational Linguistics (ACL)}}.
  \bibinfo{publisher}{Association for Computational Linguistics},
  \bibinfo{pages}{2799--2808}.
\newblock


\bibitem[\protect\citeauthoryear{Mathur, Baldwin, and Cohn}{Mathur
  et~al\mbox{.}}{2020}]%
        {acl20-honor-metrics}
\bibfield{author}{\bibinfo{person}{Nitika Mathur}, \bibinfo{person}{Timothy
  Baldwin}, {and} \bibinfo{person}{Trevor Cohn}.}
  \bibinfo{year}{2020}\natexlab{}.
\newblock \showarticletitle{Tangled up in {BLEU:} Reevaluating the Evaluation
  of Automatic Machine Translation Evaluation Metrics}. In
  \bibinfo{booktitle}{\emph{Proceedings of the 58th Annual Meeting of the
  Association for Computational Linguistics, (ACL)}}.
  \bibinfo{publisher}{Association for Computational Linguistics},
  \bibinfo{pages}{4984--4997}.
\newblock


\bibitem[\protect\citeauthoryear{McDaniel}{McDaniel}{2004}]%
        {mcdaniel2004modal}
\bibfield{author}{\bibinfo{person}{Kris McDaniel}.}
  \bibinfo{year}{2004}\natexlab{}.
\newblock \showarticletitle{Modal realism with overlap}.
\newblock \bibinfo{journal}{\emph{Australasian Journal of Philosophy}}
  \bibinfo{volume}{82}, \bibinfo{number}{1} (\bibinfo{year}{2004}),
  \bibinfo{pages}{137--152}.
\newblock


\bibitem[\protect\citeauthoryear{Mellor}{Mellor}{2006}]%
        {mellor2006wholes}
\bibfield{author}{\bibinfo{person}{David~Hugh Mellor}.}
  \bibinfo{year}{2006}\natexlab{}.
\newblock \showarticletitle{Wholes and parts: The limits of composition}.
\newblock \bibinfo{journal}{\emph{South African Journal of Philosophy}}
  \bibinfo{volume}{25}, \bibinfo{number}{2} (\bibinfo{year}{2006}),
  \bibinfo{pages}{138--145}.
\newblock


\bibitem[\protect\citeauthoryear{Moon, Cho, and Park}{Moon
  et~al\mbox{.}}{2020}]%
        {revisiting20}
\bibfield{author}{\bibinfo{person}{Jihyung Moon}, \bibinfo{person}{Hyunchang
  Cho}, {and} \bibinfo{person}{Eunjeong~L. Park}.}
  \bibinfo{year}{2020}\natexlab{}.
\newblock \showarticletitle{Revisiting Round-Trip Translation for Quality
  Estimation}, In \bibinfo{booktitle}{the 22nd Annual Annual Conference of the
  European Association for Machine Translation (EAMT)}.
\newblock \bibinfo{journal}{\emph{CoRR}}, \bibinfo{pages}{91--104}.
\newblock


\bibitem[\protect\citeauthoryear{Okrent}{Okrent}{2016}]%
        {LittelTranslation16}
\bibfield{author}{\bibinfo{person}{Arika Okrent}.}
  \bibinfo{year}{2016}\natexlab{}.
\newblock \showarticletitle{9 Little Translation Mistakes That Caused Big
  Problems}.
\newblock  (\bibinfo{year}{2016}).
\newblock
\urldef\tempurl%
\url{http://mentalfloss.com/article/48795/9-little-translation-mistakes-
  caused-big-problems}
\showURL{%
\tempurl}


\bibitem[\protect\citeauthoryear{Ong}{Ong}{2017}]%
        {fbApologize17}
\bibfield{author}{\bibinfo{person}{Thuy Ong}.} \bibinfo{year}{2017}\natexlab{}.
\newblock \showarticletitle{Facebook apologizes after wrong translation sees
  Palestinian man arrested for posting ’good morning’}.
\newblock  (\bibinfo{year}{2017}).
\newblock
\urldef\tempurl%
\url{https://www.theverge.com/us-
  world/2017/10/24/16533496/facebook-apology-wrong-translation-palestinian-
  arrested-post-good-morning}
\showURL{%
\tempurl}


\bibitem[\protect\citeauthoryear{Papineni, Roukos, Ward, and Zhu}{Papineni
  et~al\mbox{.}}{2002}]%
        {BLEU02}
\bibfield{author}{\bibinfo{person}{Kishore Papineni}, \bibinfo{person}{Salim
  Roukos}, \bibinfo{person}{Todd Ward}, {and} \bibinfo{person}{Wei{-}Jing
  Zhu}.} \bibinfo{year}{2002}\natexlab{}.
\newblock \showarticletitle{Bleu: a Method for Automatic Evaluation of Machine
  Translation}. In \bibinfo{booktitle}{\emph{Proceedings of the 40th Annual
  Meeting of the Association for Computational Linguistics (ACL)}}.
  \bibinfo{pages}{311--318}.
\newblock


\bibitem[\protect\citeauthoryear{Park, Ko, Cognetta, and Han}{Park
  et~al\mbox{.}}{2019}]%
        {softregex19}
\bibfield{author}{\bibinfo{person}{Jun{-}U. Park}, \bibinfo{person}{Sang{-}Ki
  Ko}, \bibinfo{person}{Marco Cognetta}, {and} \bibinfo{person}{Yo{-}Sub Han}.}
  \bibinfo{year}{2019}\natexlab{}.
\newblock \showarticletitle{SoftRegex: Generating Regex from Natural Language
  Descriptions using Softened Regex Equivalence}. In
  \bibinfo{booktitle}{\emph{Proceedings of the 2019 Conference on Empirical
  Methods in Natural Language Processing and the 9th International Joint
  Conference on Natural Language Processing {EMNLP-IJCNLP}}}.
  \bibinfo{publisher}{Association for Computational Linguistics},
  \bibinfo{pages}{6424--6430}.
\newblock


\bibitem[\protect\citeauthoryear{Partee, ter Meulen, and Wall}{Partee
  et~al\mbox{.}}{1990}]%
        {partee1990mathematical}
\bibfield{author}{\bibinfo{person}{Barbara~BH Partee}, \bibinfo{person}{AG ter
  Meulen}, {and} \bibinfo{person}{R Wall}.} \bibinfo{year}{1990}\natexlab{}.
\newblock \bibinfo{booktitle}{\emph{Mathematical Methods in Linguistics}}.
  Vol.~\bibinfo{volume}{30}.
\newblock \bibinfo{publisher}{Springer Science \& Business Media}.
\newblock


\bibitem[\protect\citeauthoryear{Pesu, Zhou, Zhen, and Towey}{Pesu
  et~al\mbox{.}}{2018}]%
        {monte18}
\bibfield{author}{\bibinfo{person}{Daniel Pesu}, \bibinfo{person}{Zhi~Quan
  Zhou}, \bibinfo{person}{Jingfeng Zhen}, {and} \bibinfo{person}{Dave Towey}.}
  \bibinfo{year}{2018}\natexlab{}.
\newblock \showarticletitle{A Monte Carlo Method for Metamorphic Testing of
  Machine Translation Services}. In \bibinfo{booktitle}{\emph{3rd {IEEE/ACM}
  International Workshop on Metamorphic Testing {MET}}}.
  \bibinfo{publisher}{{ACM}}, \bibinfo{pages}{38--45}.
\newblock


\bibitem[\protect\citeauthoryear{Peters, Westerstahl, and Westerstahl}{Peters
  et~al\mbox{.}}{2006}]%
        {peters2006quantifiers}
\bibfield{author}{\bibinfo{person}{Stanley Peters}, \bibinfo{person}{Dag
  Westerstahl}, {and} \bibinfo{person}{Dag Westerstahl}.}
  \bibinfo{year}{2006}\natexlab{}.
\newblock \bibinfo{booktitle}{\emph{Quantifiers in language and logic}}.
\newblock \bibinfo{publisher}{Oxford University Press}.
\newblock


\bibitem[\protect\citeauthoryear{Pezzelle, Steinert{-}Threlkeld, Bernardi, and
  Szymanik}{Pezzelle et~al\mbox{.}}{2018}]%
        {ACL18some}
\bibfield{author}{\bibinfo{person}{Sandro Pezzelle}, \bibinfo{person}{Shane
  Steinert{-}Threlkeld}, \bibinfo{person}{Raffaella Bernardi}, {and}
  \bibinfo{person}{Jakub Szymanik}.} \bibinfo{year}{2018}\natexlab{}.
\newblock \showarticletitle{Some of Them Can be Guessed! Exploring the Effect
  of Linguistic Context in Predicting Quantifiers}. In
  \bibinfo{booktitle}{\emph{Proceedings of the 56th Annual Meeting of the
  Association for Computational Linguistics, {ACL} 2018, Melbourne, Australia,
  July 15-20, 2018, Volume 2: Short Papers}}. \bibinfo{publisher}{Association
  for Computational Linguistics}, \bibinfo{pages}{114--119}.
\newblock


\bibitem[\protect\citeauthoryear{Popovic}{Popovic}{2015}]%
        {chrF15}
\bibfield{author}{\bibinfo{person}{Maja Popovic}.}
  \bibinfo{year}{2015}\natexlab{}.
\newblock \showarticletitle{chrF: character n-gram F-score for automatic {MT}
  evaluation}. In \bibinfo{booktitle}{\emph{Proceedings of the Tenth Workshop
  on Statistical Machine Translation, {(WMT@EMNLP)}}}. \bibinfo{publisher}{The
  Association for Computer Linguistics}, \bibinfo{pages}{392--395}.
\newblock


\bibitem[\protect\citeauthoryear{Potts}{Potts}{2007}]%
        {potts2007logic}
\bibfield{author}{\bibinfo{person}{Christopher Potts}.}
  \bibinfo{year}{2007}\natexlab{}.
\newblock \bibinfo{title}{Logic for linguists. Course for LSA institute}.
\newblock
\newblock


\bibitem[\protect\citeauthoryear{Programm}{Programm}{2020}]%
        {policy20}
\bibfield{author}{\bibinfo{person}{United Nations World~Food Programm}.}
  \bibinfo{year}{2020}\natexlab{}.
\newblock \bibinfo{booktitle}{\emph{Policy Brief: The Impact of COVID-19 on
  children}}.
\newblock \bibinfo{type}{{T}echnical {R}eport}. \bibinfo{institution}{United
  Nations World Food Programm}.
\newblock
\newblock
\shownote{https://www.wfp.org/publications/policy-brief-impact-covid-19-children}.


\bibitem[\protect\citeauthoryear{Ranta}{Ranta}{1998}]%
        {ranta1998multilingual}
\bibfield{author}{\bibinfo{person}{Aarne Ranta}.}
  \bibinfo{year}{1998}\natexlab{}.
\newblock \showarticletitle{A multilingual natural-language interface to
  regular expressions}. In \bibinfo{booktitle}{\emph{Finite State Methods in
  Natural Language Processing}} \emph{(\bibinfo{series}{FSMNLP '09})}.
  \bibinfo{pages}{79–90}.
\newblock


\bibitem[\protect\citeauthoryear{Reimers and Gurevych}{Reimers and
  Gurevych}{2019}]%
        {SBERT19}
\bibfield{author}{\bibinfo{person}{Nils Reimers} {and} \bibinfo{person}{Iryna
  Gurevych}.} \bibinfo{year}{2019}\natexlab{}.
\newblock \showarticletitle{Sentence-BERT: Sentence Embeddings using Siamese
  BERT-Networks}. In \bibinfo{booktitle}{\emph{Proceedings of the 2019
  Conference on Empirical Methods in Natural Language Processing and the 9th
  International Joint Conference on Natural Language Processing, {EMNLP-IJCNLP}
  2019, Hong Kong, China, November 3-7, 2019}}. \bibinfo{publisher}{Association
  for Computational Linguistics}, \bibinfo{pages}{3980--3990}.
\newblock


\bibitem[\protect\citeauthoryear{Ristad and Yianilos}{Ristad and
  Yianilos}{1997}]%
        {leven97}
\bibfield{author}{\bibinfo{person}{Eric~Sven Ristad} {and}
  \bibinfo{person}{Peter~N. Yianilos}.} \bibinfo{year}{1997}\natexlab{}.
\newblock \showarticletitle{Learning String Edit Distance}. In
  \bibinfo{booktitle}{\emph{Proceedings of the Fourteenth International
  Conference on Machine Learning {(ICML)}}}. \bibinfo{publisher}{Morgan
  Kaufmann}, \bibinfo{pages}{287--295}.
\newblock


\bibitem[\protect\citeauthoryear{Shannon}{Shannon}{1948}]%
        {shannon1948mathematical}
\bibfield{author}{\bibinfo{person}{Claude~E Shannon}.}
  \bibinfo{year}{1948}\natexlab{}.
\newblock \showarticletitle{A mathematical theory of communication}.
\newblock \bibinfo{journal}{\emph{The Bell system technical journal}}
  \bibinfo{volume}{27}, \bibinfo{number}{3} (\bibinfo{year}{1948}),
  \bibinfo{pages}{379--423}.
\newblock


\bibitem[\protect\citeauthoryear{Sharvy}{Sharvy}{1980}]%
        {sharvy1980more}
\bibfield{author}{\bibinfo{person}{Richard Sharvy}.}
  \bibinfo{year}{1980}\natexlab{}.
\newblock \showarticletitle{A more general theory of definite descriptions}.
\newblock \bibinfo{journal}{\emph{The philosophical review}}
  \bibinfo{volume}{89}, \bibinfo{number}{4} (\bibinfo{year}{1980}),
  \bibinfo{pages}{607--624}.
\newblock


\bibitem[\protect\citeauthoryear{Shigenobu}{Shigenobu}{2007}]%
        {BT07}
\bibfield{author}{\bibinfo{person}{Tomohiro Shigenobu}.}
  \bibinfo{year}{2007}\natexlab{}.
\newblock \showarticletitle{Evaluation and Usability of Back Translation for
  Intercultural Communication}. In \bibinfo{booktitle}{\emph{Usability and
  Internationalization. Global and Local User Interfaces, Second International
  Conference on Usability and Internationalization, (UI-HCII)}}
  \emph{(\bibinfo{series}{Lecture Notes in Computer Science},
  Vol.~\bibinfo{volume}{4560})}. \bibinfo{publisher}{Springer},
  \bibinfo{pages}{259--265}.
\newblock


\bibitem[\protect\citeauthoryear{Snover, Dorr, Schwartz, Micciulla, and
  Makhoul}{Snover et~al\mbox{.}}{2006}]%
        {snover2006study}
\bibfield{author}{\bibinfo{person}{Matthew Snover}, \bibinfo{person}{Bonnie
  Dorr}, \bibinfo{person}{Richard Schwartz}, \bibinfo{person}{Linnea
  Micciulla}, {and} \bibinfo{person}{John Makhoul}.}
  \bibinfo{year}{2006}\natexlab{}.
\newblock \showarticletitle{A study of translation edit rate with targeted
  human annotation}. In \bibinfo{booktitle}{\emph{Proceedings of association
  for machine translation in the Americas}}, Vol.~\bibinfo{volume}{200}.
  Cambridge, MA.
\newblock


\bibitem[\protect\citeauthoryear{Solt}{Solt}{2011}]%
        {solt2011vagueness}
\bibfield{author}{\bibinfo{person}{Stephanie Solt}.}
  \bibinfo{year}{2011}\natexlab{}.
\newblock \showarticletitle{Vagueness in quantity: Two case studies from a
  linguistic perspective}.
\newblock \bibinfo{journal}{\emph{Understanding vagueness. logical,
  philosophical and linguistic perspectives}} (\bibinfo{year}{2011}),
  \bibinfo{pages}{157--174}.
\newblock


\bibitem[\protect\citeauthoryear{Somers}{Somers}{2005}]%
        {goodfor05}
\bibfield{author}{\bibinfo{person}{Harold~L. Somers}.}
  \bibinfo{year}{2005}\natexlab{}.
\newblock \showarticletitle{Round-trip Translation: What Is It Good For?}. In
  \bibinfo{booktitle}{\emph{Proceedings of the Australasian Language Technology
  Workshop, {ALTA} 2005, Sydney, Australia, December 10-11, 2005}}.
  \bibinfo{publisher}{Australasian Language Technology Association},
  \bibinfo{pages}{127--133}.
\newblock


\bibitem[\protect\citeauthoryear{Srivastava, Labutov, and Mitchell}{Srivastava
  et~al\mbox{.}}{2018}]%
        {ACL18quantification}
\bibfield{author}{\bibinfo{person}{Shashank Srivastava}, \bibinfo{person}{Igor
  Labutov}, {and} \bibinfo{person}{Tom~M. Mitchell}.}
  \bibinfo{year}{2018}\natexlab{}.
\newblock \showarticletitle{Zero-shot Learning of Classifiers from Natural
  Language Quantification}. In \bibinfo{booktitle}{\emph{Proceedings of the
  56th Annual Meeting of the Association for Computational Linguistics, {ACL}
  2018, Melbourne, Australia, July 15-20, 2018, Volume 1: Long Papers}}.
  \bibinfo{publisher}{Association for Computational Linguistics},
  \bibinfo{pages}{306--316}.
\newblock


\bibitem[\protect\citeauthoryear{Stock}{Stock}{2010}]%
        {semanticRelation2010}
\bibfield{author}{\bibinfo{person}{Wolfgang~G. Stock}.}
  \bibinfo{year}{2010}\natexlab{}.
\newblock \showarticletitle{Concepts and semantic relations in information
  science}.
\newblock \bibinfo{journal}{\emph{J. Assoc. Inf. Sci. Technol.}}
  \bibinfo{volume}{61}, \bibinfo{number}{10} (\bibinfo{year}{2010}),
  \bibinfo{pages}{1951--1969}.
\newblock


\bibitem[\protect\citeauthoryear{Sun, Zhang, Harman, Papadakis, and Zhang}{Sun
  et~al\mbox{.}}{2020}]%
        {transrepair20}
\bibfield{author}{\bibinfo{person}{Zeyu Sun}, \bibinfo{person}{Jie~M. Zhang},
  \bibinfo{person}{Mark Harman}, \bibinfo{person}{Mike Papadakis}, {and}
  \bibinfo{person}{Lu Zhang}.} \bibinfo{year}{2020}\natexlab{}.
\newblock \showarticletitle{Automatic testing and improvement of machine
  translation}. In \bibinfo{booktitle}{\emph{{ICSE} '20: 42nd International
  Conference on Software Engineering, Seoul, South Korea, 27 June - 19 July,
  2020}}, \bibfield{editor}{\bibinfo{person}{Gregg Rothermel} {and}
  \bibinfo{person}{Doo{-}Hwan Bae}} (Eds.). \bibinfo{publisher}{{ACM}},
  \bibinfo{pages}{974--985}.
\newblock


\bibitem[\protect\citeauthoryear{Szymanik}{Szymanik}{2016}]%
        {Szymanik16quantifiers}
\bibfield{author}{\bibinfo{person}{Jakub Szymanik}.}
  \bibinfo{year}{2016}\natexlab{}.
\newblock \bibinfo{booktitle}{\emph{Quantifiers and Cognition - Logical and
  Computational Perspectives}}. \bibinfo{series}{Studies in linguistics and
  philosophy}, Vol.~\bibinfo{volume}{96}.
\newblock \bibinfo{publisher}{Springer}.
\newblock
\showISBNx{978-3-319-28747-8}


\bibitem[\protect\citeauthoryear{Tiedemann}{Tiedemann}{2012}]%
        {Newscommentary}
\bibfield{author}{\bibinfo{person}{J{\"{o}}rg Tiedemann}.}
  \bibinfo{year}{2012}\natexlab{}.
\newblock \showarticletitle{Parallel Data, Tools and Interfaces in {OPUS}}. In
  \bibinfo{booktitle}{\emph{Proceedings of the Eighth International Conference
  on Language Resources and Evaluation, {LREC} 2012, Istanbul, Turkey, May
  23-25, 2012}}, \bibfield{editor}{\bibinfo{person}{Nicoletta Calzolari},
  \bibinfo{person}{Khalid Choukri}, \bibinfo{person}{Thierry Declerck},
  \bibinfo{person}{Mehmet~Ugur Dogan}, \bibinfo{person}{Bente Maegaard},
  \bibinfo{person}{Joseph Mariani}, \bibinfo{person}{Jan Odijk}, {and}
  \bibinfo{person}{Stelios Piperidis}} (Eds.). \bibinfo{publisher}{European
  Language Resources Association {(ELRA)}}, \bibinfo{pages}{2214--2218}.
\newblock


\bibitem[\protect\citeauthoryear{V{\'a}zquez, Scherrer, Virpioja, and
  Tiedemann}{V{\'a}zquez et~al\mbox{.}}{2021}]%
        {vazquez2021helsinki}
\bibfield{author}{\bibinfo{person}{Ra{\'u}l V{\'a}zquez}, \bibinfo{person}{Yves
  Scherrer}, \bibinfo{person}{Sami Virpioja}, {and} \bibinfo{person}{J{\"o}rg
  Tiedemann}.} \bibinfo{year}{2021}\natexlab{}.
\newblock \showarticletitle{The Helsinki submission to the AmericasNLP shared
  task}. In \bibinfo{booktitle}{\emph{Proceedings of the First Workshop on
  Natural Language Processing for Indigenous Languages of the Americas}}.
  \bibinfo{pages}{255--264}.
\newblock


\bibitem[\protect\citeauthoryear{Veanes, de~Halleux, and Tillmann}{Veanes
  et~al\mbox{.}}{2010}]%
        {Rex10}
\bibfield{author}{\bibinfo{person}{Margus Veanes}, \bibinfo{person}{Peli de
  Halleux}, {and} \bibinfo{person}{Nikolai Tillmann}.}
  \bibinfo{year}{2010}\natexlab{}.
\newblock \showarticletitle{Rex: Symbolic Regular Expression Explorer}. In
  \bibinfo{booktitle}{\emph{Third International Conference on Software Testing,
  Verification and Validation, {ICST}}}. \bibinfo{publisher}{{IEEE} Computer
  Society}, \bibinfo{pages}{498--507}.
\newblock


\bibitem[\protect\citeauthoryear{Voita, Sennrich, and Titov}{Voita
  et~al\mbox{.}}{2019}]%
        {context-aware-acl19}
\bibfield{author}{\bibinfo{person}{Elena Voita}, \bibinfo{person}{Rico
  Sennrich}, {and} \bibinfo{person}{Ivan Titov}.}
  \bibinfo{year}{2019}\natexlab{}.
\newblock \showarticletitle{When a Good Translation is Wrong in Context:
  Context-Aware Machine Translation Improves on Deixis, Ellipsis, and Lexical
  Cohesion}. In \bibinfo{booktitle}{\emph{Proceedings of the 57th Conference of
  the Association for Computational Linguistics, (ACL)}}.
  \bibinfo{publisher}{Association for Computational Linguistics},
  \bibinfo{pages}{1198--1212}.
\newblock


\bibitem[\protect\citeauthoryear{Wang and Stolee}{Wang and Stolee}{2018}]%
        {wild18}
\bibfield{author}{\bibinfo{person}{Peipei Wang} {and}
  \bibinfo{person}{Kathryn~T. Stolee}.} \bibinfo{year}{2018}\natexlab{}.
\newblock \showarticletitle{How well are regular expressions tested in the
  wild?}. In \bibinfo{booktitle}{\emph{Proceedings of the 2018 {ACM} Joint
  Meeting on European Software Engineering Conference and Symposium on the
  Foundations of Software Engineering, {ESEC/SIGSOFT} {FSE}}}.
  \bibinfo{publisher}{{ACM}}, \bibinfo{pages}{668--678}.
\newblock


\bibitem[\protect\citeauthoryear{Wang, Zheng, Liu, Zhang, Zeng, Deng, Yang, He,
  and Xie}{Wang et~al\mbox{.}}{2019a}]%
        {DBLP:conf/dsn/WangZLZZDYHX19}
\bibfield{author}{\bibinfo{person}{Wenyu Wang}, \bibinfo{person}{Wujie Zheng},
  \bibinfo{person}{Dian Liu}, \bibinfo{person}{Changrong Zhang},
  \bibinfo{person}{Qinsong Zeng}, \bibinfo{person}{Yuetang Deng},
  \bibinfo{person}{Wei Yang}, \bibinfo{person}{Pinjia He}, {and}
  \bibinfo{person}{Tao Xie}.} \bibinfo{year}{2019}\natexlab{a}.
\newblock \showarticletitle{Detecting Failures of Neural Machine Translation in
  the Absence of Reference Translations}. In \bibinfo{booktitle}{\emph{49th
  Annual {IEEE/IFIP} International Conference on Dependable Systems and
  Networks, {DSN} (Industry Track) 2019, Portland, OR, USA, June 24-27, 2019}}.
  \bibinfo{publisher}{{IEEE}}, \bibinfo{pages}{1--4}.
\newblock


\bibitem[\protect\citeauthoryear{Wang, Zheng, Liu, Zhang, Zeng, Deng, Yang, He,
  and Xie}{Wang et~al\mbox{.}}{2019b}]%
        {wechat19}
\bibfield{author}{\bibinfo{person}{Wenyu Wang}, \bibinfo{person}{Wujie Zheng},
  \bibinfo{person}{Dian Liu}, \bibinfo{person}{Changrong Zhang},
  \bibinfo{person}{Qinsong Zeng}, \bibinfo{person}{Yuetang Deng},
  \bibinfo{person}{Wei Yang}, \bibinfo{person}{Pinjia He}, {and}
  \bibinfo{person}{Tao Xie}.} \bibinfo{year}{2019}\natexlab{b}.
\newblock \showarticletitle{Detecting Failures of Neural Machine Translation in
  the Absence of Reference Translations}. In \bibinfo{booktitle}{\emph{49th
  Annual {IEEE/IFIP} International Conference on Dependable Systems and
  Networks, {DSN} (Industry Track)}}. \bibinfo{publisher}{{IEEE}},
  \bibinfo{pages}{1--4}.
\newblock


\bibitem[\protect\citeauthoryear{Wright, Gaskell, and O'Muircheartaigh}{Wright
  et~al\mbox{.}}{1994}]%
        {wright1994much}
\bibfield{author}{\bibinfo{person}{Daniel~B Wright}, \bibinfo{person}{George~D
  Gaskell}, {and} \bibinfo{person}{Colm~A O'Muircheartaigh}.}
  \bibinfo{year}{1994}\natexlab{}.
\newblock \showarticletitle{How much is ‘quite a bit’? Mapping between
  numerical values and vague quantifiers}.
\newblock \bibinfo{journal}{\emph{Applied Cognitive Psychology}}
  \bibinfo{volume}{8}, \bibinfo{number}{5} (\bibinfo{year}{1994}),
  \bibinfo{pages}{479--496}.
\newblock


\bibitem[\protect\citeauthoryear{Yamashita and Ishida}{Yamashita and
  Ishida}{2006}]%
        {DBLP:conf/cscw/YamashitaI06}
\bibfield{author}{\bibinfo{person}{Naomi Yamashita} {and} \bibinfo{person}{Toru
  Ishida}.} \bibinfo{year}{2006}\natexlab{}.
\newblock \showarticletitle{Effects of machine translation on collaborative
  work}. In \bibinfo{booktitle}{\emph{Proceedings of the 2006 {ACM} Conference
  on Computer Supported Cooperative Work (CSCW)}}. \bibinfo{publisher}{{ACM}},
  \bibinfo{pages}{515--524}.
\newblock


\bibitem[\protect\citeauthoryear{Zhang, Utiyama, Sumita, Neubig, and
  Nakamura}{Zhang et~al\mbox{.}}{2018}]%
        {DBLP:conf/naacl/ZhangUSNN18}
\bibfield{author}{\bibinfo{person}{Jingyi Zhang}, \bibinfo{person}{Masao
  Utiyama}, \bibinfo{person}{Eiichiro Sumita}, \bibinfo{person}{Graham Neubig},
  {and} \bibinfo{person}{Satoshi Nakamura}.} \bibinfo{year}{2018}\natexlab{}.
\newblock \showarticletitle{Guiding Neural Machine Translation with Retrieved
  Translation Pieces}. In \bibinfo{booktitle}{\emph{Proceedings of the 2018
  Conference of the North American Chapter of the Association for Computational
  Linguistics: Human Language Technologies, {NAACL-HLT}}}.
  \bibinfo{publisher}{Association for Computational Linguistics},
  \bibinfo{pages}{1325--1335}.
\newblock


\bibitem[\protect\citeauthoryear{Zhao, Dua, and Singh}{Zhao
  et~al\mbox{.}}{2018}]%
        {DBLP:conf/iclr/ZhaoDS18}
\bibfield{author}{\bibinfo{person}{Zhengli Zhao}, \bibinfo{person}{Dheeru Dua},
  {and} \bibinfo{person}{Sameer Singh}.} \bibinfo{year}{2018}\natexlab{}.
\newblock \showarticletitle{Generating Natural Adversarial Examples}. In
  \bibinfo{booktitle}{\emph{6th International Conference on Learning
  Representations, {ICLR} 2018, Vancouver, BC, Canada, April 30 - May 3, 2018,
  Conference Track Proceedings}}. \bibinfo{publisher}{OpenReview.net}.
\newblock
\urldef\tempurl%
\url{https://openreview.net/forum?id=H1BLjgZCb}
\showURL{%
\tempurl}


\bibitem[\protect\citeauthoryear{Zhong, Guo, Yang, Peng, Xie, Lou, Liu, and
  Zhang}{Zhong et~al\mbox{.}}{2018a}]%
        {semregex18}
\bibfield{author}{\bibinfo{person}{Zexuan Zhong}, \bibinfo{person}{Jiaqi Guo},
  \bibinfo{person}{Wei Yang}, \bibinfo{person}{Jian Peng}, \bibinfo{person}{Tao
  Xie}, \bibinfo{person}{Jian{-}Guang Lou}, \bibinfo{person}{Ting Liu}, {and}
  \bibinfo{person}{Dongmei Zhang}.} \bibinfo{year}{2018}\natexlab{a}.
\newblock \showarticletitle{SemRegex: {A} Semantics-Based Approach for
  Generating Regular Expressions from Natural Language Specifications}. In
  \bibinfo{booktitle}{\emph{Proceedings of the 2018 Conference on Empirical
  Methods in Natural Language Processing {EMNLP}}}.
  \bibinfo{publisher}{Association for Computational Linguistics},
  \bibinfo{pages}{1608--1618}.
\newblock


\bibitem[\protect\citeauthoryear{Zhong, Guo, Yang, Xie, Lou, Liu, and
  Zhang}{Zhong et~al\mbox{.}}{2018b}]%
        {arewethere18}
\bibfield{author}{\bibinfo{person}{Zexuan Zhong}, \bibinfo{person}{Jiaqi Guo},
  \bibinfo{person}{Wei Yang}, \bibinfo{person}{Tao Xie},
  \bibinfo{person}{Jian{-}Guang Lou}, \bibinfo{person}{Ting Liu}, {and}
  \bibinfo{person}{Dongmei Zhang}.} \bibinfo{year}{2018}\natexlab{b}.
\newblock \showarticletitle{Generating Regular Expressions from Natural
  Language Specifications: Are We There Yet?}. In \bibinfo{booktitle}{\emph{The
  Workshops of the The Thirty-Second {AAAI} Conference on Artificial
  Intelligence}} \emph{(\bibinfo{series}{{AAAI} Workshops},
  Vol.~\bibinfo{volume}{{WS-18}})}. \bibinfo{publisher}{{AAAI} Press},
  \bibinfo{pages}{791--794}.
\newblock


\bibitem[\protect\citeauthoryear{Zhou, Xiang, and Chen}{Zhou
  et~al\mbox{.}}{2016}]%
        {ZhouXC16}
\bibfield{author}{\bibinfo{person}{Zhiquan Zhou}, \bibinfo{person}{Shaowen
  Xiang}, {and} \bibinfo{person}{Tsong~Yueh Chen}.}
  \bibinfo{year}{2016}\natexlab{}.
\newblock \showarticletitle{Metamorphic Testing for Software Quality
  Assessment: {A} Study of Search Engines}.
\newblock \bibinfo{journal}{\emph{{IEEE} Trans. Software Eng.}}
  \bibinfo{volume}{42}, \bibinfo{number}{3} (\bibinfo{year}{2016}),
  \bibinfo{pages}{264--284}.
\newblock


\bibitem[\protect\citeauthoryear{Zhou and Sun}{Zhou and Sun}{2018}]%
        {MT4MT}
\bibfield{author}{\bibinfo{person}{Zhi~Quan Zhou} {and} \bibinfo{person}{Liqun
  Sun}.} \bibinfo{year}{2018}\natexlab{}.
\newblock \showarticletitle{Metamorphic Testing for Machine Translations:
  {MT4MT}}. In \bibinfo{booktitle}{\emph{Proceedings of the 25th Australasian
  Software Engineering Conference (ASWEC)}}. \bibinfo{publisher}{{IEEE}
  Computer Society}, \bibinfo{pages}{96--100}.
\newblock


\end{thebibliography}

\end{document}